\documentclass[journal]{IEEEtran}
\ifCLASSINFOpdf
\else
  \usepackage[dvips]{graphicx}
\fi

\newcommand{\dif}{{\rm d}}
\newcommand{\dvol}{{\rm d}^3{\bf r}}
\newcommand{\dsur}{{\rm d}{\bf s}}

\usepackage{color}
\usepackage{tikz}
\usetikzlibrary{intersections}
\usepackage{graphicx,amsmath,multirow}
\usepgflibrary{arrows} 
\usepgflibrary[arrows] 
\usetikzlibrary{arrows} 
\usetikzlibrary[arrows] 

\usepackage{fancyhdr}
\renewcommand{\footskip}{20 pt}

\begin{document}
%
\title{Computation of Losses in HTS Under the Action of Varying Magnetic Fields and Currents}
%
%
%

\author{Francesco Grilli, Enric Pardo,~\IEEEmembership{Member,~IEEE}, Antti Stenvall, Doan N. Nguyen, Weijia Yuan, and Fedor G\"om\"ory,~\IEEEmembership{Member,~IEEE} 
\thanks{F. Grilli is with the Institute for Technical Physics, Karlsruhe Institute of Technology, Karlsruhe, Germany. E-mail contact: francesco.grilli@kit.edu. E. Pardo and F. G\"om\"ory are with the Institute of Electrical Engineering, Slovak Academy of Sciences, Bratislava, Slovak Republic. A. Stenvall is with the Tampere University of Technology, Tampere, Finland. D. N. Nguyen is with the Los Alamos National Laboratory, Los Alamos, NM, USA. W. Yuan is with the University of Bath, Bath, UK.}
\thanks{The authors would like to acknowledge the following financial support: 
Helmholtz-University Young Investigator Grant VH-NG-617 (F. Grilli);
Structural Funds (SF) of the European Union (EU) through the Agency for the SF of the EU from the Ministry of Education, Science, Research and Sport of the Slovak Republic (contract number 26240220028) (E. Pardo); 
Academy of Finland, project number 131577, the Foundation for Technology Promotion in Finland, the Emil Aaltonen Foundation (A. Stenvall). Los Alamos National Laboratory LDRD grant 20120603ER (D. Nguyen).}
\thanks{Manuscript received on September 24, 2012.}}

\markboth{IEEE TRANSACTIONS ON APPLIED SUPERCONDUCTIVITY}%
{Shell \MakeLowercase{\textit{et al.}}: Bare Demo of IEEEtran.cls for Journals}
%

\maketitle

\begin{abstract}
Numerical modeling of superconductors is widely recognized as a powerful tool for interpreting experimental results, understanding physical mechanisms and predicting the performance of high-temperature superconductor (HTS) tapes, wires and devices. This is especially true for ac loss calculation, since a sufficiently low ac loss value is imperative to make these materials attractive for commercialization. In recent years, a large variety of numerical models, based on different techniques and implementations, have been proposed by researchers around the world, with the purpose of being able to estimate ac losses in HTSs quickly and accurately. This article presents a literature review of the methods for computing ac losses in HTS tapes, wires and devices. Technical superconductors have a relatively complex geometry (filaments, which might be twisted or transposed, or layers) and consist of different materials. As a result, different loss contributions exist. In this paper, we describe the ways of computing such loss contributions, which include hysteresis losses, eddy current losses, coupling losses, and losses in ferromagnetic materials. We also provide an estimation of the losses occurring in a variety of power applications.
\end{abstract}



%

\section{Introduction}
%
%
%
%
\IEEEPARstart{T}{he} capability of computing ac losses in high-temperature superconductors (HTS) is very important for designing and manufacturing marketable devices such as cables, fault current limiters, transformers and motors. As a matter of fact, in many cases the prospected ac loss value is too high to make the application attractive on the market and numerical calculations can help to find solutions for reducing the losses.

In the past decades several analytical models for computing the losses in superconductor materials have been developed. In general, these models can provide the loss value for a given geometry in given working conditions by means of relatively simple formulas. While these models are undoubtedly fast and very useful for a basic understanding of the loss mechanisms and for predicting the loss value for a certain number of geometries and tape arrangements, they have several important limitations, which restrict their usefulness for an accurate estimation of the ac losses in real HTS devices. Numerical models, on the other hand, can overcome these limitations and can simulate geometries and situations of increasing complexity. This increased computing capability comes at the price of more complex software implementation and longer computation times. Two other papers of this special issue are specifically dedicated to reviewing both the analytical and numerical models that can be found in the literature and to point out their strengths and limitations. This contribution focuses on methods for computing ac losses by means of analytical and numerical techniques.

\begin{figure}[t!]
\centering
\includegraphics[width=\columnwidth]{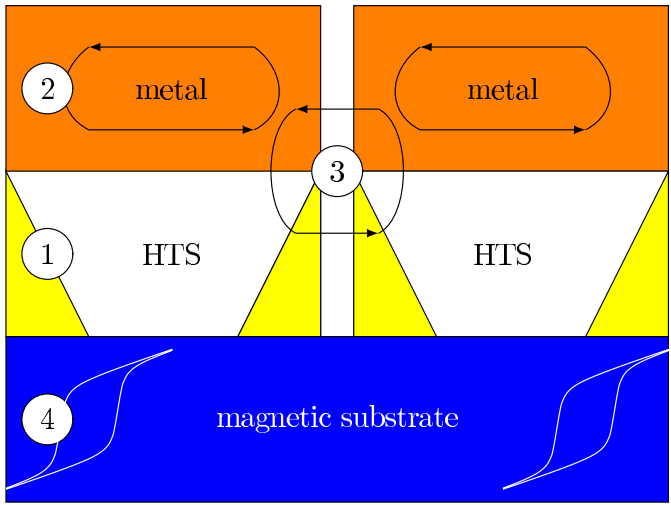}
\caption{\label{fig:loss_contributions_fil}Schematic illustration of the different loss contributions in a  technical HTS conductor, e.g. ReBCO coated conductor (tape's cross-section displayed): 1) Hysteresis losses in the superconductor; 2) Eddy current losses in the normal metal stabilizer; 3) Coupling losses between filaments (e.g. through conducting paths due to imperfections of the striation process); 4) Ferromagnetic losses in the substrate.}
\end{figure}

Technical superconductors consist of different materials beside the superconductor itself (metal, isolating buffers, magnetic materials, etc.), all of which, depending on the operating conditions, can give a significant contribution to the total losses. These loss contributions can be divided into four categories, summarized in the following list and schematically illustrated in Fig.~\ref{fig:loss_contributions_fil}.
\begin{enumerate}
\item{Hysteresis losses -- caused by the penetration of the magnetic flux in the superconducting material;}
\item {Eddy current losses -- caused by the currents induced by a magnetic field and circulating in the normal metal parts of a superconducting tape;}
\item{Coupling losses -- caused by the currents coupling two or more superconducting filaments via normal metal regions separating them;}
\item{Ferromagnetic losses -- caused by the hysteresis cycles in magnetic materials.}
\end{enumerate}

The paper is structured as follows: sections from~\ref{sec:hysteresis} to \ref{sec:ferromagnetic} focus each on a different loss contribution, as in the list above. Section~\ref{sec:power} gives the loss estimation for various power applications. The two appendices contain the details of calculations used in section~\ref{sec:JHA}.

\section{Hysteresis losses}\label{sec:hysteresis}
This section is divided into different parts, as follows. First we describe the two most commonly used models for describing a superconductor with the purpose of ac loss computation; then we explain how to solve the electromagnetic quantities; afterwards, we illustrate how to compute the ac losses, once the electromagnetic quantities are known; finally, we show the solutions for several particular cases relevant for applications.
\subsection{Models for the superconductor}
\label{sec:modelSC}
The reason why hard superconductors are able to carry large electrical currents is that the magnetic field (penetrating type II superconductors in the form of super-current vortices each carrying the same amount of magnetic flux) is pinned in the volume of the superconducting material. Due to this mechanism, the vortices do not move under the action of the Lorentz force pushing them in the direction perpendicular to both  the flow of electrical current and the magnetic field.  Also a gradient in the density of vortices is reluctant to any rearrangement. Thus, once a dc current is established in a hard superconductor, it will persist also after switching the driving voltage off (resistance-less circulation of an electrical current). However, in ac regime the vortices must move to follow the change of the magnetic field: the pinning force represents an obstacle and superseding it is an irreversible process. The accompanied dissipation is called the hysteresis loss in hard superconductors. 

It is not easy to link the interaction between a current vortex (and the involved pinning centers) and the material's properties that can be used in electromagnetic calculations. However, in the investigation of low-temperature superconductors (LTS) like NbTi or $\rm Nb_3Sn$, it was found that for practical purposes one can obtain very useful predictions utilizing the phenomenological description introduced by Bean~\cite{Bean:PRL62, London:PL63}, commonly known as critical state model (CSM).
The model is valid on a macroscopic scale that neglects the details of electrical current distribution in individual vortices and replaces it with an average taken over a large number of vortices. The critical state model states that in any (macroscopic) part of a hard superconductor one can find either no electrical current, or a current with density $J$ equal to the so-called critical current density, $J_c$.  

In the original formulation,  $J_c$  is constant and it fully characterizes  the properties of the material in the processes of magnetic field variation. In addition it was formulated for bodies of high symmetry such as infinitely long cylinders and slabs. Its value is controlled by the magnetic history, according to these two general principles:
\begin{itemize}
\item no current flows in the regions not previously penetrated by the flux vortices;
\item in the rest of superconductor the flux vortices arrange with a gradient of density that could be of different direction but always the same magnitude.
\end{itemize}
Since the penetration of vortices is accompanied by a change of local magnetic flux density $B$ that is directly proportional to the density of vortices, the mathematical formulation of the principle of critical state with constant $J_c$ is as follows:
\begin{align}\label{eq:csm1}
\left| J \right| = \left\{ {\begin{array}{*{20}c}
    {0\quad {\rm{in}}\;{\rm{regions}}\;{\rm{where}}\;B = 0}  \\
    {J_c \quad {\rm{elsewhere}}}.  \\
\end{array}} \right.
\end{align}
One should note that the direction of the  current density is not defined by this formula. Fortunately, many problems of practical importance can be simplified to a 2-D formulation, for example in devices made of straight superconducting tapes or wires, which can be considered infinitely long. The advantage of a 2-D formulation of the critical state is that the current density is always parallel to the longitudinal direction. Therefore, also in this section we assume that the task is to find a 2-D distribution. 

\begin{figure}[t!]
\centering
\includegraphics[width=\columnwidth]{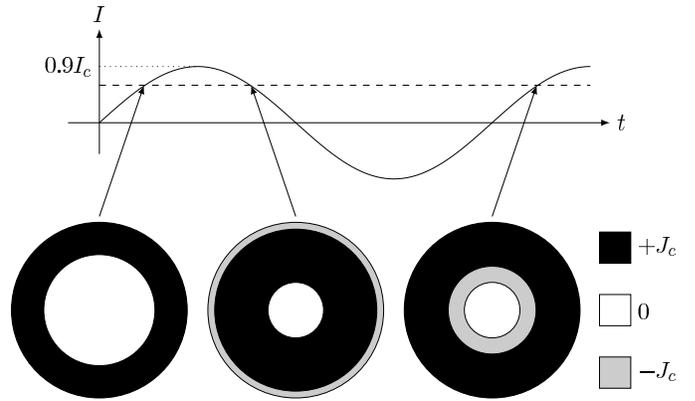}
\caption{\label{fig:fedor_csm1}Current density distribution in the cross-section of a round superconducting wire carrying a current $0.6I_c$, taken at different instants of a sinusoidal signal of amplitude $0.9I_c$.}
\end{figure}

We now illustrate the use of (\ref{eq:csm1}) for the problem of a round wire carrying a sinusoidal transport current of amplitude $0.9I_c$. The wire's critical current $I_c$ is simply obtained by multiplying the critical current density $J_c$ by the wire's cross-section. Figure~\ref{fig:fedor_csm1} shows the current density distributions when the transport current equals $0.6I_c$, taken at three different instants of the sinusoidal signal.
The significant difference of the current density distribution is due to the ``history'' of the arrangement of the flux vortices in the superconductor during the sinusoidal signal.

The solutions shown in Fig.~\ref{fig:fedor_csm1} are derived by utilizing~(\ref{eq:csm1}) and by considering that, due to the circular symmetry and according to Maxwell equations, in the region with $B=0$ also $J=0$ (no current can flow without generating a magnetic field). In his subsequent work~\cite{Bean:RMP64}, Bean transformed his formulation using the fact that any change of magnetic field is linked to the appearance of an electrical field, $E$. Then he stated that in a hard superconductor there exists a limit of the macroscopic current density it can carry, and any electromotive force, however small, will induce this current to flow. This formulation of critical state can be written as:
\begin{align}\label{eq:csm2}
\left| J \right| = \left\{ {\begin{array}{*{20}c}
   {0\quad {\rm{in}}\;{\rm{regions}}\;{\rm{where}}\;E = 0\;{\rm in}\;{\rm all}\;{\rm past}\;{\rm history}}  \\
   {J_c \quad {\rm{elsewhere}}}.  \\
\end{array}} \right.
\end{align}
Interestingly, the most general $J(E)$ relation of the critical state model for infinitely long conductors is~\cite{Bossavit:TMAG94}
\begin{equation}\label{eq:csm3}
\left \{ 
\begin{array}{l}
|J| \le J_c \qquad {\rm if }\ E=0 \\
|J|=J_c \qquad {\rm otherwise}.
\end{array}
\right .
\end{equation}
For conductors of finite thickness, this equation is equivalent to (2) [39]. The generalization above is essential for thin strips, which present regions where $0 < |J| < J_c$. In addition, it is also useful for numerical purposes.

Establishing the correspondence between the current density and the electrical field is a standard technique in electromagnetic calculations. On the other hand, (\ref{eq:csm2}) is multi-valued because for $|J|=J_c$ any value of $E$ is possible. The value of $E$ is determined by the electromagnetic history of the whole sample. This hinders the direct use of this $E-J$ relation in calculations. 
Nevertheless, with this approach the principal analytical formulas for the estimation of ac losses in LTS devices have been derived~\cite{Norris:JPDAP70, Halse:JPDAP70, Campbell:Cryo82, Wilson83, Carr83} and early numerical techniques proposed~\cite{Ashkin:JAP79}. The critical state model still represents the first approximation that allows predicting important features of HTS materials and devices, such as current density and magnetic field profiles.

Thanks also to the rapid development of computer technology in both computing power and price affordability, we can now include various refinements of the $E-J$ relation, which in certain cases are quite important. These include:
\begin{enumerate}
\item  thermal activation leading to substantial flux creep~\cite{Yeshurun:PRL88};
\item  dependence of $J_c$ on the amplitude and orientation of the magnetic field;
\item spatial variation of $J_c$ (e.g. due to non-uniformities related to the manufacturing processes).
\end{enumerate}
The first item of this list is of fundamental importance because it provides a new formulation of the model for the electromagnetic properties of hard superconductors. The superconductor's behavior described by~(\ref{eq:csm1}) or (\ref{eq:csm2}) is independent of the time derivative of the electromagnetic quantities -- this e.g. means that any part of the time axis shown in the upper part of Fig.~\ref{fig:fedor_csm1} could be stretched or compressed without influencing the resulting distributions. Only the {\it existence} of an electrical field, $E$, but not its magnitude, matters. On the other hand, thermal activation is a process happening on a characteristic time scale. Thus, if thermal activation is taken into account, the current density distribution depends on the rate of change -- in the case illustrated in Fig.~\ref{fig:fedor_csm1} on the frequency -- and on the shape of the waveform of the transport current.

\begin{figure}[t!]
\centering
\includegraphics[width=\columnwidth]{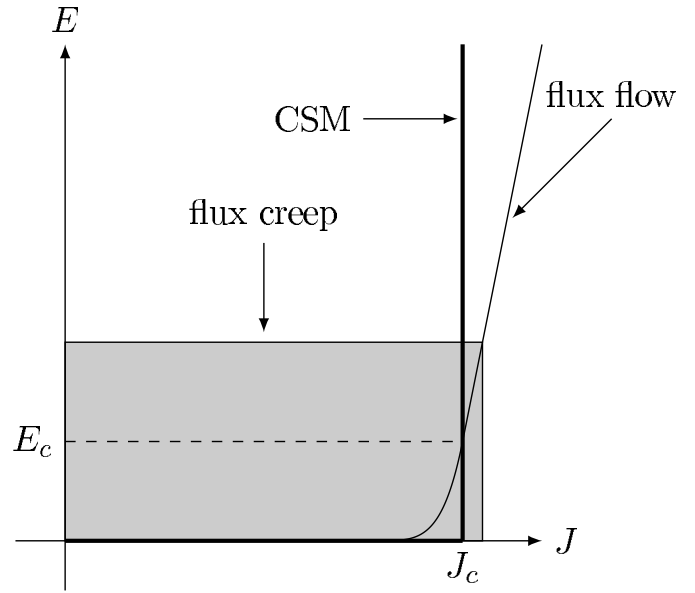}
\caption{\label{fig:fedor_csm2}
Typical $E(J)$ characteristics for HTS (thin line). Differently from the critical state model (thick line, equation~(\ref{eq:csm3})) there is always ${\rm d}E/{\rm d}J >0$ (not visible at low currents because of the used linear scale) caused by significant flux creep. The critical state model is a good approximation in the shaded area. It does not consider the flux-flow regime that would prevail at electrical fields much higher than $E_c$.}
\end{figure}

Experimental observations of strongly non-linear current-voltage characteristics of hard superconductors (e.g. the work of Kim et al.~\cite{Kim:PR65}) led to a formulation where the link between the current density and the electrical field is expressed as
\begin{equation}\label{eq:power-law}
J = J_c \left( {\frac{E}{{E_c }}} \right)^{1/n} 
\end{equation}
where $E_c$ is the characteristic electrical field (usually set equal to $\rm 10^{-4}~V/m$) that defines the current density $J_c$, and the power exponent $n$ characterizes the steepness of the current-voltage curve. The asymptotic behavior for $n \rightarrow \infty$ leads in practice to the same dependence between current density and electrical field as in (\ref{eq:csm2}). Nevertheless, there is one substantial difference: in (\ref{eq:power-law}) $J$ depends on the {\it actual} value of $E$ at the same given instant. This feature enables the incorporation of hard superconductors in electromagnetic calculations by considering it an electrically conductive and non-magnetic  material with a conductivity that depends on the electrical field:
\begin{equation}
\sigma _{SC}  = \frac{J}{E} = \frac{{J_c }}{{E_c^{1/n} }}E^{\frac{{1 - n}}{n}}.
\end{equation}
This description is valid until the current density  reaches the level causing a  microscopic driving force on the vortices that overcomes the pinning force. Then the movement  of vortices enters the so-called flux flow regime, which can be defined by a differential resistivity as~\cite{Kim:PR65}
\begin{equation}
\rho _{FF}  = \frac{{\Delta E}}{{\Delta J}} \approx \rho _n \frac{B}{{B_{c2} }},
\end{equation}
where $\rho_n$  is the normal state resistivity and $B_{c2}$ is the upper critical magnetic field. The different regimes and the interval where the critical state model is applicable are schematically illustrated in Fig.~\ref{fig:fedor_csm2}. 

In the calculation of critical current or ac losses in HTS materials it is desirable to take into account also the dependence of the critical current density on the magnetic field and its orientation. In the case of Bi-2223 multifilamentary tapes the use of a modified Kim's formula~\cite{Kim:PR65} was quite successful in reproducing experimental data~\cite{Gomory:SST07}. In the so-called elliptical approximation $J_c$ depends on the  components of the magnetic field parallel and perpendicular to the flat face of the tape by means of the anisotropy parameter $\gamma$  (with $\gamma>1$):
\begin{equation}\label{eq:JcB}
J_c (B_\perp  ,B_\parallel ) = \frac{{J_{c0} }}{{\left( {1 + \frac{{\sqrt {B_ \perp ^2  + {B_\parallel^2 }/{\gamma ^2 }} }}{{B_0 }}} \right)^\beta  }}.
\end{equation}
The cut-off value at low fields and the rapidity of the $J_c$ reduction are given by $B_0$ and $\beta$, respectively. 

An alternative way to express the asymmetric scaling of $J_c$ with respect to the perpendicular and parallel magnetic field components is by introducing the angular dependence of the characteristic field $B_0$~\cite{Majoros:SST01}. With the introduction of artificial pinning centers in the superconductor, the simple description given by the four-parameter expression~(\ref{eq:JcB}) is no longer adequate; more elaborated expressions are necessary to describe the experimentally observed behavior of the samples~\cite{Pardo:SST11}.

\begin{figure}[t!]
\centering
\includegraphics[width=\columnwidth]{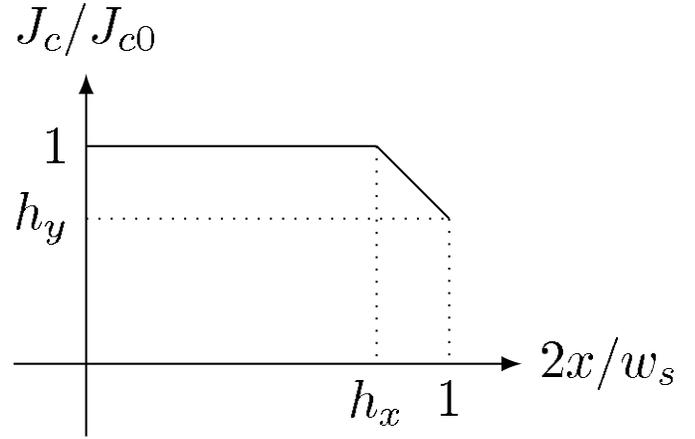}
\caption{\label{fig:fedor_csm3}Simplified model describing the non-uniformity of the current  density across the tape's width (in the direction of the coordinate $x$) by the piece-wise linear profile of $J_c(x)$ defined by  (\ref{eq:Jcx}). The tape width is $w_s$, the profile is symmetric with respect to the tape center $x = 0$.}
\end{figure}

The issue of possible variations of the superconductor's critical current density with respect to spatial coordinates is quite important in the case of HTS conductors. The longitudinal uniformity is assessed by the determination of critical current variation along the conductor length. Since the loss at full penetration is proportional to the critical current, the change of ac losses due to longitudinal non-uniformity is only marginal for commercial tapes, which typically present a fluctuation of the critical current below 10\%. On the contrary, the non-uniformity across the tape's width, investigated in detail for Bi-2223 multifilamentary tapes in~\cite{Pashitski:APL95, Grasso:PhysC95}, has been proposed as explanation for anomalous loss behavior of coated conductor tapes~\cite{Tsukamoto:SST05, Nishioka:TAS05, Grilli:TAS07a}. This effect is of particular importance for low-loss configurations like bifilar coils used in resistive fault current limiters. Due to the manufacturing process, the critical current density is lower at the tape's edges than in the center, where it is usually quite uniform. As a first approximation to take this non-uniformity into account, a symmetric piece-wise linear profile as that shown in Fig.~\ref{fig:fedor_csm3} can be used. Its mathematical form is
\begin{equation}\label{eq:Jcx}
J_c (x) = J_{c0}  \times \left\{ {\begin{array}{*{20}c}
   {1\quad \quad \quad \quad \quad \quad \quad \quad {\rm if}{\kern 1pt} \quad x \le h_x \frac{{w_S }}{2}}  \\
   {1 - \left( {1 - h_y } \right)\frac{{2\left| {\frac{x}{{w_S }}} \right| - h_x }}{{1 - h_x }}\quad {\rm if}{\kern 1pt} \quad x > h_x \frac{{w_S }}{2}}.  \\
\end{array}} \right.\quad 
\end{equation}

\subsection{How to solve the electromagnetic quantities}
\label{sec:JHA}

The main step to calculate the ac losses in superconductors is to solve the electromagnetic state variable. Depending on the simulation method, the state variable can be the current density ${\bf J}(t,{\bf r})$, the magnetic field ${\bf H}(t,{\bf r})$, the pair composed of the current vector potential and the magnetic scalar potential ${\bf T}(t,{\bf r})$-${\Omega}(t,{\bf r})$ \cite{Carpenter:PIEE77,Preston:TMAG82}, or the vector potential ${\bf A}(t,{\bf r})$, where $t$ and ${\bf r}$ are the time and the position vector, respectively. Additionally, the scalar potential $\phi(t,{\bf r})$ may be an extra variable. For formulations using the vector potential, the gauge is usually the Coulomb's gauge (see Appendix \ref{s.A}). All the methods reviewed in this article assume that the displacement current is negligible, i.e. that $|{\bf J}|\gg|\partial{\bf D}/\partial t|$ where ${\bf D}$ is the displacement vector.

Once the state variable is solved for the time-varying excitation, the power loss can be calculated by using the methods in section \ref{s.hystloss}. The time-varying excitation can be cyclic or not, such as a ramp increase in dc magnets. It can be a transport current $I(t)$, an applied field ${\bf H}_a(t,{\bf r})$ (which can be non-uniform and with an orientation varying in time -- see section \ref{s.hystsolutions} for particular solutions), or a combination of them. 

\subsubsection{Cross-section methods}
\label{s.modsection}

In the following, we outline the methods that solve the state variable for an arbitrary combination of current and applied field. Other methods optimised for particular situations and their results are summarized in section \ref{s.hystsolutions}. These methods solve the cross-section of infinitely long conductors and bodies with cylindrical symmetry, and hence they are mathematically 2-D problems. Simulations for bi-dimensional surfaces and 3-D bodies are treated in \ref{s.mod2D} and \ref{s.mod3D}, respectively.

A superconductor with a smooth $E(J)$ relation can be solved by calculating different electromagnetic quantities. Brandt's method \cite{Brandt:PRB96,Brandt:PRB98a}, generalized for simultaneous currents and fields by Rhyner and Yazawa {\it et al.} \cite{Rhyner:PhysC98,Yazawa:PhysC98}, solves $\bf J$ and $\phi$ in the superconductor volume. Alternatively, finite element method (FEM) models solve ${\bf H}$ \cite{Stavrev:TAS02,Kajikawa:TAS03,Grilli:TAS07a,Hong:SST06}, ${\bf T}$ and $\Omega$ \cite{Amemiya:PhysC98b,Enomoto:PhysC04,Grilli:TAS05a,Stenvall:SST10a} or ${\bf A}$ and $\phi$ \cite{Hauser:TMAG97,Tonsho:TAS03,Stenvall:SST10b} in a finite volume containing the superconductor and the surrounding air, forming the simulation volume. All FEM models require setting the boundary conditions of the state variable on the boundary of the simulation volume. For most of the methods, the boundary conditions are for asymptotic values, hence the simulation volume is much larger than the superconducting one. However, for the implementation of Kajikawa {\it et al.}, the boundary conditions only require to contain the superconductor volume \cite{Kajikawa:TAS03}, thus reducing the simulation volume and computation time. In this sense, Brandt's method is also advantageous because it simulates the superconductor volume only. FEM simulations are suitable for commercial software, which simplifies the implementation and analysis. The computing time for all the methods mentioned above dramatically increases for a power-law $E(J)$ relation with a high exponent. Variational methods may also be applied to solve problems with a smooth $E(J)$ relation \cite{Bossavit:TMAG94,Maslouh:TAS97,Prigozhin:TAS97,Prigozhin:SST11}, although they are mostly used for the critical state model.

Critical-state calculation models are ideal to simulate superconductors with a large exponent in the power-law $E(J)$ relation. Moreover, they are usually faster than the simulations with a smooth $E(J)$ relation also for relatively low $n$. This improvement in speed may justify the sacrifice in accuracy caused by using the critical-state approximation. However, the critical state model cannot describe relaxation effects or over-current situations.

Most of the existing methods for a general current and applied field imply variational methods. They were firstly proposed by Bossavit \cite{Bossavit:TMAG94}, although their most important contribution is from Prigozhin, who developed the $J$ formulation \cite{Prigozhin:JCP96,Prigozhin:TAS97}. Later, Badia {\it et al.} proposed the $\bf H$ formulation \cite{Badia:PRB02}. An alternative numerical implementation to minimize the functional in the $\bf J$ formulation and to set the current constrains is the Minimum Magnetic Energy Variation (MMEV) (this method sets the current constraints directly, not through the electrostatic potential). The general formulation is described in \cite{Pardo:SST07,Souc:SST09}, although it was firstly introduced in \cite{Sanchez:PRB01} for magnetization cases. Actually, the early works on MMEV implicitly assume that the current fronts penetrate monotonically in a half cycle. This occurs in some practical cases but not in general~\cite{Pardo:SST07}.

In addition, FEM models with the $A$ formulation can solve the critical state situation, as shown by G\"om\"ory {\it et al.} \cite{Gomory:SST09,Gomory:SST10}, by means of the $A$ variation method. This technique is inspired by Campbell's $A$ formulation for superconductors with both reversible and irreversible contributions to pinning \cite{Campbell:SST07,Campbell:SST09}. Another method based on the $A$ formulation was developed by Barnes {\it et al.} \cite{Barnes:SST99}, who merged the FEM technique with the critical-state constraint on $J$, $|J|\le J_c$. Methods assuming a smooth $E(J)$ relation can also approximate the critical-state model by considering an $E(J)$ relation with pice-wise linear segments or a power law with a high exponent \cite{Takeda:Cryo94,Maslouh:TAS97,Kajikawa:TAS03}. 

All the methods above, both for a smooth $E(J)$ relation and the critical-state model, can take the magnetic field dependence of the critical current density into account and, in principle, also a position dependence.

\subsubsection{On the ${\bf A}-\phi$ and ${\bf T}-\Omega$ formulations}

For completeness, we next discuss the meaning of the scalar and vector potentials for the ${\bf A}-\phi$ and ${\bf T}-\Omega$ formulations. Further details can be found in a dedicated paper in this issue~\cite{IEEE-TAS_paper3}.
First, we consider the ${\bf A}-\phi$ formulation and later the ${\bf T}-\Omega$ one.

In the Coulomb's gauge, the scalar potential $\phi$ is the electrostatic potential created by the electrical charges (Appendix \ref{s.A}). In practice, the scalar potential needs to be taken into account in the ${\bf A}-\phi$ formulation in the following cases. 

First, when there is a net transport current. In the Coulomb's gauge the current distribution creates a non-zero $\partial{\bf A}/{\partial t}$ at the current-free core (see figure 2 for a typical current distribution), while the electrical field ${\bf E}=-\partial{\bf A}/{\partial t}-\nabla \phi$ vanishes. Therefore, it is necessary a certain $\nabla \phi$ in order to compensate $\partial{\bf A}/{\partial t}$. One may include $\nabla \phi$ in $\bf A$ as a gauge, ${\bf A}'={\bf A}+\nabla \phi$, where $\bf A$ is in th e Coulomb's gauge. As a result, ${\bf A}'=0$ at the current-free core. However, $\phi$ has still to be taken into account in the boundary conditions far away from the superconductor. Explicitly, ${\bf A}'\approx \nabla \phi$ and ${\bf A}'\approx -\mu_0I\ln{r}/(2\pi)+\nabla \phi$ for 3-D and infinitely long 2-D geometries, respectively, while in the Coulomb's gauge the boundary conditions are ${\bf A}=0$ and ${\bf A}'\approx -\mu_0I\ln{r}/(2\pi)$, respectively ($r$ is the radial distance from the superconductor electrical center). In this case, $\phi$ is the electrostatic potential generated by the current source in order to keep a constant current $I$. The physical source of $\phi$ are surface electrical charges at both ends of the superconductor wire and on the lateral surfaces. 

Second, $\phi$ is necessary when a multi-filamentary body with isolated filaments is submitted to a changing applied magnetic field. In order to ensure a zero net current in all filaments, there needs to appear a scalar potential $\phi$ between the filaments. The reason is the following. The electric field in the current free zone is zero (see figure 10 for a typical current distribution in this case). In order achieve zero electric field there, it is necessary $\nabla \phi$ because $\partial{\bf A}/\partial t$ is non-zero at the current-free zone because there is a net flux between the filaments, which equals to $\oint {\bf A}\cdot{\rm d}{\bf r}$. A similar situation appears when the filaments are connected by a normal conducting material.

Finally, the scalar potential is necessary for the general 3-D case and also 2-D thin surfaces. A general body submitted to a magnetic field created by a round winding produces $\bf A$ with cylindrical symmetry (in the Coulomb's gauge). Then, if the superconducting body does not match this cylindrical symmetry, for example because it has corners, there needs to be a certain $\nabla \phi$ that ``corrects" the direction of ${\bf E}$ in order to allow current parallel to the superconductor surface~\cite{Campbell:SST09}. This is clear when $\bf E$ and $\bf J$ are parallel but a certain $\nabla \phi$ should also be present for an arbitrary tensor relation between $\bf E$ and $\bf J$.

The meaning of the $\bf T$ and $\Omega$ potentials is the following. The $\bf T$ potential is such that ${\bf J}=\nabla \times {\bf T}$. Although it follows that ${\bf J}=\nabla \times {\bf H}$, $\bf T$ is not necessarily $\bf H$. Since $\nabla \times ({\bf T}-{\bf H})=0$, the quantity ${\bf T}-{\bf H}$ can be written as a gradient of a scalar function, ${\bf T}-{\bf H}=\nabla \Omega$ and ${\bf H}={\bf T}-\nabla\Omega$. Note that $\bf T$ is subjected to gauge invariance. That is, ${\bf T}'={\bf T}-\nabla \varphi$ generates the same current density ${\bf J}=\nabla \times {\bf T}=\nabla \times {\bf T}'$, where $\varphi$ is any scalar function. As a consequence, the meaning of $\Omega$ depends on the gauge of $\bf T$. When the gauge function $\varphi$ is zero, $\nabla \cdot{\bf H}=-\nabla^2\Omega$; and thence $\Omega$ is the magnetic scalar potential due to the magnetic pole density. For this gauge, $\bf T$ is the magnetic field created by the currents. In addition, other gauges are also used in computations \cite{Carpenter:PIEE77,Preston:TMAG82}.

\subsubsection{Methods for 2D surfaces}
\label{s.mod2D}

An obviously mathematically 2-D geometry are thin films with finite length. 

The advantage of flat 2-D surfaces is that the current density can be written as a function of a scalar field as ${\bf J}(x,y)=\nabla \times {\hat{\bf z}}g(x,y)$, where the $z$ direction is perpendicular to the surface \cite{Brandt:PRL95}. Actually, $g$ is the $z$ component of the current potential $\bf T$, since ${\bf J}={\nabla \times {\bf T}}$. It could also be regarded as an effective density of magnetic dipoles $\bf M$. For a smooth $E(J)$ relation, this quantity can be computed by numerically inverting the integral equation for the vector potential in equation (\ref{ACou}) \cite{Brandt:PRL95} or the Biot-Savart law \cite{Vestgarden:PRB08}. Naturally, FEM models based on the ${\bf T}$ formulation are useful for these surfaces \cite{Amemiya:JAP06}, as are those based on the ${\bf H}$ formulation \cite{Grilli:Cryo12,Zhang:SST12}. Variational principles can solve the situation of the critical-state model \cite{Prigozhin:JCP98,Navau:JAP08}. 

A more complex situation is when the surfaces are bent in three dimensions, like twisted wires, spiral cables, and Roebel cables (this latter with important simplifications). The models developed so far use the FEM with the ${\bf T}$ \cite{Amemiya:JAP06,Takeuchi:SST11, Nii:SST12} and ${\bf H}$ formulations \cite{Zhang:SST12}.

\subsubsection{Methods for fully 3-D problems}
\label{s.mod3D}

We define a fully 3-D situation as the case when the body of study is a 3-D domain and the problem cannot be reduced to a mathematically 2-D domain. Thus, we exclude problems with cylindrical symmetry, as well as thin surfaces with a 3-D bending (section \ref{s.mod2D}). 

For a fully 3-D situation (and flat surfaces in 3-D bending), the magnetic field is not perpendicular to the current density in certain parts of the sample. Then, the modeling should take flux cutting~\cite{Campbell:AP72} into account. This can be done by using a dependence of $J_c$ on the angle between the electric field, $\bf E$, and the current density, $\bf J$, \cite{Badia:PRB02,RomeroSalazar:APL03,BadiaMajos:PRB09}. Experimental evidence reveals an elliptical dependence \cite{Clem:SST11}.

The implementation of 3-D models can be done as follows. In principle, variational methods can describe the critical state model in 3-D \cite{Bossavit:TMAG94}, although this has not yet been brought into practice. FEM models have successfully calculated several 3-D situations, although in all cases the model assumes that $\bf J$ is parallel to $\bf E$ and a  smooth $E(J)$ relation. These calculations are for the ${\bf T}-\Omega$ \cite{Grilli:TAS05a}, ${\bf A}-\phi$ \cite{Grilli:TAS05a,Lousberg:SST09,Campbell:SST09} or $\bf H$ \cite{Grilli:Cryo12,Zhang:SST12} formulations. An alternative model based on the $\bf H$ formulation that does not use the Garlekin method for FEM can be found in \cite{Zehetmayer:SST06}.

References for examples of 3-D calculations are in section \ref{s.hystsol3D}.


\subsection{How to calculate ac losses, once the electromagnetic quantities are known}
\label{s.hystloss}

This section deduces several formulas to calculate ac losses in superconductors, once the electromagnetic quantities are known. These can be obtained by any method described in section \ref{sec:JHA}. 

First, section \ref{s.pJE} justifies that the density of local dissipation is $p={\bf J}\cdot{\bf E}$. Section \ref{s.QCSM} outlines how to calculate the ac losses for the critical-state model. Section \ref{s.Qmag} presents simplified formulas for the magnetization ac losses. Finally, section \ref{s.Qsource} discusses how to calculate the ac losses from the energy delivered by the power sources.

\subsubsection{Local instantaneous dissipation in type-II superconductors}
\label{s.pJE}

The local instantaneous power dissipation, $p$, in a type-II superconductor is
\begin{equation}           
\label{pJE}
p={\bf J}\cdot{\bf E}.
\end{equation}
In the following, we summarize the mechanisms that lead to this relation \cite{Kim:PR65,Blatter:RMP94,Brandt:RPP95}. The driving force per unit length, ${\bf F}_d$, on a vortex is ${\bf F}_d={\bf J}\times{\bf \Phi}_0$,
where $|{\bf \Phi}_0|$ is the vortex flux quantum, with value $\pi\hbar/e$, and the direction of ${\bf \Phi}_0$ follows the direction of the vortex. Then, the local rate of work per unit volume, $p$, on the vortices is
\begin{equation}
\label{pPhivJ}
p=({\bf J}\times{\bf \Phi}_0n)\cdot {\bf v}=({\bf \Phi}_0n\times {\bf v})\cdot {\bf J},
\end{equation}
where $n$ and ${\bf v}$ are the vortex density and velocity, respectively. From electromagnetic analysis, it can be shown that the moving vortices with a speed $\bf v$ create an electric field (Appendix \ref{s.Evortex})
\begin{equation}
\label{BvEp}
{\bf E}={\bf \Phi}_0n\times {\bf v}={\bf B}\times{\bf v}.
\end{equation}
The fields $\bf E$ and $\bf B$ above are the average in a volume containing several vortices. Finally, by inserting equation (\ref{BvEp}) into (\ref{pPhivJ}) we obtain equation (\ref{pJE}). 

Equation (\ref{pJE}) also applies to normal conductors (see \cite{Jackson}, page 258). Then, this equation is also valid to calculate the loss due to the eddy and coupling currents. However, the loss mechanism in normal conductors and superconductors is very different. In normal conductors, ${\bf E}$ creates work on the charge carriers, induces movement in them and this movement manifests in ${\bf J}$. In contrast, in superconductors ${\bf J}$ is the quantity that exerts work and induces movement on the vortices. This vortex movement manifests in ${\bf E}$.

For a given local $\bf J$, $\bf E$ does not change in time, which means that $\bf v$ is constant. This is because several dissipation mechanisms cause a viscous flow of the vortices \cite{Kim:PR65,Brandt:RPP95}. For power applications, the relation between $E$ and $J$ does not depend on the speed of the change of the current density. K\"otzler {\it et al.} experimentally showed that the response to ac fields for a wide range of frequencies, 3 mHz to 50 MHz, is consistent with a unique $E(J)$ relation of the studied sample \cite{Kotzler:PRB94,Brandt:RPP95}.

\subsubsection{Application to the critical state model}
\label{s.QCSM}

The application of equation (\ref{pJE}) for the local loss dissipation requires to know $\bf J$ and $\bf E$. Once the state variable is known, the current density can be found by means of one of the following relations: $\nabla \times {\bf H}={\bf J}$, $\nabla \times {\bf T}={\bf J}$, $\nabla \times ( \nabla \times {\bf A})={\bf J}$. Then, by assuming a smooth ${\bf E}({\bf J})$ relation for the superconductor, $\bf E$ is simply obtained from that relation. For the critical-state model, the relation between $\bf E$ and $\bf J$ is multi-valued, and hence $\bf E$ cannot be found in this way. 

For the critical state model, $\bf E$ is calculated from
\begin{equation}
\label{EAphi}
{\bf E}=-\partial {\bf A}/\partial t-\nabla \phi .
\end{equation}
The vector potential in the Coulomb's gauge is obtained from $J$ by using equation (\ref{ACou}). For this gauge, $\phi$ becomes the electrostatic potential (Appendix \ref{s.A}). The electrostatic potential can be solved as a variable during the computation of ${\bf J}$ \cite{Prigozhin:TAS97,Prigozhin:SST11} or can be calculated from ${\bf A}$ from the following relation \cite{Pardo:SST07,Gomory:SST10}
\begin{equation}
\label{phiA0t}
\nabla \phi (t) = - \frac{\partial {\bf A}(t,{\bf r}_0)}{\partial t} ,
\end{equation} 
where ${\bf r}_0$ is a position in the superconductor where $|{\bf J}|<J_c$, and hence ${\bf E}=0$, see equation~(\ref{eq:csm3}). In long conductors and bodies with cylindrical symmetry, the single component of $\bf J$, $J$, vanishes at the current-free core and at the current fronts (boundaries between $J=+J_c$ and $J=-J_c$). In the thin-film approximation, $|J|<J_c$ in the under-critical region. It should be noted that the time derivative in equation (\ref{phiA0t}) is the {\em partial} time derivative, and thus the time dependence of ${\bf r}_0$ should not be taken into account in evaluating the time derivative. From equations (\ref{pJE}) and (\ref{phiA0t}), the instantaneous power loss density is
\begin{equation}
\label{PJAtPhi}
p(t,{\bf r})={ J}\left[\frac{\partial { A}(t,{\bf r}_0)}{\partial t}-\frac{\partial { A}(t,{\bf r})}{\partial t}\right] .
\end{equation}

The equation above can be simplified for monotonic penetration of the current fronts, constant $J_c$ and infinitely long conductors or cylindrical symmetry. The deduction below follows the concepts from \cite{Norris:JPDAP70,Rhyner:PhysC02,Pardo:SST07}. In the following, we present a deduction assuming any combination of transport current and uniform applied field proportional to each other. With monotonic penetration of the current fronts, the current density in the reverse and returning curves in an ac cycle, $J_{\rm rev}$ and $J_{\rm ret}$, are related to the one in the initial curve, $J_{\rm in}$, \cite{Bean:RMP64,London:PL63,Brandt:EPL93} as
\begin{eqnarray}
\label{JinJrevJret}
J_{\rm rev}(u) & = & J_{{\rm in},m}-2J_{\rm in}[(1-u)/2] \nonumber \\
J_{\rm ret}(u) & = & -J_{{\rm in},m}+2J_{\rm in}[(1+u)/2] ,
\end{eqnarray}
where $J_{{\rm in},m}$ is the current distribution at the end of the initial curve and $u$ is defined as follows. For a transport current, $u(t)$ is such that $I(t)=u(t)I_m$, where $I_m$ is the amplitude of the transport current. If there is no transport current but the applied field is created by a winding with a linear current-field characteristic, we define $u(t)$ as ${\bf H}_a(t,{\bf r})=u(t){\bf H}_m({\bf r})$, where ${\bf H}_m({\bf r})$ is the maximum applied field at position $\bf r$. This relation is also valid for the simultaneous action of a transport current and an in-phase applied field. For monotonic current penetration, there is always at least one point, called kernel, where $J=0$ in the whole cycle. Then, ${\bf r}_0$ in equation (\ref{PJAtPhi}) becomes constant. Integrating the loss power density of equation (\ref{PJAtPhi}) in the volume and integrating by parts for the time variable, we find that
\begin{equation}
\label{QAJtA}
Q=-\oint \dif t \int_V \frac{\partial { J}}{\partial t}  [ { A}(t,{\bf r}_0) - { A}(t,{\bf r}) ] \dvol,
\end{equation}
where we took into account that $J$ in each position changes once in each half-cycle and that this change is instantaneous from $J_c$ to $-J_c$ (or opposite). The time derivative of this discontinuous dependence is $\partial J(t,{\bf r})/\partial t=-2J_c\delta[t-t_c({\bf r})]$, where $\delta$ is Dirac's delta and $t_c({\bf r})$ is the time of the shift of $J$ at the position $\bf r$. The time integration results in
\begin{equation}
\label{QJcAtc}
Q=4\int_V J_{{\rm in},m}({\bf r})  \{ { A}[t_c({\bf r}),{\bf r}_0] - { A}[t_c({\bf r}),{\bf r}] \} \dvol .
\end{equation}
The relation of equation (\ref{JinJrevJret}) is also valid for the vector potential $A$ in the Coulomb's gauge. Then, 
\begin{equation}
\label{AinArevAret}
A({\bf r},t)=\pm A_{{\rm in},m}({\bf r})\mp 2A_{\rm in}[{\bf r},(1\mp u(t))/2], 
\end{equation}
where the top sign is for the reverse curve and the bottom one is for the returning one. The current density shifts its sign at the current front, see Fig.~\ref{fig:fedor_csm1}. For the initial curve, the current front encloses the flux-free region, where ${\bf B}={\bf J}=0$ and hence $A$ is uniform. Then, $A$ at the current front is the same as in the kernel. By taking this into account and using equation (\ref{AinArevAret}), the loss per cycle from equation (\ref{QJcAtc}) becomes
\begin{equation}
\label{QJcA}
Q=4 \int_V {J_{{\rm in},m}({\bf r})}  [ A_{{\rm in},m}({\bf r}_0) - A_{{\rm in},m}({\bf r}) ] \dvol .
\end{equation}
For a pure transport current, the equation above turns into
\begin{equation}
\label{QJcAtran}
Q=4J_c \int_V [ A_{{\rm in},m}({\bf r}_0) - A_{{\rm in},m}({\bf r}) ] \dvol
\end{equation}
because at the zone where $J_{{\rm in},m}$ vanishes $A_{{\rm in},m}({\bf r})=A_{{\rm in},m}({\bf r}_0)$. Equation (\ref{QJcAtran}) corresponds to the original formula from Norris \cite{Norris:JPDAP70}. The general equation (\ref{QJcA}) reduces to the one given by Rhyner for a uniform applied field \cite{Rhyner:PhysC02}.

\subsubsection{Magnetization loss}
\label{s.Qmag}

When the only time-dependent excitation is the applied magnetic field ${\bf H}_a(t,{\bf r})$, the expression for the ac losses per cycle can be simplified. Next, we take into account the ac losses in the {\em superconductor only}. In the following, we deduce several formulas for the magnetization loss. From equation (\ref{pJE}), the density of the instantaneous power dissipation in the whole superconductor is
\begin{equation}
\label{pJsE}
p={\bf J}_s\cdot {\bf E},
\end{equation}
where ${\bf J}_s$ is the current distribution in the superconductor. Then, the instantaneous power dissipation is $P=\int_V {\bf J}_s\cdot {\bf E} \ \dvol$, where $V$ is any volume containing the superconductor and $\dvol$ is the volume differential. The loss per cycle $Q$ is the time integration in one cycle.

Using $\nabla\times{\bf H}_s={\bf J}_s$, $\nabla\times{\bf E}=-\partial{\bf B}/\partial t$, the differential vector relation $(\nabla\times{\bf H}_s)\cdot{\bf E}=\nabla\cdot({\bf H}_s\times{\bf E})+(\nabla\times{\bf E})\cdot{\bf H}_s$ and the divergence theorem we find that
\begin{equation}
\label{PPoynting}
P=- \int_V {\bf H}_s\cdot\frac{\partial{\bf B}}{\partial t} \ \dvol +\oint_{\partial V} {\bf H}_s\times{\bf E} \cdot \dsur,
\end{equation}
where ${\bf H}_s$ is the magnetic field created by the superconductor currents, $\partial V$ is the surface of the volume $V$ and $\dsur$ is the surface differential. When the volume $V$ contains not only the superconductor but also the sources of the applied magnetic field (the current or magnetic pole densities) and $V$ approaches infinity, the second term of equation (\ref{PPoynting}) vanishes. This can be seen as follows. Let us take $V$ as a sphere of radius $r$. With large $r$, the largest contribution to the fields is their dipolar component. Then, $|{\bf H}_s|\sim 1/r^3$ and since ${\bf E}=-\partial{\bf A}/\partial t-\nabla\phi$ with $|{\bf A}|\sim 1/r^2$ and $|\nabla \phi|\sim 1/r^3$ it follows $|{\bf H}_s\times{\bf E}|\sim 1/r^5$. Since the surface only increases as $r^2$, the integration vanishes with $r\to\infty$. Then, the instantaneous power loss is
\begin{equation}
\label{PHBp}
P=- \int_{V_\infty} {\bf H}_s\cdot\frac{\partial{\bf B}}{\partial t} \ \dvol,
\end{equation}
where $V_\infty$ is the whole space. Note that equation ($\ref{PHBp}$) for the total power loss is general, although its integrand is no longer the local rate of power dissipation.

The ac losses per cycle, $Q$, as a response to a periodic ac field can be further simplified. If the whole system is in the void, ${\bf B}=\mu_0{\bf H}=\mu_0({\bf H}_a+{\bf H}_s)$ and ${\bf H}_s\cdot{\partial{\bf B}}/{\partial t}=\mu_0[{\bf H}_s\cdot {\partial{\bf H}_a}/{\partial t}+(1/2)\partial|{\bf H}_s|^2/\partial t]$. When integrating in one cycle the second term trivially vanishes, resulting in
\begin{equation}
\label{QHsHa}
Q=-\mu_0\oint \dif t \int_{V_\infty} {\bf H}_s\cdot\frac{\partial{\bf H}_a}{\partial t} \ \dvol .
\end{equation}
Note that the applied field ${\bf H}_a(t,{\bf r})$ in the equation above is not necessarily uniform and it may rotate during the cycle. In addition, the volume integral in equation (\ref{QHsHa}) is not the instantaneous power loss.

For uniform ${\bf H}_a$, we can use the fact that $\int_{V_\infty} {\bf H}_s \dvol=(1/2)\int_{V_\infty} {\bf r}\times{\bf J}_s	 \dvol={\bf m}$, and hence equation (\ref{QHsHa}) becomes
\begin{equation}
\label{QmHap}
Q=-\mu_0\oint {\bf m}\cdot\frac{\partial{\bf H}_a}{\partial t} \ \dif t,
\end{equation}
where ${\bf H}_a$ may still rotate in one cycle. For an applied field in a constant direction, the ac losses per cycle are
\begin{equation}
\label{QmHa}
Q=-\mu_0\oint m_z \dif H_a,
\end{equation}
where we choose the direction of the $z$ axis as the direction of the applied field.

Next, we re-write equation (\ref{QHsHa}) in a more useful way for calculations using the ${\bf J}$ or $\bf A$ formulations. Using $\mu_0{\bf H}_s=\nabla\times{\bf A}_s$, the differential vector equality $(\nabla\times{\bf A}_s)\cdot{\bf H}_a=\nabla\cdot({\bf A}_s\times{\bf H}_a)+(\nabla\times{\bf H}_s)\cdot{\bf A}_a$ and the divergence theorem we find that
\begin{equation}
\label{QJApsur}
Q=-\oint \dif t \int_{V_\infty} {\bf J}_s\cdot\frac{\partial{\bf A}_a}{\partial t} \ \dvol +\oint_{\partial V_\infty} {\bf A}_s\times\frac{\partial{\bf H}_a}{\partial t} \cdot \dsur .
\end{equation}
Using the same arguments as for equation (\ref{PPoynting}), the term with the surface integral vanishes. Note that even if the applied field may be uniform in the superconductor, far away from its sources it has to decay with the distance. Then, equation (\ref{QJApsur}) becomes
\begin{equation}
\label{QJA}
Q=-\oint \dif t \int_{V_\infty} {\bf J}_s\cdot\frac{\partial{\bf A}_a}{\partial t} \ \dvol.
\end{equation}

For magnetic materials originated by a microscopic density of magnetic dipoles ${\bf M}$, it is practical to use formulas independent of the loss mechanism associated to the change of the local density of dipoles because they can be applied to any magnetic material. The variation of the free energy of a magnetic material is $\delta \mathcal{F}=-\int_V {\bf M}\cdot{\delta {\bf H}}_a$ (page 116 of \cite{landau}). The loss per cycle is the integration of the free energy in a cycle,
\begin{equation}
\label{QMHap}
Q=-\mu_0\oint \dif t \int_{V} {\bf M}\cdot\frac{\partial{\bf H}_a}{\partial t} \ \dvol .
\end{equation}
From this formula, it is trivial to reproduce equations (\ref{QmHap}) and (\ref{QmHa}) for uniform applied fields. This formula can also be written in terms of the total magnetic field $\bf H$, $Q=-\mu_0\oint \dif t \int_{V} {\bf M}\cdot{\partial{\bf H}}/{\partial t}$. For soft magnetic materials, $\bf M$ is practically parallel to $\bf H$. By taking this into account and neglecting the loss due to a rotation in the magnetic field, the loss per cycle is
\begin{equation}
\label{QMsoft}
Q\approx-\mu_0 \int_{V}  \oint {M}{\rm d}{H}=\mu_0 \int_{V} q(H_m,H_{\rm bias}) \ \dvol .
\end{equation}
The equation above assumes that the magnitude of the local magnetic field oscillates with amplitude $H_m$ around a value $H_{\rm bias}$. The quantity $q(H_m,H_{\rm bias})$ is the density of hysteresis loss corresponding to this amplitude, which can be experimentally obtained. If the applied magnetic field oscillates with no dc component, $H_{\rm bias}=0$ for the whole volume. The density of hysteresis loss can also be written as a function of the magnetic flux density, $B_m$ and $B_{\rm bias}$.

For infinitely long wires and tapes, the power loss and loss per cycle per unit length, $P_l$ and $Q_l$, follow equations (\ref{PHBp}),(\ref{QHsHa}),(\ref{QJA})-(\ref{QMsoft}), after the replacement of the volume integrations with the cross-section surface integrations. The same applies to equations (\ref{QmHap}) and (\ref{QmHa}) but with the replacement of the magnetic moment $\bf m$ with the magnetic moment per unit length ${\bf m}_l$.

\subsubsection{AC losses from the point of view of the power source}
\label{s.Qsource}

By continuity of the energy flow, the loss per cycle dissipated in the superconductor equals  the energy per cycle supplied by the current sources. In general, a superconductor may be under the effect of a transport current or a magnetic field created by an external winding, which implies two current sources \cite{Ashworth:PhysC99a,Vojenciak:SST06}.

If the model solves a single conductor in pure transport current or a complete superconducting coil, there is only one power source. The voltage delivered by the power source corresponds to the difference of the electrostatic potential $\Delta \phi$. Then, the ac losses per cycle in the coil are
\begin{equation}
\label{QIphi}
Q=-\oint I\Delta \phi \ \dif t .
\end{equation}
For the critical state model, the drop of the electrostatic potential in each turn can be calculated from the scalar potential from equation (\ref{phiA0t}) and by multiplying by the length of the conductor \cite{Pardo:SST08,Souc:SST09}. The total voltage drop is the sum of the drop in all the turns. 

Although equation (\ref{QIphi}) for the loss per cycle is valid, the instantaneous power dissipation in the coil not necessarily equals the instantaneous power delivered by the source, $I(t)\Delta \phi(t)$. The cause is the (non-linear) inductance of the coil. This can be seen as follows. From equations (\ref{EAphi}) and (\ref{pJsE}) the instantaneous power ac losses are
\begin{equation}
\label{PJphi}
P(t)=-\int_V \left({\bf J}_s\cdot \nabla \phi + {\bf J}_s\cdot\frac{\partial{\bf A}}{\partial t}\right) \dvol .
\end{equation}
Since there are no external sources of magnetic field, $\bf A$ in the Coulomb's gauge only depends on ${\bf J}_s$ through equation (\ref{ACou}). As a consequence, $\int_V{\bf J}_s\cdot{\partial{\bf A}}/{\partial t}\ \dvol = \int_V \partial{\bf J}_s/\partial t \cdot{\bf A}\ \dvol$. Then, $\int_V{\bf J}_s\cdot \partial{\bf A}/{\partial t}\ \dvol$ is the time derivative of the magnetic energy $U=(1/2) \int_V {\bf J}_s\cdot{\bf A}\,\dvol$. Thus, the time integral in one cycle of the second term in equation (\ref{PJphi}) vanishes. The remaining term is equation (\ref{QIphi}) because in straight conductors $\nabla \phi$ is uniform \cite{Carr:SST06b}, as well as in each turn of circular coils \cite{Pardo:SST08}.

If there is also an applied magnetic field, the total loss per cycle from equations (\ref{EAphi}) and (\ref{pJsE}) becomes
\begin{equation}
\label{PJphiAa}
Q=-\oint I\Delta \phi \ \dif t - \oint \dif t \int_{V_\infty} {\bf J}_s\cdot\frac{\partial{\bf A}_a}{\partial t} \ \dvol,
\end{equation}
where we follow the same steps as the reasoning after equation (\ref{PJphi}). The second term of equation (\ref{PJphiAa}) is the magnetization loss from equation (\ref{QJA}). That formula is equivalent to equation (\ref{QHsHa}), which reduces to equation (\ref{QmHap}) for uniform applied fields {\em in the superconductor}. Under this condition, the total loss in the superconductor is
\begin{equation}
\label{PJphiHa}
Q=-\oint I\Delta \phi \ \dif t - \mu_0\oint {\bf m}\cdot\frac{\partial{\bf H}_a}{\partial t} \ \dif t .
\end{equation}
The first term is usually called ``transport loss", while the second one is called ``magnetization loss". If the applied magnetic field and the transport current are in phase, the transport and magnetization losses correspond to the loss covered by the transport and magnetization sources, respectively. This is because the sources are only coupled inductively, and hence the energy transfer from one source to the other in one cycle vanishes. Equation (\ref{PJphiHa}) is universal, and thus it is also valid when the sources are not in phase to each other. However, in that case, the two terms in equation (\ref{PJphiHa}) cannot be always attributed to the loss covered by each source separately~\cite{Vojenciak:SST06}.


\subsection{Solutions for particular cases}
\label{s.hystsolutions}

The purpose of this section is to gather a reference list of the hysteresis loss classified by topics. The number of published articles in the field is very vast. Although we made the effort of making justice to the bulk of published works, a certain degree of omission is inevitable, and we apologize in advance to the concerned authors. Readers interested in a particular topic may go directly to the section of interest according to its title. Below, we do not regard the case of levitation. A review on that problem can be found in another article of this issue \cite{levitation}. 

\subsubsection{AC applied magnetic field}
\label{s.hystsolfield}

The response to an applied magnetic field strongly depends  on the geometry of the superconducting body, as discussed in the books by Wilson and Carr \cite{Wilson83,Carr83}.

The first works were for geometries infinitely long in the field direction because there are no demagnetizing effects. Bean introduced the critical-state model for slabs \cite{Bean:PRL62} and cylinders \cite{Bean:RMP64}, and calculated the magnetization and the ac losses in certain limits. Later, Goldfarb and Clem obtained formulas for the ac susceptibility for any field amplitude \cite{Goldfarb91,Clem91}. 
The formula for the ac losses per unit volume is
\begin{eqnarray}
\label{Xppslab}
Q_v & = & \frac{2\mu_0H_m^3}{3 H_p} \qquad {\rm if}\ H_m<H_p \nonumber \\ 
 & = & \frac{\mu_0H_p}{3}\left( { 6H_m-4H_p } \right) \qquad {\rm if}\ H_m\ge H_p \nonumber \\ 
 & \approx & 2\mu_0H_pH_m \qquad {\rm if}\ H_m\gg H_p.
\end{eqnarray}
In the equation above $H_m$ is the amplitude of the applied magnetic field and $H_p=J_cw/2$, where $w$ is the thickness of the slab. To obtain this equation we used the fact that the ac losses per unit volume, $Q_v$, are related to the imaginary part of the ac susceptibility $\chi ''$ as $Q_v=\mu_0\pi H_m^2\chi ''$. The imaginary part of the ac susceptibility (and the loss factor $Q/H_m^2$) presents a peak at an amplitude, $H_{\rm peak}$,
\begin{equation}
H_{\rm peak}=\frac{2}{3} J_c w.
\end{equation}
This formula is useful to obtain $J_c$ from magnetization ac loss measurements.

There are extensive analytical results for the critical state with a magnetic field dependence of the critical current density, $J_c(B)$, for infinite geometries. Most articles only calculate the current distribution and magnetization in slabs \cite{Fietz:PR64,Watson:JAP68}. The most remarkable analytical work for the ac losses is from Chen and Sanchez \cite{Chen:JAP91}, who studied rectangular bars. 

Comparisons between the critical state model and a power-law $E(J)$ relation are in \cite{Amemiya:TAS97} for a slab and in \cite{Chen:APL06} for a cylinder. The latter article also finds that the frequency for which the ac susceptibility from the critical state model and a power-law $E(J)$ relation are the most similar, at least for field amplitudes above that of the peak in $\chi''$. This frequency is
\begin{equation}
\label{f_c}
f_c=E_c/(2\pi\mu_0a^2J_c),
\end{equation}
where $a$ is the radius of the cylinder. This frequency is based on an extension of the scaling law by Brandt \cite{Brandt:PRB97}. This scaling law applied to $\chi''$ reads
\begin{equation}
\label{scalingXpp}
\chi''_{PL}(H_m,f) = \chi''_{PL}(H_mC^\frac{1}{n-1},Cf),
\end{equation}
where $\chi''_{PL}$ is $\chi''$ for a power-law $E(J)$ relation, $C$ is any arbitrary constant and $n$ is the exponent of the power-law. Combining the scaling law in (\ref{scalingXpp}) with equation (\ref{f_c}), one can find a relation between $\chi''$ for the critical-state model, $\chi''_{CSM}$, and that for the power-law:
\begin{eqnarray}
\label{XppPLandCSM}
\chi''_{PL}(H_m,f) & = & \chi''_{PL}\left[{H_m\left({\frac{f_c}{f}}\right)^\frac{1}{n-1},f_c}\right] \nonumber \\
& \approx & \chi''_{CSM}\left[{H_m\left({\frac{f_c}{f}}\right)^\frac{1}{n-1}}\right].
\end{eqnarray}

The first geometries with demagnetizing effects to be solved were thin strips and cylinders. For these geometries, conformal mapping techniques allowed to find analytical solutions for the critical state model. Halse and Brandt {\it et al.} obtained the case of a thin strip in a perpendicular applied field \cite{Halse:JPDAP70,Brandt:EPL93}, although an earlier work from Norris developed the conformal mapping technique for a transport current \cite{Norris:JPDAP70}. The ac losses per unit volume are
\begin{eqnarray}
\label{Qlstrip}
Q_v & = & \mu _0wJ_cH_m\left( { \frac{2}{x}\ln\cosh x-\tanh x }\right) \nonumber \\
 & \approx & \mu_0 wJ_c H_m \qquad \rm{if}\ H_m \gg H_c 
\end{eqnarray}
where $x$ is defined as $x=H_m/H_c$ with $H_c=J_cd/\pi$, and $d$ and $w$ are the thickness and width of the strip, respectively. For the same dimension $w$ and low $H_m$, the ac losses per unit volume in a slab are smaller than in a thin film. However, the limit of high applied field amplitudes is the same for the strip and the slab, $Q_v\approx \mu_0 w J_cH_m$ [from equations (\ref{Xppslab}) and (\ref{Qlstrip})]. The peak of $\chi ''$ (and $Q/H_m^2$) is at an amplitude \cite{Pardo:SST04b}
\begin{equation}
H_{\rm peak}= 0.7845 J_c d.
\end{equation}
The ac losses in a thin disk with constant $J_c$ were obtained by Clem and Sanchez \cite{Clem:PRB94}. The solutions of the current and field distributions (as well as the magnetization) are analytical, although the expression for ac losses is in integral form.

In general, to solve the geometries with intermediate thickness requires numerical techniques. The text below first summarizes the works on geometries for wires in a transverse applied field (rectangular, circular, elliptical and tubular), then several of these wires parallel to each other (multiple conductors), afterwards the case of wires under an arbitrary angle with the surface of the wires and, finally, solutions for several cases with rotational symmetry subjected to a magnetic field in the axial direction (cylinders, rings, spheres, spheroids and others).

Rectangular wires were studied in detail by Brandt, who calculated the current distribution and other electromagnetic quantities for a power-law $E(J)$ relation \cite{Brandt:PRB96}. For the critical-state model, Prigozhin computed the current distribution for the critical state model \cite{Prigozhin:TAS97} and Pardo {\it et al.} discussed the ac susceptibility and presented tables in \cite{Pardo:SST04b}.

Circular and elliptical wires are geometries often met in practice. The main articles for the critical-state model are the following. The earliest work is from Ashkin, who numerically calculated the current fronts and the ac losses in round wires \cite{Ashkin:JAP79}. Later, several authors calculated the elliptical wire by analytical approximations, either by assuming elliptical flux fronts \cite{Wilson83,Krasnov:PhysC91,Gomory:SST02a} or very low thickness \cite{Bhagwat:PhysC95}. Numerical solutions beyond these assumptions are in \cite{Prigozhin:JCP96,Gomory:SST02a,Chen:SST05}, the latter containing tables for the ac susceptibility. Actually, earlier work already solved this situation for a power-law $E(J)$ relation \cite{Amemiya:PhysC98a,Nibbio:TAS01}.

The most complete work on tubular wires is from Mawatari \cite{Mawatari:PRB11}. That article presents analytical solutions for the current and flux distribution, as well as the ac losses, in a thin tubular wire. The current distribution in a tube with finite thickness had previously been published in \cite{Prigozhin:JCP96}.

For the hysteresis loss in multiple conductors, it is necessary to distinguish between the uncoupled and fully coupled cases. These cases are the following. If several superconducting conductors (wires or tapes) are connected to each other by a normal conductor, either at the ends or along their whole length, there appear coupling currents and, as a consequence, the ac losses per cycle depend on the frequency (section \ref{sec:coupling}). The ac  losses become hysteretic not only for the limit of high inter-conductor resistance (or low frequency) but also for low resistance (or high frequency), as shown by measurements \cite{Fukumoto:APL95} and simulations \cite{Amemiya:JAP06}. These limits are the ``uncoupled" and ``fully coupled" cases that we introduced above. It is possible to simulate these two cases by setting different current constrains: for the uncoupled case the net current is zero in all the conductors, while the fully coupled one allows the current to freely distribute among all the tapes \cite{Pardo:PRB03}. For thin strips, there exist analytical solutions. Mawatari found the ac losses in vertical and horizontal arrays of an infinite number of strips \cite{Mawatari:PRB96} for the uncoupled case and Ainbinder and Maksimova solved two parallel strips for both the uncoupled and fully coupled cases in \cite{Ainbinder:SST03}. The study of an intermediate number of tapes requires numerical calculations. Simulations for vertical arrays can be found in \cite{Tebano:PhysC02,Grilli:PhysC06} for a power-law $E(J)$ relation and in \cite{Pardo:PRB03} for the critical-state model. Complicated cross-sections for the fully coupled case can be found in \cite{Stavrev:PhysC03}. Matrix arrays for both the coupled and uncoupled cases are in \cite{Pardo:PRB03,Grilli:SST10b}. A wire with uncoupled filaments is in \cite{Rostila:TAS11}. A systematic study of the initial susceptibility in multi-filamentary tapes with uncoupled filaments is in \cite{Fabbricatore:PRB00}.

All the results above for single and multiple conductors are for applied magnetic fields parallel to either the thin or wide dimension of the conductor cross-section. Solutions for intermediate angles can be found in the following articles. Mikitik and Brandt analytically studied a thin strip but they did not calculate the ac losses \cite{Mikitik:PRB04,Brandt:PRB05}. The ac losses in single conductors are in \cite{Rhyner:PhysC98,Ichiki:PhysC04,Enomoto:TAS05,Stavrev:SST05} for a power-law $E(J)$ relation and in \cite{Karmakar:PRB01,Gomory:SST10} for the critical-state model. Solutions for arrays of strips for both a power-law $E(J)$ relation and the critical-state model are in \cite{Pardo:SST12}. The latter reference also reports that the ac losses in a thin strip with a magnetic field dependence of $J_c$ may depend on both the parallel and perpendicular components of the applied field, therefore the ac loss analysis cannot be always reduced to only the variation on the perpendicular component of the applied field.

Next, we list the works for bodies with rotational symmetry. Apart from the analytical solutions in \cite{Zhu:PhysC93,Clem:PRB94} for thin disks, the main works for cylinders subjected to a uniform applied field are the following. Brandt studied in detail several electromagnetic properties of cylinders with finite thickness in \cite{Brandt:PRB98a} and the ac susceptibility in \cite{Brandt:PRB98b} for a power-law $E(J)$ relation. Later, Sachez an Navau studied the same geometry for the critical-state model \cite{Sanchez:PRB01} and D.-X. Chen {\it et al.} discussed the ac susceptibility and publish tables \cite{Chen:SST05}. For cylindrical rings, the ac susceptibility and ac losses were calculated in \cite{Brandt:PRB97} for thin rings and in \cite{Navau:PRB05} for finite thickness. Pioneering works on the current distribution and ac losses in spheres and spheroids in the critical-state model can be found in \cite{Navarro:PRB91,Navarro:SST92,Telschow:PRB94}. Finally, the current distribution for several bodies of revolution can be found in \cite{Prigozhin:JCP96,Prigozhin:TAS97}.

All the numerical models discussed in section \ref{sec:modelSC} can solve the general situation of a conductor with an arbitrary cross-section, either infinitely long or with rotational symmetry.

\subsubsection{AC transport current}

The effect of an alternating transport current in a superconducting wire (or tape) also depends on the wire geometry. 

The simplest situation is a round wire because of symmetry. London calculated this situation in the critical-state model \cite{London:PL63}. Actually, London ideated the critical-state model in parallel to Bean \cite{Bean:PRL62}. The ac losses per unit length, $Q_l$, in a round wire is
\begin{eqnarray}
\label{norrisellipse}
\frac{2\pi Q_l}{\mu_0I_c^2} & = & (2-i_m)i_m+2(1-i_m)\ln (1-i_m) \nonumber \\
& \approx & \frac{i_m^3}{3} \qquad {\rm if}\ i_m\ll 1,
\end{eqnarray}
where the expression on the left is adimensional, $i_m=I_m/I_c$ and $I_m$ is the current amplitude. Norris found that this formula is also valid for wires with elliptical cross-section (elliptical wires), based on complex mathematical arguments \cite{Norris:JPDAP70}. He also developed the technique of conformal mapping to calculate the current distribution in thin strips \cite{Norris:JPDAP70}. The resulting ac losses in a thin strip are
\begin{eqnarray}
\label{norrisstrip}
\frac{2\pi Q_l}{\mu_0I_c^2} & = & 2[(1+i_m)\ln (1+i_m) \nonumber \\
 & & +(1-i_m)\ln (1-i_m)-i_m^2] \nonumber \\
& \approx & \frac{i_m^4}{3} \qquad {\rm if}\ i_m\ll 1.
\end{eqnarray}
Later, Clem {\it et al.} calculated the magnetic flux outside the strip \cite{Clem:CJP96}. They found the key result that it is necessary to use C-shaped loops closing at least the width of the strip in order to measure properly the ac loss. Contrary to the case of an applied magnetic field, thin films cannot be simplified as a slab with critical current density penetrating in a uniform depth across the film width. The formula for a slab from Hancox \cite{Hancox:PIEE66} can only be applied to estimate the ac losses due to the current penetration from the wide surfaces of the film, also known as ``top" and ``bottom" losses. Curiously, the formula for a slab exactly corresponds to the limit of low current amplitudes in equation (\ref{norrisellipse}).

Numerical methods are usually necessary to calculate the ac losses for other geometries or an $E(J)$ relation.

For an elliptical wire, Amemiya {\it et al.} compared the ac loss for the critical-state model and a power-law $E(J)$ relation \cite{Amemiya:PhysC98a}. Later, Chen and Gu further discussed this kind of comparison for round wires in \cite{Chen:APL05}, where tables for the ac losses are presented. They also found a scaling law for the ac losses, as for the magnetization case. The scaling law for the loss per unit length, $Q_{l,PL}$, is
\begin{equation}
\frac{Q_{l,PL}}{I_m^2}(I_m,f)=\frac{Q_{l,PL}}{I_m^2}(I_mC^\frac{1}{n-1},Cf),
\end{equation}
where $I_m$ is the amplitude of the current and $C$ is an arbitrary constant. A study of the effect of the field dependence in an elliptical wire can be found in \cite{Nibbio:TAS01}.

The effect of non-homogeneity in the cross-section of a round wire and thin strip are discussed in \cite{Gomory:PhysC97} and \cite{Tsukamoto:SST05}, respectively. In summary, degraded superconductor at the edges of the conductor increases the ac loss, and {\it vice versa}. A similar effect appears if the strips are thinner close to the edges \cite{Kajikawa:SST04}.

The main results for rectangular wires with finite thickness are the following. Norris was the first to numerically calculate the current fronts and ac losses for the critical-state model \cite{Norris:JPDAP71}. Later, several authors published more complete works \cite{Fukunaga:APL98,Daumling:SST98,Pardo:SST04b}, where~\cite{Pardo:SST04b} also presents tables and a fitting formula. Results for a power-law $E(J)$ relation can be found in \cite{Gu:TAS05}.

The following articles calculate the current distribution and ac losses in multiple wires and tapes connected in parallel, and thus the current can distribute freely among the conductors (the case of wires connected in series corresponds to coils, which is discussed in section \ref{s.hystsolcoils}). The only analytical solution is for two co-planar thin strips in the critical state \cite{Ainbinder:SST03}. Always in the critical state, numerical results for Bi2223 multi-filamentary tapes with several geometries are in \cite{Fukunaga:TAS99,Oota:PhysC03}. The effect of arranging rectangular tapes in horizontal, vertical and rectangular arrays is studied in \cite{Pardo:PRB05}. A comparison between the critical-state model and a power-law $E(J)$ relation in a matrix array of coated conductors is in \cite{Grilli:SST10b}. Valuable earlier work using a power-law for several kinds of multi-filamentary wires and tapes is in \cite{Stavrev:PhysC03,Stavrev:TAS03b}.

We review cables transporting alternating current in sections \ref{s.hystsolcables} and \ref{s.hystsolroebel}.

\subsubsection{Simultaneous alternating transport current and applied field}

This section outlines the results for superconducting tapes under the action of both an alternating transport current and uniform applied field. First, we summarize the case of in-phase current and field and, afterwards, with an arbitrary phase shift.

For a current and field in phase with each other, analytical solutions only exist for slabs and strips in the critical state model. Carr solved the situation of a slab and presented a formula for the ac losses \cite{Carr:TMAG79} (the reader can find the same formula in SI units in \cite{Pardo:SST07}). Later, Brandt {\it et al.} and Zeldov {\it et al.} simultaneously obtained the magnetic field and current distributions for a thin strip \cite{Brandt:PRB93,Zeldov:PRB94} and Schonborg deduced the ac losses from them \cite{Schonborg:JAP01}. However, as discussed in the original two articles \cite{Brandt:PRB93,Zeldov:PRB94}, the formulas are only valid when the critical region penetrates monotonically in the initial curve. This is ensured at least for the high-current-low-field regime, which appears for $I\ge I_c \tanh (H_a/H_c)$ with $H_c=J_cd/\pi$. The range of applicability of the slab and strip formulas compared to numerical calculations is discussed in \cite{Pardo:SST07}. In that reference, the current penetration process in a rectangular wire with finite thickness is studied in detail. However, the earliest numerical calculations are for a power-law $E(J)$ relation \cite{Yazawa:PhysC98}. Elliptical wires with a power law are computed and discussed in detail by several authors \cite{Amemiya:PhysC98a,Zannella:TAS01,Tonsho:TAS03,Stavrev:TAS02}. The latter reference also compares different shapes: round, elliptical and rectangular wires. The main features for the thin geometry of a coated conductor is discussed in \cite{Enomoto:PhysC04}.

In electrical machines, the current and magnetic field are not always in phase with each other. Analytical solutions for an arbitrary phase shift only exist for slabs \cite{Takacs:SST07} and strips \cite{Mawatari:APL07} in the critical state model and for low applied magnetic fields. These calculations predict that the ac loss maximum  is when the alternating current and field are in phase with each other. However, simulations for slabs \cite{Kajikawa:TAS01} and elliptical wires \cite{Nguyen:JAP05,Vojenciak:SST06} show that for large applied fields, the ac loss maximum is at intermediate phase shifts for both a power-law $E(J)$ relation and the critical-state model. Experiments confirm this \cite{Nguyen:JAP05,Vojenciak:SST06}.

\subsubsection{Power-transmission cables}
\label{s.hystsolcables}

\begin{figure}[htb]
\centering
\includegraphics[width=\columnwidth]{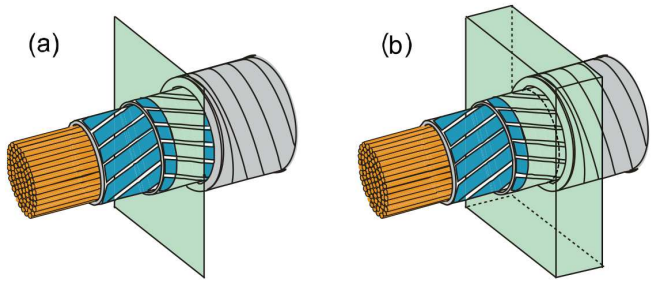}
\caption{Schematic view of a two-layer HTS power cable and of 2-D and 3-D regions to be considered for simulation. Figure taken from~\cite{Takeuchi:SST11} and reproduced with permission from IOP.}
\label{fig:power_cable}
\end{figure}

The cross-section of state-of-the-art power-transmission cables is of superconducting tapes lying on a cylindrical former, in order to minimize the ac loss. These tapes are spirally wound on the former to allow bending the cable (figure \ref{fig:power_cable}).

To estimate the loss, the cross-section is often approximated as a cylindrical shell (or monoblock) because its solution in the critical-state model is analytical \cite{Hancox:PIEE66}. However, other analytical approximations may also be applicable \cite{Vellego:SST95}.

Most of the models for the real cross-section assume that the tapes are straight, neglecting the effect of the spiral shape -- this corresponds to considering just the cross-section in Fig.~\ref{fig:power_cable}a and to assuming it extends straight along the cable's axis. The error of this approximation is negligible for single-layer cables. Moreover, thanks to symmetry, the cross-section in the computations can be reduced to a single circular sector \cite{Grilli:TAS03,Grilli:SST04,Klincok:JOPCS06}. Simulations for tapes with a ferromagnetic substrate show that its presence always increases the ac losses \cite{Miyagi:PhysC07,Amemiya:PhysC08}. In multi-layer cables, the twist pitch of the tapes strongly influences the current sharing between layers \cite{Rostila:PhysC10}. One reasonable approximation is to assume that the twist pitch is optimum, so that all the tapes carry the same current independently of the layer that they belong to \cite{Jiang:SST08b,Nakahata:SST08,Amemiya:SST10}. All the numerical calculations above are for a smooth $E(J)$ relation, although it is also possible to assume the critical-state model \cite{Inada:JOPCS08}. These computations show that non-uniformity in the tapes increases ac losses \cite{Inada:JOPCS08}. Analytical solutions are more recent than numerical ones, due to the complexity of the deductions. These have been found for cables made of thin strips in the critical-state model. They show that if the tapes are bent, and hence the cable cross-section is a tube with slits, the ac losses are lower than if the tapes are straight, forming a polygon \cite{Mawatari:SST10}.

The real spiral geometry is more complicated to compute. The first step has been to calculate the critical current \cite{Ohmatsu:TAS04}. Ac loss calculations for cables made of Bi-2223 tapes require 3-D models (see Fig.~\ref{fig:power_cable}b), resulting in a high computing complexity so that only coarse meshes are considered \cite{Nakamura:TAS05,Grilli:TAS05a}. More recent models for coated conductors improve the accuracy by reducing the problem to a mathematically 1-D \cite{Siahrang:TAS10} or 2-D \cite{Takeuchi:SST11,Grilli:Cryo12} one.	

\subsubsection{Roebel cables}
\label{s.hystsolroebel}

\begin{figure}[htb]
\centering
\includegraphics[width=0.8\columnwidth]{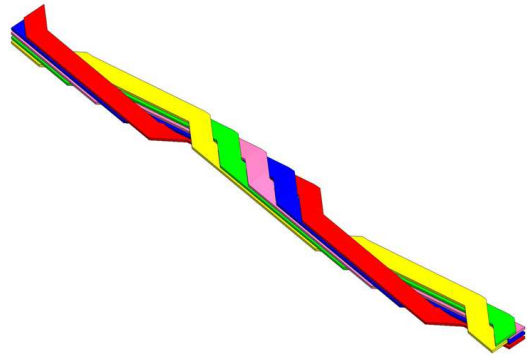}
\caption{Typical structure of a Roebel cable composed of meander-shaped strands obtained from ReBCO coated conductors. Only half of the stands are shown for clarity.}
\label{f.roebel}
\end{figure}

Roebel cables are compact transposed cables for windings that require high currents (figure \ref{f.roebel}). Thanks to transposition, coupling loss is minimized.

The complicated structure of Roebel cables can be simplified with a 2-D approximation. If the transposition length is much larger than the cable width, the loss in the crossed strands is much smaller than in the straight parts. As a consequence, the cable can be modelled as a matrix of tapes (see \cite{Terzieva:SST10,Grilli:SST10b} for more details). For an applied magnetic field, the computations distinguish between the ``uncoupled" and ``fully coupled" cases, as for multiple tapes (section \ref{s.hystsolfield}), not only if the applied field is perpendicular to the strand surface \cite{Terzieva:SST10,Grilli:SST10b} but also for an arbitrary angle \cite{Pardo:SST12}. The equivalent can also be done for the transport loss, with the interpretation that the ``uncoupled" and ``fully coupled" cases are for transposed and untransposed strands \cite{Terzieva:SST10,Grilli:SST10b,Jiang:SST11}. The ``uncoupled" case requires that the net current in all the strands is the same, while the ``fully coupled" one allows the current to freely distribute among strands.

Until now, only two articles have taken the real 3-D bending into account~\cite{Nii:SST12,Zermeno:SST13}. The work by Nii {\it et al.} is based on the $T$ formulation and assumes the strands to be infinitely thin~\cite{Nii:SST12}. The authors calculate the current density and ac loss distribution in the ``uncoupled case" for one example of transport current, applied magnetic field and a combination of both. That work shows that there appear points of high loss density close to the crossing strands. This increases the ac loss comparing to cross-sectional models. The increase is important for low ac applied fields (and no transport current) because the current penetration in the crossed strands is much larger than in the straight parts. Otherwise, cross-sectional models do not introduce an error larger than 10\% concerning the total loss. Zermeno {\it et al.} used the $H$ formulation to develop a full 3-D electromagnetic model of a Roebel cable~\cite{Zermeno:SST13}. The use of periodic boundary conditions allowed simulating a periodic cell, in this way keeping the number of degrees of freedom at a manageable level. This model too reveals high loss density regions near the crossing strands in the case of applied perpendicular field. In the transport current case, the losses are mostly localized along the long edges of the strands, which supports the possibility of using a 2-D model for this case. Results are also compared to experimental data, showing a fairly good agreement; the observed mismatch for the magnetization losses is probably due to different reasons: for example, the fact that the experimental setup does not grant the perfect uncoupling of the strands (assumed in the simulations), and that the model does not take into account the dependence of $J_c$ on the magnetic field.

\subsubsection{Coils}
\label{s.hystsolcoils}

The ac losses in the tape (or wire) that composes a superconducting coil is subjected to both a transport current and a magnetic field created by the rest of the turns. Since real coils are made of hundreds or thousands of turns, computations are complex.

Early calculations computed only the critical current, which already provided valuable information \cite{Pitel:SST02}. 

One possible simplification for the ac losses is to approximate that the effect of the whole coil in a certain turn is the same as an applied magnetic field. This applied field is computed by assuming that the current density is uniform in the rest of the turns. Afterwards, the ac losses in the turn of study are estimated by either measurements in a single tape \cite{Oomen:SST03,Kawagoe:TAS04} or by numerical calculations \cite{Tonsho:TAS04}. The problem of this approximation is that the neighboring turns shield the magnetic field from the coil \cite{Pardo:PRB03,Grilli:PhysC06}. This situation is the worst if the winding consists of stacks of pancake coils because the whole pancake shields the magnetic field, as it is the case for stacks of tapes \cite{Pardo:PRB03,Grilli:PhysC06}.

There are several ways to take into account the non-uniform current distribution in all the turns simultaneously. The first way is to assume infinite stacks of tapes, describing only the ac losses at the center of a single pancake coil, either by analytical calculations for the critical state model \cite{Muller:PhysC97b} or by numerical ones for a smooth $E(J)$ relation \cite{Brambilla:SST09}. The second is to simply solve all the turns simultaneously for single pancake coils (pancakes) \cite{Grilli:SST07,Pardo:SST08,Souc:SST09,Prigozhin:SST11, Zhang:JAP12}, double pancakes \cite{Ainslie:SST11} and stacks of pancakes \cite{Pardo:SST12}. For this, it is possible to calculate the circular geometry \cite{Grilli:SST07,Pardo:SST08,Souc:SST09,Pardo:SST12, Zhang:JAP12} or assume stacks of infinitely long tapes \cite{Prigozhin:SST11,Ainslie:SST11}. Finally, for closely packed turns it is possible to approximate the pancake cross-section as a continuous object~\cite{Yuan:SST09,Prigozhin:SST11}. Semi-analytical approaches assume that the region with critical current is the same for all the turns \cite{Claassen:APL06,Clem:SST07}. This condition can be relaxed by assuming a parabolic shape of the current front \cite{Yuan:SST09}. However, advanced numerical simulations do not require any assumption about the shape of these current fronts \cite{Prigozhin:SST11}. 

The presence of ferromagnetic materials strongly influences the ac losses in the superconductor. Magnetic diverters reduce the ac losses in pancake coils \cite{Shevchenko:PhysC98,Pardo:SST09}, while a ferromagnetic substrate increases it \cite{Ainslie:SST11,Ainslie:PhysC12}.

\subsubsection{Non-inductive windings}

Non-inductive windings are the typical configurations for resistive superconducting fault-current limiters. The reason is the low inductance and low ac losses.

Modeling of this situation starts with two parallel tapes on top of each other transporting opposite current (antiparallel tapes). For thin strips, it is possible to obtain approximated analytical formulas because the ac losses are dominated by the penetration from the edges \cite{Norris:JPDAP70}. However, if the thickness of the rectangular wire is finite, numerical computations show that for low current amplitudes, the ac losses is dominated by the flux penetration from the wide surfaces \cite{Norris:JPDAP71}. This contribution of the ac losses is called ``top and bottom'' loss~\cite{Clem:PRB08}. For practical superconductors, the aspect ratio is high (at least 15 and 1000 for Bi2223 and ReBCO tapes, respectively). For this cases, the ac losses for anti-parallel tapes is much lower than if the tapes are placed far away from each other. Most of this ac loss reduction is lost if the tapes are not well aligned \cite{Nguyen:SST09}. Degradation at the edges of the coated conductors also worsens the low-loss performance of the anti-parallel configuration \cite{Souc:SST12}. 

Bifilar windings (windings made of anti-parallel tapes) reduce even further the ac loss, either if they are made of elliptical wires \cite{Majoros:TAS07} or coated conductors \cite{Clem:PRB08,Souc:SST12}. The most favorable situation concerning ac losses is for bifilar pancake coils \cite{Souc:SST12}. For strips in this configuration, there exist valuable analytical formulas \cite{Clem:PRB08}. There also exist analytical formulas for the equivalent of a bifilar coil but with all the strips on the same plane \cite{Mawatari:APL99}. The ac losses for this configuration are larger than if the tapes are far away from each other, opposite to bifilar pancake coils.

\subsubsection{Rotating fields}
\label{s.hystsolrot}

Modeling the effect of rotating fields is very different if the magnetic field is perpendicular to the current density or not.

For infinitely long wires or tapes in transverse applied magnetic field, the local current density is always perpendicular to the magnetic field. All the numerical methods from section \ref{s.modsection} can solve this situation. The main feature of the current distribution is that it becomes periodic only after the first half cycle that follows from the initial curve, as seen for cylinders \cite{Prigozhin:JCP96}, square bars \cite{Prigozhin:JCP96} and rectangular bars \cite{Maslouh:TAS97}. This feature has an effect on the magnetization \cite{BadiaMajos:PRB07}. If the sample has been previously magnetized, a rotating field partially demagnetizes the sample after a few cycles \cite{BadiaMajos:PRB07,Vanderbemden:SST07}.

If the magnetic field is not perpendicular to the current density in part of the ac cycle, there appear flux cutting mechanisms in addition to vortex depinning, and thus the situation is more complex. This situation can be described by a double critical state model with two characteristic critical current densities, one for depinning and another one for flux cutting \cite{Clem:PRB82}. This model allows calculating analytical solutions of the whole electromagnetic process, including the ac losses, for a slab submitted to a rotating magnetic field parallel to its surface \cite{Clem:PRB84}. An extension of this model takes into account a continuous variation of the critical current density as a function of the orientation of the electric field with respect to the current density \cite{Badia:PRB02,Ruiz:SST10}. This model requires numerical methods to solve the current distribution \cite{Badia:PRB02,Ruiz:SST10,Karmakar:PhysC04}. Experimental evidence suggests an elliptical dependence between the electric field and the critical current density \cite{Clem:SST11}. 

\subsubsection{Pulsed applied field}
\label{s.hystsolpulse}

This section first summarizes the main modeling issues for bulk applications (slabs and cylinders) and, afterwards, for coils in pulse mode.

Calculations for slabs show that, essentially, the response to a pulsed applied field is very similar to a periodic ac field \cite{Meerovich:SST96,Takacs:SST05}. If the pulse starts from 0 applied field, then increases to a certain maximum $H_m$ and finally decreases to 0 again, the response for the periodic case in the increasing and decreasing parts of the pulses is the same as the initial and reverse curves for a periodic ac field, respectively. For the critical state model, the response is exactly the same, although for a smooth $E(J)$ relation the instantaneous current and flux distribution depends on the shape of the pulse, as seen for slabs \cite{Meerovich:SST96}. In case that there is a bias dc field super-imposed to the pulse, the movement of the current fronts is exactly the same as for an ac field except for the first increase of current. In addition, as a consequence of the dc field, there is a region with critical current density inside the sample. This has been analytically studied in~\cite{Takacs:SST05}.

The geometry of finite cylinders is more realistic to model bulk samples but it requires numerical computations. The process of the current and flux penetration has been studied for the critical state model \cite{Kajikawa:TAS08} and for a power-law $E(J)$ relation \cite{Fujishiro:SST10}. The latter work also takes into account the self-heating effects caused by the changing magnetic field.

For the design of dc magnets, it is essential to consider the dissipation in pulse mode~\cite{Kawagoe:TAS03,Lee:TAS06}. There are several ways to model the loss in pulse mode. The first one is to use the measured ac losses of the cable in a pulsed applied field as a function of the peak field. The total energy loss is calculated from computations of the magnetic field of the coil by assuming uniform current density in the superconductor and by taking the corresponding measured loss in the cable into account \cite{Kawagoe:TAS03}. The second way is to assume the slab model for the wire \cite{Lee:TAS06}, which is a good approximation for large magnetic fields. The last way is to calculate the detailed current distribution in the whole magnet, as explained in section \ref{s.hystsolcoils}.

\subsubsection{AC ripple on DC background}

The problems that a researcher can encounter in the modeling of the ac losses in a dc background are very different if the dc background is a transport current or an applied magnetic field. The case of a dc applied magnetic field perpendicular to the ac excitation is outlined in section \ref{s.hystsolshake}.

The situation of an ac applied magnetic field superimposed to a dc component in the same direction is basically the same as pulsed applied fields (section \ref{s.hystsolpulse}). After the increase of the applied field to the peak, the process of flux penetration is the same as a pure ac field. The only difference is that the critical current density is lower because of the presence of the dc component, as seen for slabs \cite{Takacs:SST05} and stacks of tapes in perpendicular applied field \cite{Hong:TAS11b}. This concept is exploited to estimate the magnetization ac losses in complex coils due to current fluctuations around the working current. One possibility is to use analytical approximations for the ac losses in one wire \cite{Bottura:TAS07}. 

An interesting case is when the dc applied field is in the same direction as a certain alternating transport current. For this situation, the ac transport current overwhelmes the dc magnetization, approaching to the reversible value after a few cycles \cite{Giordano:SST06}. These calculations require advanced models because the magnetic field has a parallel component to the current density (section \ref{s.hystsolrot}).

For a wire with a dc transport current and a transverse applied ac field, the most interesting feature is the appearance of a dynamic resistance \cite{Ogasawara:Cryo76}. This dynamic resistance appears for amplitudes of the applied magnetic field above a certain threshold. Analytical formulas exist for slabs with either a constant $J_c$ \cite{Ogasawara:Cryo76} or a field dependent one \cite{Oomen:SST99}. This case has also been numerically calculated for stacks of tapes, in the modeling of a pancake coil \cite{Hong:TAS11b}. However, the applied fields are small compared to the penetration field of the stack, and hence the dynamic resistance may not be present.

A dc transport current may increase the loss due to an ac oscillation, as shown by analytical approximations for slabs and strips in the critical state model \cite{Schonborg:TAS01}. These formulas also take an alternating applied field into account.
Extended equations including also the effect of a dc magnetic field can be found in \cite{Wang:TAS11}, where they are applied to optimize the design of a dc power transmission cable.

\subsubsection{AC applied magnetic field perpendicular to DC background field}
\label{s.hystsolshake}

Given a superconductor magnetized by a dc magnetic field, a relatively small ac magnetic field perpendicular to the dc one strongly reduces the magnetization (collapse of magnetization), as experimentally shown in \cite{Fisher:PhysC97}. The explanation is different if the sample is slab-like with both applied fields parallel to the surface or if either the dc field is perpendicular to the sample surface or the sample has a finite thickness. 

For a slab with both applied fields parallel to the surface of a slab, the collapse of the magnetization can only be explained by flux cutting mechanisms \cite{Fisher:PhysC97}. By assuming the double critical state model, solutions can be found analytically for a limited amplitude of the ac field \cite{Fisher:PhysC97}. General solutions for a continuous angular dependence between the critical current density and the electric field require numerical methods \cite{Badia:PRL01,RomeroSalazar:APL03}.

For infinitely long wires in a transverse applied field, the current density is always perpendicular to the magnetic field independently of the angle with the surface. Then, flux cutting cannot occur. The collapse of magnetization is explained by the flux penetration process. This can be seen by physical arguments and analytical solutions for thin strips \cite{Brandt:PRL02} and by numerical calculations for rectangular wires \cite{Vanderbemden:PRB07}. Collapse of magnetization is also present when the ac applied field is parallel to the infinite direction of a strip (and the dc one is perpendicular to the surface) \cite{Mikitik:PRB03}.

\subsubsection{3D calculations}
\label{s.hystsol3D}

In this article, we consider 3-D calculations as those where the superconductor is a 3-D object that cannot be mathematically described by a 2-D domain. A summary of the requirements and existing models for the 3-D situation is in section \ref{s.mod3D}.

The calculated examples are power-transmission cables made of tapes with finite thickness \cite{Nakamura:TAS05,Grilli:TAS05a,Zhang:SST12}, multi-filamentary twisted wires \cite{Grilli:Cryo12}, two rectangular prisms connected by a normal metal \cite{Grilli:TAS05a}, a permanent magnet moving on a thick superconductor \cite{Zehetmayer:SST06}, a finite cylinder in a transverse applied field \cite{Campbell:SST09}, a cylinder with several holes \cite{Lousberg:SST09} and an array of cylinders or rectangular prisms \cite{Zhang:SST12}.

\section{Eddy currents}\label{sec:eddy}
Eddy current computations go beyond problems involving superconductivity. They are important for modeling electric machines, induction heaters, eddy current brakes, electromagnetic launching and biomedical apparatuses, to name just a few applications~\cite{Kriezis:PIEEE92}. Thus we begin this section by shortly reviewing the key research done by the computational electromagnetism community. Then we discuss  the specific numerical methods to solve eddy current problems and derive losses. Finally, we review some of the numerical loss computations related to non-superconducting and non-magnetic but conducting parts present in technical superconductors.
\subsection{Eddy Currents and Conventional Electromagnetism}
Numerical simulations of eddy current problems in conventional computational electromagnetism have been under investigation since the early 1970s~\cite{Nedelec:SIAMJNA78,Verite:IJEPES79}. With the exception of some work done at the very beginning~\cite{Chari:TPAS74,McWhirter:TMAG79},  most of the early research concentrated on 3-D computations~\cite{Bossavit:TMAG82,Coulomb:TMAG81,Verite:IJEPES79} and theoretical background for numerical eddy current simulations is mostly established in 3-D space~\cite{Biro:TMAG90,Bossavit:CMAME81,Bossavit:IEOT82,Trowbridge:TMAG82}. The need for 3-D modeling comes from practical electrotechnical devices, such as electric motors. It is necessary to model them in three dimensions to guarantee accurate simulations which also include end effects~\cite{Carpenter:TMAG75,Coulomb:TMAG81,Yamazaki_PIEMD99}.

In order to compute the eddy currents, the majority of researchers solved partial differential equations with the finite element method~\cite{Biblecombe:TMAG82}. But since the computing capacity in the 1980s was relatively low, hybrid methods, combining the good sides of integral equation and finite element methods, have also been widely utilized from various perspectives~\cite{Bossavit:TMAG82,Kettunen:TMAG92,Lari:TMAG83,Salon:TMAG82,Tarhasaari:TMAG98}. 

The trend to analyze only 3-D geometries in conventional eddy current problems characteristically differs from the approaches to model hysteresis losses in superconductors by treating them as conventional electric conductors with non-linear power-law resistivity in two dimensions~\cite{IEEE-TAS_paper3}. Typically, a 2-D analysis of superconductors is adequate to solve a wide class of problems, but in certain cases a new and more accurate analysis requires modeling in three dimensions also in case of superconductors~\cite{Elliott:IMAJNA06,Takeuchi:SST11}.

The analysis of a 2-D geometry results in simpler formalism and program implementation than the analysis in a 3-D one, since for a solution of eddy current problem in two dimensions, it is not mandatory to solve all the components of a vector field~\cite{Carpenter:TMAG75}. However,  edge elements for vector-valued unknowns have been used for more than twenty years~\cite{Biro:CMAME99,Bossavit:IEEEPA88,Bossavit:TMAG88,Webb:TMAG90}, and their usability in 3-D eddy current problems based on the vector potential is well known and demonstrated~\cite{Kameari:TMAG90,Preis:TMAG92}, as well as in 2-D problems based on the $H$-formulation~\cite{Brambilla:SST07}. Nowadays there are several publicly available or commercial software packages for eddy current simulations~\cite{Flux,Opera,Comsol,Geuzaine:IJNME09}, the visualization in post-processing and skills to use the software as well as the computation time are the major differences between 2-D and 3-D eddy current simulations.

However, in 3-D eddy current computations of conventional conductor materials, such as copper, typically not much attention is paid to the conductivity, which is expected to be constant. Naturally, this is a modeling decision, since for example copper suffers from anisotropic magnetoresistance~\cite{Launay:JPCS59,Lengeler_JLTP70}. Thus, the problem is  closer to the problem of utilizing power-law in 3-D geometries for superconductor resistivity~\cite{IEEE-TAS_paper1} than the literature related to computations suggests. Anyhow, the good correspondence between the simulations and the experiments still suggests that the approximation of constant conductivity does not lead to severe errors in loss computations as it would in the case of superconductor simulations~\cite{Takagi:TMAG90,Wang:TMAG99}.

\subsection{Computation Methods for Eddy Currents}

The methods to compute eddy current losses from solved current density and/or magnetic field distributions do not differ from the methods used to solve superconductor hysteresis loss from a solution obtained with an eddy current solver utilizing a non-linear resistivity for the superconductor. Thus, the main methods are the utilization of Joule heating and domain integrals and integration of Poynting vector through surfaces surrounding the domain under consideration~\cite{Brambilla:SST07,Ferreira:TEDU88,Oomen:Thesis00}. However, when one needs to separate the magnetization loss in superconductor from the eddy current loss, it is difficult to use the Poynting vector approach since these two domains are typically in direct contact with each other. 

The main numerical methods for eddy current simulations are finite element method (FEM), integral equation method (IEM) and their combinations~\cite{Kriezis:PIEEE92}. Integral equation methods can be combined with finite element methods in two ways: the boundary terms are computed either by the volume integration (VIM)~\cite{Kalimov:TMAG97} or by the boundary integration method (BEM)~\cite{Salon:TMAG85}. Integration methods typically suffer from dense system matrices, but fast multipole acceleration increases their speed considerably~\cite{Darve:JCP00}. However, to the best of our knowledge, these sophisticated methods are not available in commercial software packages.

All these methods can work with several formulations. Within these formulations there is often the possibility to make different modeling decisions: for example, in the so-called $T-\Omega$ formulation, the field $T$ can be solved by Biot-Savart integration~\cite{Preis:TMAG92} or by using the so-called co-tree gauge~\cite{Albanese:TM88}. Table~\ref{tab:EC_formulations} summarizes the main formulations used for eddy current simulations with the most typical methods. Several other formulations with slight variations have been reviewed in~\cite{Biro:CMAME99}.

\begin{table}[ht!]
\caption{\label{tab:EC_formulations}Most common formulations and solution methods for numerical eddy current computations.}
\centering
\begin{tabular}{@{}lp{2.4cm}p{4cm}@{}}
\hline
{\bf Formulation} & {\bf Solution method} & {\bf Notes}\\
\hline
$A-V$ 
& FEM~\cite{Biblecombe:TMAG82,Kameari:TMAG90,Preis:TMAG92}
\\
& FEM-BEM~\cite{Salon:TMAG82}
\\ 
\hline
$T-\Omega$ 
& FEM~\cite{Albanese:TM88,Preston:TMAG82} & 
\multirow{2}{*}{\begin{tabular}{@{}p{4cm}@{}}
Two different gauges for $T$-field: co-tree and Biot-Savart
\end{tabular}}
\\
& FEM-BEM~\cite{Bossavit:TMAG82}
\\
\hline
$H$ & FEM~\cite{Brambilla:SST07} & 
\multirow{3}{*}{\begin{tabular}{@{}p{4cm}@{}}
Primarily used only in superconductivity community and with hysteresis loss computations of superconductors, not eddy current computations in normal metals
\end{tabular}}
\\
& FEM-VIM~\cite{Kalimov:TMAG97}
\\
& VIM~\cite{Kettunen:TMAG92}
\\
\\
\end{tabular}
\end{table}

\subsection{Computation of Eddy Current Losses in Superconductors}

Even though the primary concern for the computation of losses in technical superconductors is the hysteresis loss of superconducting materials,  a numerical estimation of eddy current losses is also important. Furthermore,  methods to reduce eddy current losses, at the expense of stability however, are well known and low ac loss conductors have thus increased matrix resistivity~\cite{Barnes:Cryo05,Vase:SST00}. The contribution of eddy current loss at power frequencies (below 200~Hz) to the total loss of a superconductor~\cite{Stavrev:PhysC98b,Stavrev:PhysC98a} or a superconducting coil~\cite{Paasi:TMAG96} is generally reported to be low and many studies neglect these losses~\cite{Magnusson:PhysC01,Majoros:SST07}. This applies to both the silver sheathed Bi-based conductors~\cite{Paasi:TMAG96,Stavrev:PhysC98b} as well as to copper stabilized layered YBCO coated conductors~\cite{Duckworth:TAS05b,Muller:PhysC97b}.

For a 1~cm wide YBCO tape having YBCO layer thickness of 0.9~$\mu$m, copper stabilization of 50~$\mu$m and self-field critical current of 260~A at 77~K, the hysteresis loss at 60~Hz sinusoidal current with peak current of critical current is 380~mW/m. The corresponding eddy current loss in the copper stabilizer can be estimated to be in the order of 5~mW/m~\cite{Duckworth:TAS05b}. However, if the tape is brought to 4.2~K, the copper resistivity still decreases and the eddy current loss increases. The eddy current power per unit length $P_{eddy}$ in a copper stabilizer layer of a YBCO tape can be estimated as~\cite{Muller:PhysC97b}
\begin{equation}
\label{equ:EC_coppereddycurrent}
P_{eddy}=\frac{4\mu_0^2}{\pi}\frac{twf^2}{\rho}I_c^2h(i),
\end{equation}
where $t$, $w$ and $\rho$ are the copper stabilizer thickness, width and resistivity, respectively. $h$ is a complex function of normalized operation current $i$. At $i=1$ it receives value 0.03 and at $i=0.5$ value 10$^{-5}$. See details in~\cite{Muller:PhysC97b}. Thus, from this formula we can conclude  that the eddy current loss in the stabilizer, or the matrix, is proportional to the second power of the frequency and inversely proportional to the resistivity of the normal metal; in addition, it also largely depends on the current amplitude.

If one wants to go a bit deeper in the analysis, one should note that a realistic operation current is considerably below the critical current and thus the eddy current losses can be on the level of hysteresis losses. In Bi-based Ag sheathed tapes the contribution of eddy current losses is from 30\% down to 1\% when the operation current increases from 0.1$I_c$ to 0.8$I_c$ at a frequency of 60~Hz and a temperature of 77~K~\cite{Stavrev:Thesis02}. In addition, it should be noted that at low currents the filament arrangement can change the contribution of eddy current losses at least from 5\% to 30\% of the overall losses. Thus, one should keep in mind that eddy current losses in the matrix cannot be totally forgotten, even though in many cases they seem to give little contribution to the overall losses at power frequencies~\cite{Sumption:SST05}.

Eddy current losses can be estimated from overall loss measurements since their frequency dependence is different from hysteretic and resistive losses. The measured total average loss power $P_{tot}$ of a superconductor under periodic excitation can be expressed as
\begin{equation}
P_{tot} = \underbrace{P_{res}}_{RI_{rms}^2}+\underbrace{P_{hyst}}_{fQ_{hyst}}+
\underbrace{P_{eddy}}_{\propto f^2},
\end{equation}
where $P_{res}$ and $P_{hyst}$ are the average contributions of resistive and hysteretic losses in the sample, respectively, $R$ is the sample's resistance along resistive paths where part of the current flows, $I_{rms}$ is the root mean square value of the excitation current and $Q_{hyst}$ is the superconductor hysteresis loss over one cycle. Thus, the loss in one cycle $Q_{cycle}$ can be expressed as~\cite{Friend:TAS99}
\begin{equation}
\label{equ:EC_lossdependence}
Q_{cycle}(f)=Q_{hyst}+\frac{P_{res}}{f}+C_{eddy}f,
\end{equation}
where $C_{eddy}$ is a frequency-independent eddy current loss constant (J/Hz). The constants $C_{hyst}$, $P_{res}$ and $P_{eddy}$ can be extracted form measurements performed with the same amplitudes but varying frequencies.

In YBCO and Bi-based tapes, materials with high conductivity, like silver, need to be used to allow high quality superconducting regions. Then, the eddy current loss in these materials can pose a real challenge. In~\cite{Friend:TAS99} the authors measured Bi-2223/Ag tapes with filaments from one to 55 and extracted the $C_{eddy}$ factor of equation~(\ref{equ:EC_lossdependence}). They studied frequency range 59-2500~Hz. Additionally, FEM modeling was performed to estimate the losses in the silver matrix. The modeling resulted in a prediction for eddy current loss at least four times smaller than that extracted from simulations using~(\ref{equ:EC_lossdependence}). However, the measured eddy current loss had negligible contribution to $Q_{cycle}$ when the frequency was below 200~Hz and the operation current was below 0.3 of the critical one.

\section{Coupling losses}\label{sec:coupling}

There is not wide consensus about what loss components are to be put under the term coupling loss~\cite{Takacs:SST97}. This is because coupling losses and eddy current losses have similar nature since they both arise from currents in non-superconducting parts of the conductor~\cite{Hlasnik:TMAG81}. Additionally, filament coupling  influences the current penetration profiles inside a superconductor and thus it changes the hysteresis loss~\cite{Grilli:SST10b}. In this review, with the term coupling loss we indicate the loss that arises in normal conducting regions when a current loop going from a filament to another via normal metal is formed due to time varying applied magnetic field. Here we discuss first about performing numerical simulations from which it is possible to extract the coupling loss. Then, we review how filamentary coupling affects hysteresis losses and how this can be simulated.

\subsection{Simulating Filament-Coupling Effects}

Coupling loss simulations that take into account the real conductor geometry need in general to be performed with models having three spatial dimensions~\cite{Gomory:SST06a}. Important characteristics for the simulations that represent real world cases are at a minimum the modeling of superconducting domain, of the normal metal matrix and of the contact resistances between these two~\cite{Amemiya:TAS07b,Hlasnik:TMAG81}. 

For estimating coupling loss, there are no easy ways to use analytical formulas which work in general cases. The reason for this is that the actual filament distribution determines largely the coupling loss. Additionally, separation of coupling loss from other loss components, such as eddy current loss, is difficult from measurement results since these losses have similar nature. However, in general one can say that short twist pitch in conductors results in low coupling loss and low transverse resistivity results in increased coupling loss. For a given conductor, the coupling loss increases with the second power of frequency. Analytical approaches to coupling losses are reviewed in~\cite{Zola:SST04}.

So far FEM-like tools have been less frequently used to model coupling losses than hysteresis losses in superconductors. There are three main reasons for this. First, when the magnetic field is not perpendicular to the current density, it is uncertain what kind of model, or constitutive law, should be adopted to model currents in superconductor~\cite{Campbell:JSNM11}. This case is usual in 3-D geometries. CSM does not work in this case and thus it is doubtful to use also its power-law approximation. Even though power-law works in some simulations of simple geometries in three dimensions~\cite{Costa:TAS03,CostaBouzo:SST04,Grilli:SST03,Grilli:TAS05a,Lousberg:PICACOMEN08}, experimental verification of these studies are missing. However, currently there is ongoing interest in solving this problem~\cite{Clem:SST11,Mikitik:LTP10,Ruiz:SST10}. Second, in three dimensions it is not possible to obtain the same discretization level, and thus accuracy, as in two dimensions. Since superconductor modeling is highly non-linear, even solving small problems takes a considerable amount of time~\cite{Grilli:TAS05a}. In three dimensions, real world problems with low accuracy still typically require several tens of thousands of degrees of freedoms. And third, methods to reduce coupling losses are very well known for filamentary superconductors~\cite{Wilson:Cryo08}. These methods include twisting the conductor, increase of matrix transverse resistivity, and utilization of high resistive barriers around filaments. Thus, especially in LTS conductors, the coupling losses are small and from the perspective of wire development and applications  the establishment of methods for coupling loss simulations has not been crucial. However, utilization of these manufacturing methods with filamentary HTS, such as Bi-based or MgB$_2$ wires, is still an unknown issue. Additionally, these methods are not applicable for striated YBCO conductors as such and there are many recent developments aiming at reducing YBCO ac losses by filamentarizing the end product~\cite{Abraimov:SST08,Amemiya:TAS07b,Ashworth:SST06,Carr:TAS99,Tsukamoto:TAS05b}. This effort must also include considerations on coupling losses and thus their modeling is important since even in the 10~Hz regime the coupling loss in striated YBCO can be important~\cite{Polak:SST07}.

\begin{figure}[t!]
\centering
\includegraphics{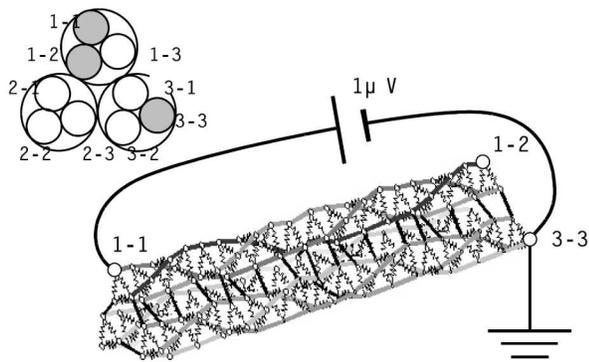}
\caption{\label{fig:CL_FigFrom_seo_2001_1_scaled}Circuit analysis model for coupling loss computation of CICC.
Reprinted with permission from Elsevier~\cite{Seo:Cryo01}.
}
\end{figure}

An alternative way of modeling coupling losses in a program that solves directly Maxwell's field equations is the utilization of equivalent circuit models~\cite{Tsukamoto:TAS05b}. This kind of models have been mainly developed for LTS cables such as current-in-conduit-conductors (CICC)~\cite{Seo:Cryo01}. These models are fairly complicated (as seen in Fig.~\ref{fig:CL_FigFrom_seo_2001_1_scaled}) and require measurement of contact resistances between many points. However, they are still applicable to coupling loss simulations of Roebel cables~\cite{Goldacker_SST09}, which also have a loose cable structure like many CICC~\cite{Pasztor:TAS04}. Also, when FEM-type simulations  target  high accuracy, the laborious characterization of interface contact resistances becomes mandatory anyway.

\begin{figure}[t!]
\centering
\includegraphics{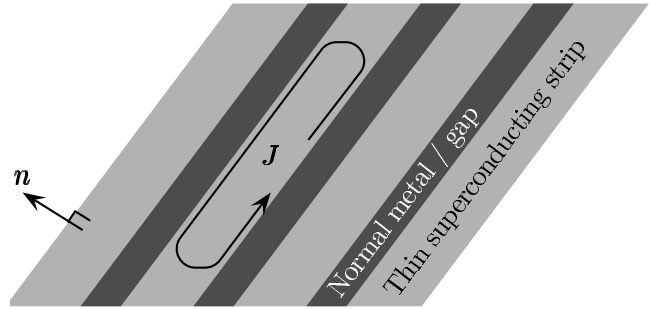}
\caption{\label{fig:CL_FigThinStripApproximationScaled}Schematic view of the thin strip approximation of a striated YBCO tape. One magnetization current loop is shown for the case of uncoupled filaments.
}
\end{figure}

When coupling losses are modeled in thin conductors by directly solving the fields, it is possible to take 2-D approaches. It can be estimated that there is no current flow from the bottom of the tape to the top or vice versa~\cite{Kasai:TMAG05}. This modeling decision is generally called as thin strip approximation~\cite{Amemiya:TAS07b}, see Fig.~\ref{fig:CL_FigThinStripApproximationScaled}. Striated YBCO conductors are examples of geometries that can be simulated with such approximation. Thus, even though the tape can have complicated 3-D geometry, at every cross-section, in local coordinates, one current density component always vanishes~\cite{Takeuchi:SST11}. Since superconductor hysteresis losses are computed simultaneously to coupling losses, we will review this method more deeply in the next subsection.

\subsection{Influence of Filamentary Coupling on Hysteresis Losses}

In 2-D hysteresis loss simulations of a multifilamentary conductor, or a cable consisting of several conductors, it is necessary to make a modeling decision concerning filament coupling effects when magnetic field is applied. However, typical programs consider only  two extreme cases: either the filaments are perfectly coupled or not coupled at all~\cite{Grilli:SST10b}. Full coupling means that the filaments interact together like a single larger filament which only has some gaps. Uncoupled situation means that filaments operate completely individually, though the fields they generate can alter current distributions in the other filaments~\cite{Kasai:TMAG05,Paasi:TAS97}.

3-D simulations are typically mandatory to study the intermediate cases, e.g. the effect of matrix resistivity and filament-to-matrix contact resistance on the current distribution in a superconductor~\cite{Grilli:Cryo12}. Naturally, these 3-D simulations also consider actual sample sizes and do not require infinite long samples. However, for high aspect ratio tapes, such as YBCO, it is possible to assume that the current density is always parallel to the tape's wide surface. Then, the problem reduces to a 2-D one, which allows different solutions with intermediate coupling where the resistivity between the filaments is varied~\cite{Kasai:TMAG05}. Naturally, in this model coupling currents can only flow from the edges of the superconductor to the normal metal, which allows using this kind of a model with Roebel cables, for example. The field problem in this thin strip approximation is formulated in terms of current density vector potential such that $\pmb{J}=\nabla\times(\pmb{n}T)$, where $\pmb{n}$ is the unit normal to the computation surface and $T$ is a scalar field. The governing equation includes also integral terms, since the field generated by the surface currents is computed by integral equations. Power-law resistivity has been used to model the electric behavior of the superconductor~\cite{Kasai:TMAG05}.

In~\cite{Kasai:TMAG05} the influence of the resistance between striations of a multifilamentary YBCO tape on hysteresis loss and coupling loss was studied with the thin strip approximation. The striated YBCO tape had 20 filaments. When the frequency was below 5000~Hz, coupling loss increased with decreasing resistance between the filaments. Roughly, one order of magnitude increase in contact resistance between filaments decreased the coupling loss by one order of magnitude. Also, the variation of hysteresis loss as a function of filament-to-filament contact resistance was studied. The difference of losses in the fully coupled case and the uncoupled case was a bit more than one order of magnitude. Losses computed with different values of contact resistances fell between these two limits. At 50~Hz, the coupling loss was still in every case much lower than the hysteresis loss and the most important factor of the filamentary coupling was the change in hysteresis loss, not the amount of coupling loss. Naturally, these results depend largely on the number of filaments as well as on the presence of a stabilization layer. Additionally, in practical applications, the end of the YBCO tape needs to be soldered to current lead. Then, the filaments are well coupled and introduction of filament transposition becomes mandatory.

\begin{figure}[ht!]
\centering
\includegraphics{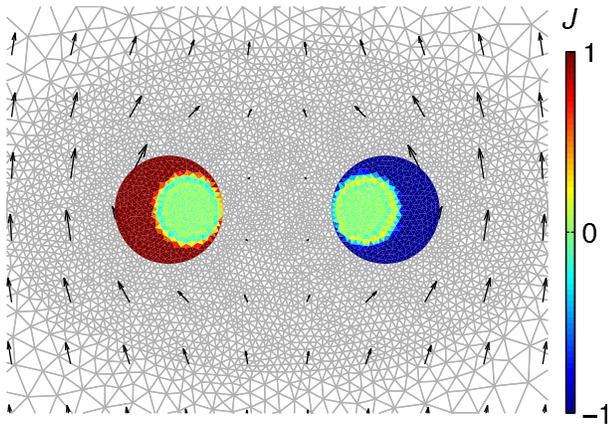}
\caption{\label{fig:CL_FigCoupled}Current penetration simulation of coupled case. Red and blue present currents towards and from the plane, respectively. Black arrows present magnetic flux density vectors. Gray background presents the mesh for numerical computation. View is displayed from the time when applied field reaches its maximum.}
\end{figure}

\begin{figure}[ht!]
\centering
\includegraphics{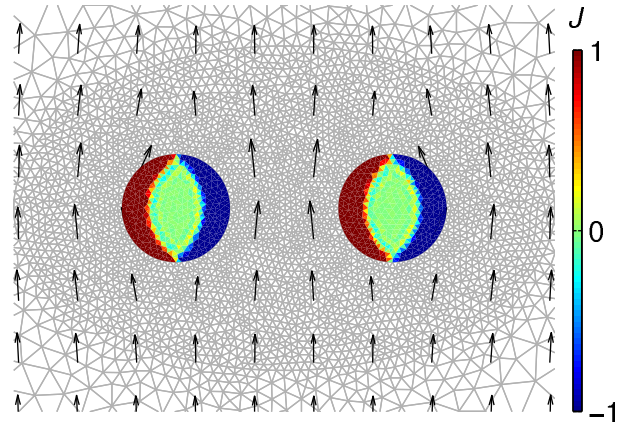}
\caption{\label{fig:CL_FigUnCoupled}Current penetration simulation of uncoupled case. Details correspond to those in Fig.~\ref{fig:CL_FigCoupled}.}
\end{figure}

In Fig.~\ref{fig:CL_FigCoupled} and~\ref{fig:CL_FigUnCoupled} we present typical current distributions for the two extreme case of fully coupled and uncoupled filaments, when an external magnetic field is applied. In these 2-D simulations the current density is a scalar quantity and it flows perpendicular to the modeling plane. We had two filaments, both having radius of 0.5~mm, separated by 1~mm. The matrix eddy current losses were neglected in this study. We applied vertical magnetic flux density $50\sin(2\pi ft)$~mT. We had field independent $J_c$ of 127~A/mm$^2$. We utilized $A-V$ formulation for the simulation~\cite{Stenvall:SST10b}. As seen, in case of uncoupled filaments, the current density distributions look like one filament alone at applied field. For the coupled case, the interaction of filaments is clearly visible. This leads to higher magnetization of the superconductor and thus to higher losses. Similar simulations for the cross-section of YBCO Roebel cable were presented in~\cite{Grilli:SST10b}.

\section{Influence of magnetic parts}\label{sec:ferromagnetic}
The presence of magnetic materials in superconducting systems is not uncommon because magnetic materials can be a part of superconducting wires or a component of superconducting devices. YBCO coated conductors manufactured by RABiTS technique typically have a ferromagnetic substrate. First generation HTS tapes (Bi-2223) or $\rm MgB_2$ wires can be manufactured with magnetic stabilizer/reinforcing layers. Superconductors and magnetic cores are the primary components of many power applications, such as generators, transformers, and motors. The magnetic components can influence the ac losses in superconducting systems in two ways: (i) they change the magnetic profile inside the superconductor (thereby affecting the superconductor losses) and (ii) they contribute additional hysteretic losses to the system. Therefore, the prediction of the effect of magnetic components on the total ac losses in a superconducting system is essential and has attracted significant effort from the research community.

In general, two types of approaches can be distinguished: analytical and numerical models. Analytical models provide relatively simple formulas to compute the losses in the superconductor; their main limitations concern the type of analyzed geometries of the superconductor and/or the magnetic parts, the physical properties of the material (typically, critical state model for the superconductor and infinite magnetic permeability for the magnetic material), and the fact that they neglect the losses in the magnetic material. On the other hand, numerical methods can overcome these limitation, although they generally require a larger computational effort.

\begin{figure}[t!]
\centering
\includegraphics[width=\columnwidth]{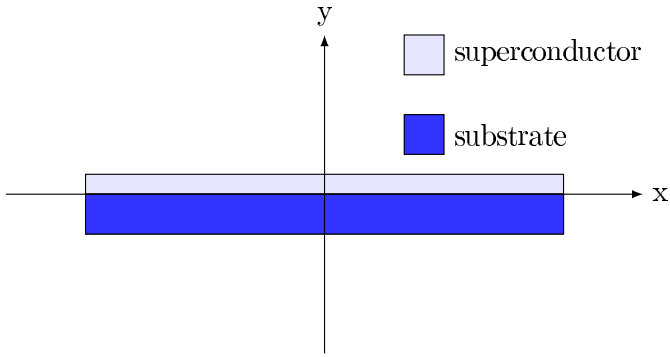}
\caption{Schematic view of the cross section of a YBCO coated conductor with magnetic substrate. Not drawn to scale.}\label{fig:xy_magnetic}
\end{figure}

In the following sections, we shortly review the main methods found in the literature. All of these models are usually solved for the 2-D cross-section of a conductor or device, which is perpendicular to the transport current. For example, these models can be used to calculate the magnetic and current density distribution in the $x-y$ plane for the cross-section of a superconductor with ferromagnetic substrate as shown in Fig.~\ref{fig:xy_magnetic}.

\subsection{Analytical Models}
Genenko pioneered analytical calculations of ferromagnetic/superconductor heterostructures by investigating the effect of an infinite magnetic barrier placed next to a current-carrying superconducting strip~\cite{Genenko:PRB00}. It was found that the effects on the current density distribution inside the superconductor mostly depend on the barrier's shape and distance from the superconductor, and only marginally on its magnetic permeability. The method was later used to propose configurations aiming at increasing the transport capability~\cite{Genenko:JAP02} and reducing the ac losses~\cite{Genenko:PhysC04} of superconducting strips. The above mentioned models, however, consider magnetic domains that are not embedded in the structure of the tape or wire, and as such they are not suitable to describe technical conductors used in applications, such as for example coated conductors with magnetic substrate or $\rm MgB_2$ wires with magnetic material.

An analytical model to calculate the magnetic distribution and ac loss of an isolated YBCO tape with magnetic substrate of infinite permeability $\mu$ was proposed in~\cite{Mawatari:PRB08} by Mawatari. This analytical model derives formulas to calculate the magnetic field, current distribution and ac loss in a superconducting strip with magnetic substrate (SC/FM strip conductors) for some special, simple cases. In the model, a two-dimensional magnetic field, $H = (H_x, H_y)$ was considered as a complex field $H(\zeta)=H_{ay}+iH_{ax}$, where $\zeta$ is a complex variable $\zeta=x+iy$. The primary idea of the model is to solve the generalized Biot-Savart law for a SC/FM strip~\cite{Mawatari:PRB08}:
\begin{equation}
H(\zeta ) = (H_{ay}  + iH_{ax} ) + \frac{1}{{2\pi }}\int\limits_{ - a}^a {dx'} \frac{{K_z (x') + i\sigma _m (x')}}{{\zeta  - x'}}
\end{equation}
where ${\mathbf H_a}=H_{ax} \hat {\mathbf x}+H_{ay} \hat {\mathbf y}$ is a uniform applied magnetic field, $K_z(x)$ is the sheet current in the superconducting strip, and $\sigma(x)$ is the effective sheet magnetic charge in the ferromagnetic substrate.  Both $K_z(x)$ and $\sigma(x)$ can be determined from Faraday's laws (equations (4,5) in~\cite{Mawatari:PRB08}) and must satisfy the following  constraints: the total net current in the superconducting strip must be equal to the transport current  and the total net magnetic charge in the substrate is zero.

In order to be able to derive the explicit results for the current and  magnetic field distribution, the model must assume that the ferromagnetic substrates is an ideal soft magnet with infinite permeability, i.e. $\mu \rightarrow \infty$. With this approximation, the  outside of the magnetic substrate has only the perpendicular component on its surface. This determines the boundary condition at the surface of a ferromagnetic substrate. 

Firstly the model was solved for a SC/FM strip conductor in the ideal Meissner state and exposed to either a perpendicular or parallel magnetic field, or a transport current. In the ideal Meissner state, the perpendicular magnetic field component on the surface of the superconducting strip must be zero. 

\begin{figure}[t]
\begin{center}
\includegraphics[width=\columnwidth]{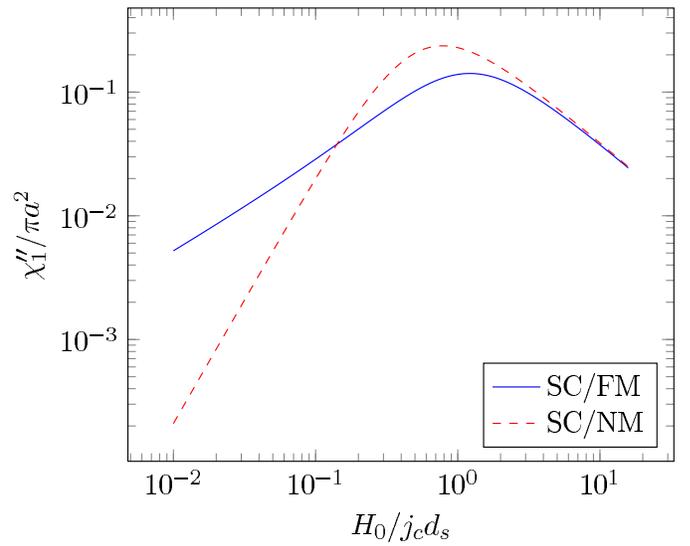}
\end{center}
\caption{Normalized imaginary part of ac susceptibility as a function of normalized applied magnetic field.}\label{fig:chi_mawa}
\end{figure}

Based on the critical state model (CSM), formulas for the case of SC/FM strip conductor exposed in a perpendicular dc or ac field were also derived. In ac perpendicular applied field, $H_a(t)=H_0\cos(\omega t)$, the time dependence of magnetic moment per unit length, $m_y(t)$, is derived explicitly and can be expressed as the Fourier series to determine the complex ac susceptibility $\chi'_n+i\chi''_n$  at $\rm n^{th}$ harmonics. The superconducting ac loss, Q, generated  per unit length of a SC/FM strip in an ac perpendicular field can therefore be calculated from the imaginary part of the first harmonic $\chi''_1$ as follows~\cite{Clem:PRB94}:
\begin{equation}
Q(H_0)=\pi \mu_0 H_0^2 \chi''_1(H_0).
\end{equation}
Figure~\ref{fig:chi_mawa} plots the normalized imaginary part $\chi''_1/\pi a^2$  ($a$ is the half width of the conductor) for the fundamental frequency ($n = 1$) of the ac susceptibility as a function of normalized applied field  $H_0/j_cd_s$ ($j_c$ and $d_s$ are the critical current density and thickness of the superconducting strip, respectively) for strip superconductors with ferromagnetic substrate (SC/FM)  and without magnetic substrate (SC/NM).
Experimental results reported by Suenaga {\it et al.} in~\cite{Suenaga:PhysC08} indicated that, in perpendicular applied field, the ferromagnetic hysteresis loss in the Ni-W substrate of a RABiTS tape would be significantly lower than that in the superconducting layer and therefore can be negligible in the total magnetization losses. The model above predicted that magnetization losses in SC/FM conductor are higher than those in SC/NM counterpart at low applied fields, $\mu_0H_0<0.14 j_cd_s$. For a conductor with $j_c = 1.2 \cdot 10^{10} \rm A/m^2$ and $d_s = 2.5~\rm \mu m$, the crossing point, $\mu_0H_0=0.14 j_cd_s=4.9~\rm mT$, is consistent with the experimental results reported in~\cite{Suenaga:PhysC08} for a YBCO tape with Ni-W substrate. That quantitative agreement suggests that the model with infinite permeability can be used to quantitatively predict ac losses for superconducting strip with magnetic substrate of large but finite permeability exposed to a perpendicular applied magnetic field. However, this model currently does not apply for other important cases which  represent better for the real operating conditions of HTS conductors in devices, such as  a SC/FM strip carrying an ac transport current or/and exposed to a perpendicular/parallel applied magnetic field. In addition, as already pointed out, this model does not provide a way to predict the ferromagnetic losses in the magnetic substrate which may be significant in some situations~\cite{Nguyen:JAP09b,Nguyen:SST10}.  Thus, the accurate prediction of the total ac losses in a SC/FM system is difficult to be obtained with this analytical model, and numerical methods are therefore needed. 

\subsection{Numerical Models}
Two numerical models have been widely used to compute losses in superconducting devices with magnetic parts and they both employ the finite-element method. While those FEM models are supposed to be suitable with soft magnetic materials whose $\mu(H)$ is considered to be approximately the same for both the upper and the lower branches of a $B-H$ loop, they also may be applied for a system with hard magnetic materials by using $\mu(H)$ approximately determined from the virgin magnetization curve. An accurate model for hard magnetic materials that is capable of considering the hysteresis in $B-H$ loop would be very complicated and has not been developed yet.  
\subsubsection{$H$-formulation FEM Model}
The $H$-formulation FEM model to compute ac losses for superconducting systems with magnetic materials was first proposed in~\cite{Nguyen:JAP09b,Nguyen:SST10}.  This model is capable of taking the field dependences of permeability and ferromagnetic loss of the magnetic materials into account. With respect to the standard version of the $H$-formulation~\cite{Hong:SST06, Brambilla:SST07}, the writing of Faraday's equation need to be modified to take into account that $\mu_r$ depends on $H$ and hence on time. As a consequence, $\mu_r$ cannot be taken out of the time derivative in the governing equation:
\begin{align}\label{eq:faraday_magn}
& \frac{{\partial (\mu _r \mu _0 {\mathbf{H}})}}{{\partial t}} + \rho \nabla  \times {\mathbf{J}} = 0
\end{align}
and the equation needs to be expanded. The details of the new formulations can be found in~\cite{Nguyen:JAP09b,Nguyen:SST10}.

The hysteresis loss in the magnetic material, $Q_{fe}$, is the area of the $B-H$ loop and is usually a function of $B_m$, the maximum magnetic field of that loop. The hysteretic loss at each location inside the magnetic material can therefore be calculated from the maximum value of the induction field $B_m$ at that location during an ac cycle. Therefore, based on the $Q_{fe}(B_m)$ relation fitted from the experimental data, the total magnetic loss, $q_{fe}$ can be estimated by taking the integral of $Q_{fe}(x,y)$ over the cross-section $A_{SUB}$ of the magnetic material~\cite{Nguyen:JAP09b,Nguyen:SST10}:
\begin{equation}
q_{fe}  = \int\limits_{A_{SUB} } {Q_{fe}} (B_m (x,y))dxdy.
\end{equation}
This model is quite versatile and can be used for HTS devices with complicated geometries, as long as the field dependences of $\mu_r$ and $Q_{fe}$ for the magnetic materials in the device are known. In fact, the model has been used extensively to calculate ac losses for superconducting systems using RABiTS coated conductors with Ni-W substrate. For this type of conductors, the field dependence of $\mu_r$ and $Q_{fe}$ of the substrate material was determined experimentally and published in~\cite{Miyagi:PhysC08}. The functions fitting the experimentally observed field dependences of $\mu_r$ and $Q_{fe}$ were entered in the model and used for the calculation of ac losses~\cite{Nguyen:JAP09b,Nguyen:SST10}.
\begin{figure}[t]
\begin{center}
\includegraphics[width=\columnwidth]{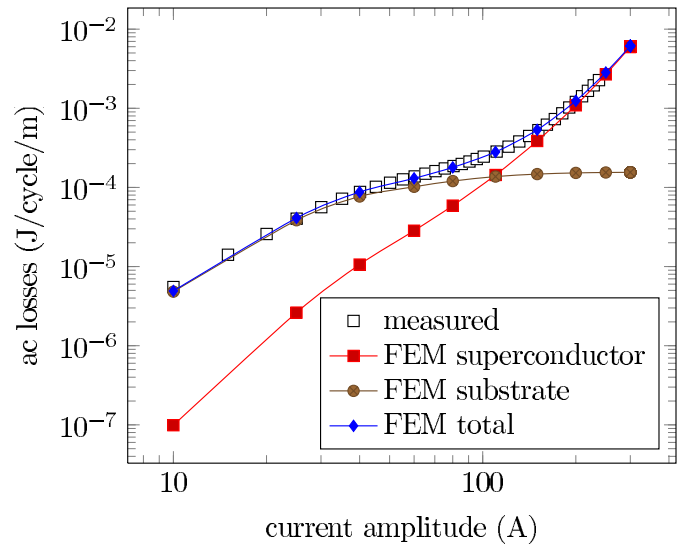}
\end{center}
\caption{FEM and experimental results for the ac loss components in a 1 cm-wide  RABiTS YBCO tape ($I_c$ = 330 A) carrying an ac transport current.}\label{fig:losses_doan}
\end{figure}
As an example, Fig.~\ref{fig:losses_doan} shows the comparison between calculated and measured total transport losses in a RABiTS tape as a function of transport current. In the simulation, the superconducting loss in the YBCO layer and ferromagnetic loss in the substrate were calculated separately and the sum of these loss components provides the total self-field loss in this sample. In general, the modeled total self-field loss agrees well with that obtained from the measurement. The good agreement between simulations and experiments was also observed for other cases of RABiTS conductors with Ni-W substrates, such as the total ac loss in parallel applied field~\cite{Nguyen:JAP09b}, the total ac losses for stacks of two parallel tapes~\cite{Nguyen:SST10}, bifilar coils~\cite{Nguyen:SST11}, racetrack coils~\cite{Ainslie:SST11}, pancake coils~\cite{Zhang:SST12b,Zhang:APL12}  or cables~\cite{Nguyen:unpublished}.

\subsubsection{$A$-formulation FEM Method}
This method is based on the following general equation:
\begin{equation}\label{eq:A-potential}
\nabla  \times \left( {\frac{1}{\mu }\nabla  \times {\bf{A}}} \right) = {\bf{j}} =  - \sigma \left( {\frac{{\partial {\bf{A}}}}{{\partial t}} + \nabla \phi } \right)
\end{equation}
where $\mathbf{A}$ is the magnetic vector potential and $\phi$ is the scalar potential. For infinitely long conductors and circular rings, $\phi$ is constant in the cross-section. Miyagi proposed a 2-D edge-based hexahedral finite element modeling technique to solve that equation with constant and variable $\mu$. At the beginning, (\ref{eq:A-potential}) was solved for conductors with substrate of constant $\mu$, but a considerable discrepancy between calculated results and experimental data was observed~\cite{Miyagi:TAS07}. The model was therefore improved to take the field dependence of permeability into account for better accuracy~\cite{Miyagi:TAS08}. The authors used the Newton-Raphson (N-R) method to handle the nonlinearity of permeability of the substrate. The detailed algorithm and techniques used to solve (\ref{eq:A-potential}) with nonlinear $\sigma(j)$ and $\mu(B)$ can be found in~\cite{Miyagi:TAS08}. This model, however, has not been validated against experiment. Although authors did not use this method to predict the ferromagnetic loss in the substrate material, this can be done by a similar way as presented in the $H$-formulation technique. 

Another numerical technique to solve (\ref{eq:A-potential}) was introduced in~\cite{Gomory:SST09}. This method is based on the critical state model and two assumptions: (i) there exists a ``neutral zone'' inside the conductor where the current density $j$ (or the electric field $E$) vanishes during the entire cycle, and (ii) the magnetic flux penetrates monotonically from the surface when the applied current or magnetic field increase monotonically.  Because the electric field is zero, the vector potential $A_c$ in the neutral zone must be uniform and satisfies: $\nabla \phi  = -\partial {\bf{A_c}}/\partial t$ . Then the authors introduced a new variable, the so-called calibrated vector potential $\mathbf{A'}=\mathbf{A}-\mathbf{A_c}$, which is zero in the neutral zone. The electric field can then be calculated by the simple equation: 
\begin{equation}
\mathbf{E}=\partial {\bf{A'}}/\partial t.
\end{equation}
Since $\mathbf{A'}$ in the neutral zone is zero, $\mathbf{A'}(x, y, t) $ is the magnetic flux per unit length between the neutral zone and the point $(x,y)$. Thus, based on CSM, the current density from one time step can be calculated analytically from the calibrated vector potential of the previous time step. The ferromagnetic hysteretic loss in the magnetic material was estimated by a similar technique as described in the $H$-formulation method. The details of the calculation procedures can be found in~\cite{Gomory:SST09}.

\begin{figure}[t]
\begin{center}
\includegraphics[width=\columnwidth]{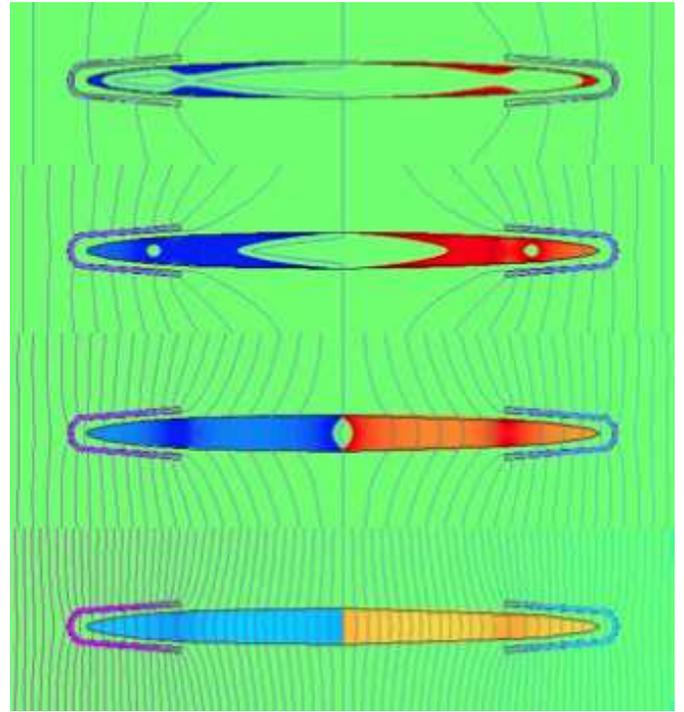}
\end{center}
\caption{The evolution of magnetic and current density profiles calculated for the initial increase of the perpendicular magnetic field. From top to bottom the applied fields are 5, 12, 22 and 50 mT, respectively. Density of electrical current is characterized by the darkness of the color, magnetic field is represented by the lines of constant vector potential. Figure taken from~\cite{Gomory:SST09} and reprinted with permission from IOP.}\label{fig:BSCCO_shield}
\end{figure}

The authors employed the model to calculate ac losses in a Bi-2223 conductor with elliptical cross-section for two cases: (i) conductor carrying an ac transport current and (ii) conductor exposed to a perpendicular applied ac field. The edges of the conductor are covered with ferromagnetic material (Ni) for ac loss reduction. As an example, Fig.~\ref{fig:BSCCO_shield} depicts the calculated evolution of magnetic and current density profiles in that conductor when the applied field changes from 5 mT to 50 mT.  The theoretical results of ac losses in this conductor were then compared to the experimental data~\cite{Gomory:SST09}. The quite good agreement between the theoretical prediction and experimental data has validated this numerical method and it can provide another option to calculate ac loss and electromagnetic behavior in a SC/FM system~\cite{Gomory:TAS12}.  
\subsubsection{Other FEM-based Models}
Other numerical models, based on the finite-element method, have recently been developed to study the interaction of superconducting and magnetic materials.

Genenko {\it et al.}~\cite{Genenko:SST09} studied the influence of the shape of magnetic shields to reduce the ac losses in superconducting strip: it was found that not only the shape of the shield, but also the magnetic permeability of the magnetic material plays a key role for that purpose if the size of the shield is comparable to or less than that of the superconductor. The same group also developed a model to simulate superconducting strips with magnetic substrate in applied ac field based on an electrostatic-magnetostatic equivalence~\cite{Genenko:APL11}.

Farinon {\it et al.} extended their adaptive resistivity algorithm~\cite{Farinon:SST10} to take into account magnetic materials~\cite{Farinon:JSNM12}. 

\section{AC Losses in HTS Power Applications}\label{sec:power}
Calculation of ac losses is essential for HTS power applications since the design of the cooling systems needs to take into account the thermal loads. The cooling system maintains the required cryogenic temperatures during the operation of HTS devices. Ac loss is a contribution source of the heat load, thus the importance of ac loss lies in two aspects: the calculation is essential for the engineering design of cooling systems; and the calculation acts as a guide to explore methods for minimizing ac losses.
This section discusses the impact of the ac losses in the most important power applications and the main contribution of the ac loss for each case.
\subsection{Cryogenic Considerations}
One important reason why HTSs are very attractive is that they can operate without liquid helium, which is a very expensive type of cryogen. Both BSCCO and YBCO-based superconductors can operate at 77~K; the price per liter of liquid nitrogen is 20-100 times lower than that of liquid helium depending on the purchase location~\cite{PriceLN2}. The latent heat of liquid nitrogen is 161 kJ/L at one atmosphere~\cite{Woodcraft:07}. Therefore in order to remove the heat $Q$ kJ produced by ac losses, one needs Q/161 L liquid nitrogen.

In order to achieve a larger critical current and higher magnetic field, HTS devices are operated at temperatures lower than 77 K with the aid of cryocoolers. One important parameter in determining the efficiency of cryocoolers is Carnot efficiency. The inverse of the Carnot efficiency is called the Carnot specific power and is the number of watts required at ambient temperature to provide 1 watt of refrigeration at the lower operating temperature.  Column 2 in Table~\ref{tab:cryo1}  presents the ideal Carnot specific power (watt input per watt lifted)~\cite{Gouge:CryoAssessRpt02}. However, commercial refrigerators run only at a fraction of Carnot efficiency, thus column 3 gives a more realistic value for the specific power (watt input at 303 K per watt lifted at $T_{op}$).
\begin{table}[t!]\renewcommand{\arraystretch}{1.3}
\caption{\label{tab:cryo1}Ideal and practical Carnot specific power for operating temperature from 4.2 to 273 K.}
\centering
\begin{tabular}{l l l}
\hline
{\bf Operating temperature} & {\bf Carnot specific} & {\bf Realistic specific}\\
{\bf $\bf T_{op}$ (K)} & {\bf power} & {\bf power} \\
\hline
273 & 0.11 & 0.4 \\
77 & 2.94	& 12-20 \\
50 &	5.06	& 25-35 \\
20 &	14.15 &100-200 \\
4.2 & 71.14 & 11000 \\
\hline
\end{tabular}
\end{table}
Thus in the cryogenic system design, the cooling power required for removing the heat load due to the ac losses can be calculated using Table~\ref{tab:cryo1}.
Table~\ref{tab:cryo2} summarizes the estimated heat load of HTS devices at the operating temperature~\cite{Gouge:CryoAssessRpt02}. 
\begin{table}[ht!]\renewcommand{\arraystretch}{1.3}
\caption{\label{tab:cryo2} Summary of  the estimated heat load of HTS devices at the operating temperature.}
\centering
\begin{tabular}{l l l}
\hline
{\bf HTS Component} & {\bf BSCCO Heat load} & {\bf YBCO Heat load}\\
\hline
Cables 	& 3-5 kW/km 	& 3-5 kW/km  \\
 (per phase) & at 65-80 K & at 65-80 K \\
 \hline
Transformers 	& 50-100 W at 25-45  K  & 50-100 W at 60-80 K  \\
(5-100 MVA) &  or  at 65-80 K &  \\
 \hline
 Motors 	& 50-200 W at 25-40 K & 50-200 W at 50-65 K \\
(1-10 MW)  &  &  \\
\hline Generators & 100-500 W at 25-40 K & 100-500 W at 50-65 K\\
(10-500) MW &  &  \\
\hline
FCLs	& 10's W at  30 K	& 1kW at 50-80 K \\
 & 750 W at 80 K& \\
\hline
SMES	& 10's W at 20-30 K	& 10-100 W at 50-65 K \\
\hline
\end{tabular}
\end{table}
\subsection{Techniques for ac loss reduction}
Ac loss calculation can be used to find solutions to reduce ac losses of HTS power devices. 

As it has been described in this paper, ac losses have different components: hysteretic, ferromagnetic, eddy current and coupling current losses. Hysteresis losses can be reduced by decreasing the filament size~\cite{Barnes:Cryo05}. Eddy current losses can be reduced by increasing the stabilizer resistivity. Coupling current losses can be reduced by increasing the matrix resistivity. Ferromagnetic losses can be eliminated by avoiding ferromagnetic materials. 

There are a few general methods for reducing ac losses of HTS applications. Controlling the amplitude and magnetic field orientation are effective ways for reducing ac losses~\cite{Hassenzahl:PIEEE04, Zushi:Cryo05}. Using flux diverters or shielding layers are common means to reduce the magnetic field, which in turn reduces ac losses. Since HTS coated conductors produce significantly higher ac losses in a perpendicular magnetic field, reducing the perpendicular component is an effective way to lower the ac losses.

The large cross-sectional aspect ratio of HTS coated conductors leads to a large hysteresis loss in external fields. Filamentization of the coated conductor proved to be effective in reducing hysteresis losses~\cite{Amemiya:SST04b,Sumption:TAS05}. Sumption {\it et al.} pointed out that the losses for a coated conductor subdivided into 10 filaments is shown to be reduced significantly~\cite{Sumption:TAS05}. However, filamentarization introduces coupling losses, caused by the relatively good conductivity of non-superconducting barriers and remnant superconducting `bridges' between neighboring strips~\cite{Marchevsky:TAS09}. Another contribution of losses in a striated conductor comes from magnetic flux coupling between filaments, because magnetic field distribution inside the superconducting filaments is influenced by the field concentration in the non-superconducting region~\cite{Marchevsky:TAS09}.

\subsection{Electrical power applications}
In this section, ac losses in various applications of HTS power applications are briefly reviewed. Typical techniques for reducing losses in different applications are described.
\subsubsection{Rotating Machines}
In an ac synchronous machine, superconducting coils can either be used as armature coils or rotor coils. If used as stator coils, superconductors will carry an ac current and will be exposed to a rotating field, thus producing large ac losses~\cite{Ainslie:COMP11, Ainslie:PhysC10,Sugimoto:TAS07}. 
If used as the rotor coils, the magnetic field of the superconducting coils is phase-locked with the stator field. Under balanced load conditions, the rotor experiences a dc field from the armature. However, non-synchronous disturbances, which are usually a result of unbalanced loads, transient events caused by system faults and noise from the grid, can produce non-synchronous ac effects that affect the superconducting coils on the rotor. The induced time-varying fields in the armature at frequencies other than the synchronous frequency induce compensating ac currents in the field windings, which in turn causes ac losses in the superconducting coils on the rotor. In a synchronous superconducting wind generator design, the superconducting coil windings on the rotor will also experience external ripple field from the power electronics excitation ripple and wind turbulence. This effect is investigated in~\cite{Chen:TAS12}. 

In the SIEMENS 4 MW demonstration motor experiment~\cite{Nick:PhysC12}, superconducting coils are used as rotor coils while conventional copper coils are used in the armature. The experiment shows that the associated rotor losses including ac losses are smaller than 120 W, which would cost 10 kW compressor power, thus the cooling system would decrease the system efficiency by 10 kW / 4 MW = 0.25\%, which is quite small for a large-scale application~\cite{Nick:PhysC12}. This power requirement is encouraging because it shows that the cooling system to remove the heat load is not a significant barrier for a large superconducting machine. 

\subsubsection{Power Transmission Cables}
In HTS power transmission cables, superconducting tapes are usually helically wound around a cylindrical former. A very simple but useful model for calculating ac loss is the `monoblock' model in which the cable is considered as a tube of superconductor~\cite{Vellego:SST95}. Although the monoblock model does not take into account the interaction between phases and the geometry of the tapes, it constitutes a frequently used engineering estimation for large-scale applications~\cite{Gouge:TAS05}.

Another useful model is the electrical circuit model~\cite{Noji:SST97, Noji:PhysC03}, which models cables with resistances and inductances connected in series and in parallel. The resistances are calculated from Norris equation, while the inductances are calculated from self and mutual  inductance formulas. This model is able to take into account the interaction between the conductor layers, the insulation layer and the shielding layer.

Clem {\it et al.} proposed that the ac loss of HTS power transmission cables could be broken down into six different mechanisms~\cite{Clem:SST10}: Bean model losses as in a Bean slab; losses in helically wound phase conductors arising from finite gaps between adjacent tapes; losses arising from the polygonal configuration of the conductor cross section if the tapes do not conform to the cylindrical former; flux cutting loss; losses when the current exceeds the critical current; and losses in the possible magnetic substrates. 

The key parameters affecting ac losses in power transmission cables are the number of conductors, the conductor's width, the gap size between adjacent tapes, the radius of the former and the current distribution. One common design of superconducting cables is the mono-layer design. Ac losses in a mono-layer cable do not increase significantly over one single tape since most of the flux lines lie outside the superconducting layers in the cable. It has been found that using narrower tapes and/or decreasing gaps between adjacent tapes are effective solutions for ac loss reduction~\cite{Amemiya:TAS07b, Malozemoff:TAS09}. Due to the mechanical requirements and architecture of the conductor, a mono-layer cable consisting of 2.5-4 mm wide tape would have minimized ac losses~\cite{Fukui:TAS06b,Malozemoff:TAS09}.

In a two-layer counter-wound cable using coated conductors, gap and polygonal losses, flux transfer losses in imbalanced two-layer cables and ferromagnetic losses for conductors with magnetic substrates are the major contribution to ac losses~\cite{Clem:SST10}. Current imbalance and ferromagnetic losses can be minimized by orienting the substrates of the inner winding inward and the outer winding outward~\cite{Clem:SST10}.

In multi-layer cables, the ac losses behave differently. It has been found that ac losses in cables with many layers are hardly affected by gaps between adjacent coated conductors as well as a lateral critical current distribution~\cite{Li:SST10}.

\subsubsection{Compact High-Current Cables}
Different from power transmission cables, superconductors are often required to be parallel or stack placed to create a high current density cable for  applications in high field magnets and busbars. Several kA to tens of kA cables are expected in these applications. 

Twisting and transposition are typical methods for reducing ac losses in parallel and stack placed superconducting cables~\cite{Barnes:Cryo05,Sumption:TAS05,Sumption:SST05,Grilli:SST10b}. This is because magnetic flux coupling can by reduced by transposing the conductors. However, it is not easy for HTS conductors be transposed as LTS wires since often HTS conductors are made into flat tapes. Recently significant efforts have been devoted to develop HTS Roebel cables, Rutherford cables and other types of twisted stack of coated conductors for high current applications~\cite{Takayasu:SST12,Staines:SST12,Schlachter:TAS11}. Filamentarization is another technique to reduce  ac losses in high current cables. In addition, decreasing the aspect ratio (width/thickness) of the cable by assembling stacks of strands instead of individual strands in cables can also reduce ac losses~\cite{Terzieva:SST10}.

\subsubsection{Fault Current Limiters}
For resistive type superconducting fault current limiters, it is preferable to contain a long length of superconductors in a small volume. One type of SFCLs uses an array or matrix of superconducting tapes~\cite{Noe:TAS01,Kudymow:TAS09,Hong:TAS11a}. For ac loss calculation, a stack of superconductors can be modeled for this type of SFCLs~\cite{Hong:TAS11a}. Another design is to use a non-inductive superconducting coil~\cite{Clem:PRB08,Nguyen:SST11}. To calculate ac losses of this type of SFCL, HTS coils need to be studied~\cite{Clem:PRB08,Nguyen:SST11}. In the dc biased iron core type SFCL, superconducting coils carry dc currents thus there are no ac losses~\cite{Noe:SST07}.  To calculate losses in the shielded iron core type SFCL, one needs to analyze the ac losses of a superconducting bulk cylinder or a tape. 

An effective way to reduce ac losses in a resistive SFCL is the bifilar design~\cite{Clem:PRB08}. This design can significantly reduce the inductance and also ac losses by decreasing the magnetic field volume~\cite{Dommerque:SST10}. Another technique for reducing ac losses is to use the two-in-hand structured wire~\cite{Nguyen:SST11}. The loss of a coil using two-in-hand structure wire is only half of that generated in an isolated tape per unit length. The ferromagnetic losses can be minimized by orienting the substrates of the inner winding layer inward and the outer winding outward (face-to-face configuration~\cite{Clem:SST10}).

In a commercial 12 kV resistive SFCL, ac losses account for 60\% of the total thermal load~\cite{Dommerque:SST10}.

\subsubsection{Transformers}
From the point of view of engineering design, ac losses in transformers can be regarded as ac losses in HTS coils. Ac losses come from the transport losses in the primary and secondary windings and the magnetization losses due to the magnetic fields, in particular, the leakage magnetic field in the end region of superconducting windings. One technique for reducing losses is to use a compensation coil to cancel the unwanted magnetic field, especially the perpendicular component. Heydari {\it et al.} discussed how to use auxiliary windings to reduce the leakage magnetic field hence the ac losses~\cite{Heydari:SST08}. 
Another way of reducing leakage magnetic field is by using flux diverters to shape the field. Al-Mosawi {\it et al.} examined the validity of this method and proved that ac losses are reduced by 40\% in a 10 kVA superconducting transformer using Bi-2223 conductors~\cite{AlMosawi:TAS01}. Low ac loss conductor designs, e.g. Roebel cables, are also proposed to be used in transformers for carrying a large transport current~\cite{Staines:SST12}. 
In an example study it is estimated that the losses only reduces the efficiency by 0.2\% in a 800 kVA transformer using BSCCO-2223 tapes, with a penalty factor of 20 taken into account~\cite{Funaki:Cryo98,Iwakuma:TAS01}. This shows that, like in a large superconducting machine, the cooling power is only a fraction of the total output power of superconducting transformers.

\subsubsection{Superconducting Energy Storage}
The energy storage devices in Superconducting Energy Storage Systems (SMES) are superconducting magnets. The ac loss associated with changing currents and magnetic fields are nearly zero if the magnets  operate in storage mode. The two significant continuous energy losses for HTS magnets are the joint resistance of HTS conductors~\cite{Tixador:TAS05} and the normal resistance in the power electronics transistors. However, ac losses may occur in the system as a result of variation in the operating current and magnetic field due to the charging or discharging of energy~\cite{Kim:TAS11a}. The ac loss per unit time might be significantly higher than in normal conditions during a discharge process since a SMES is expected to react quickly in response to a grid incident, i.e. a quick discharge to maintain grid stability~\cite{Zhu:TAS11,Zhu:PhysC11,Yuan:TAS10}.
In certain discharge circumstances, the ac loss per unit time will be much larger than a cryocooler capacity. An example study shows that for one second of the discharging time, in a 600 kJ SMES system the total ac loss is 167.2~W, which is a large amount of heat, increasing the system temperature by 3~K~\cite{Park:TAS07}. Thus the temperature rise during a discharge process should be analyzed in the design work.
The ac loss calculation is also essential for a SMES system cyclic efficiency estimation. For example, the ac loss of a 600 kJ system during one charge or discharge process is 7.53~kJ which is about 1.3\% of the total stored energy~\cite{Park:Cryo07}.

\section{Conclusion and Outlook}
We presented a review of the literature on ac loss computation in HTS tapes, wires and devices. Two-dimensional calculations by different methods are now fully developed to the point that ac losses in superconductors with geometries of arbitrary complexity can be quickly computed with good accuracy. Magnetic materials can also be taken into account, and at least with certain models, their non-linear magnetic characteristics can be modeled. In addition, situations different from pure ac excitations (such as ramps, pulses, combination of ac and dc field and currents) can be handled.
We are confident to say that all the situations relevant for realistic applications can be modeled in 2-D.

The most challenging step for the scientific community will be the extension to 3-D calculations. The challenge resides not only in the numerical difficulties related to 3-D modeling (choice of the mesh, increase of memory requirements and computation time, etc.), but also in the choice of an appropriate constitutive relation for modeling the different types of superconductors in 3-D and in its implementation in the numerical codes.

Ac losses in different electrical power applications are investigated and summarized. It can be concluded that ac loss calculation in such applications is important for cooling system design and examining innovative methods for heat load reduction.


%


\appendices

\section{Electric field created by moving vortices}
\label{s.Evortex}

This appendix outlines the main steps to obtain the macroscopic electric field created by moving vortices, equation (\ref{BvEp}).

The electric field created by a moving vortex, ${\bf E}_v$, at a certain speed, $\bf v$, is the following. This electric field is ${\bf E}_v=-\partial{\bf A}_v/\partial t-\nabla\phi_v$, where ${\bf A}_v$ is the vector potential in the Coulomb's gauge and $\phi_v$ is the scalar potential. The time dependence is due to the movement of the vortex. Then, ${\bf A}_v(t,{\bf r})={\bf A}_v[{\bf r}-{\bf r}_v(t)]$, where ${\bf r}_v$ is the position of the center of the vortex. Thus, ${\bf E}_v=({\bf v}\cdot\nabla){\bf A}_v-\nabla\phi_v$. By means of differential vector relations, this equation becomes ${\bf E}_v={\bf B}_v\times{\bf v}+\nabla({\bf v}\cdot{\bf A}_v-\phi_v)$. For a long straight vortex, the average in a volume $V$ of height $l$ and base surface $S$ (Fig.~\ref{f.vortex}) containing a portion of the vortex is ${\bf E}_{{\rm av},v}=[1/(lS)]\int_V {\bf E}_v \dvol$, where the volume integral is
\begin{equation}
\int_V {\bf E}_v \dvol = \int_V {\bf B}_v\times{\bf v} \ \dvol + \oint_{\partial V}  ({\bf v}\cdot{\bf A}_v-\phi_v) \dsur .
\end{equation}
In the equation above, $\partial V$ is the surface of the volume $V$. The second term in the equation vanishes. This is because of two reasons. First, both ${\bf A}_v$ and $\phi_v$ vanish far away from the vortex center. Second, the integral contributions for the portion close to the vortex center cancel by symmetry. {This reasoning also applies if the volume $V$ is the cell of the vortex lattice. Both the vector and scalar potentials created by the vortices in the lattice vanish on the border of the cell \cite{Brandt:RPP95,Kolacek:PRL01}. The integral of ${\bf v}\cdot{\bf A}_v$ and $\phi_v$ on the top and bottom surfaces also vanishes by symmetry. Then, the electric field created by one vortex is ${\bf E}_{{\rm av},v}=({\bf \Phi}_0\times{\bf v})/S$, and hence for $n$ vortices per unit surface the average electric field is
\begin{equation}
\label{BvE}
{\bf E}={\bf \Phi}_0n\times {\bf v}={\bf B}\times{\bf v}.
\end{equation}

Equation (\ref{BvE}) for the average electric field assumes that the dimensions of the base of the volume (Fig.~\ref{f.vortex}) are larger than the vortex separation. In practice, the vortex separation is often much smaller than the sample dimensions. Thus, ${\bf E}$ is approximately the local electric field in a ``differential" volume of the sample. For thin films in a perpendicular field, the vortices become Pearl vortices \cite{Pearl:APL64} but the above analysis is still valid.

\begin{figure}[htb]
\centering
\includegraphics[width=7 cm]{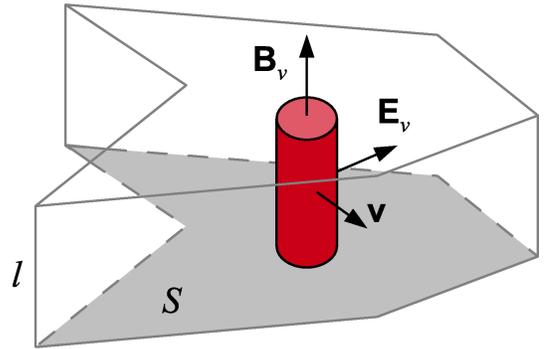}
\caption{A vortex moving at velocity $\bf v$ creates an electric field ${\bf E}_v$ due to its magnetic field ${\bf B}_v$. The cylinder represents the magnetic core of the vortex. The wire figure is an arbitrary integration volume (see Appendix \ref{s.Evortex}).}
\label{f.vortex}
\end{figure}

\section{Properties of the vector and scalar potentials and Coulomb's gauge}
\label{s.A}

Several calculation methods assume the Coulomb's gauge for the vector and scalar potentials, ${\bf A}$ and $\phi$. In this article, the discussions in sections \ref{s.QCSM} and \ref{s.Qsource} assume the Coulomb's gauge. For the reader's convenience, this appendix lists the main properties of the Coulomb's gauge and summarizes their deduction. 

Since $\nabla\cdot {\bf B}=0$, the flux density can be written as a rotational of a function, $\bf A$, as ${\bf B}=\nabla \times {\bf A}$. This defines the vector potential ${\bf A}$. From the differential Amp\`ere's law with negligible displacement current, $\nabla \times {\bf B}=\mu_0{\bf J}$, the vector potential is
\begin{equation}
\label{Agen}
{\bf A}(t,{\bf r})=\frac{\mu_0}{4\pi}\int_{V'} \frac{{\bf J}(t,{\bf r}')}{|{\bf r}-{\bf r}'|} \ \dvol '+\nabla\Psi(t,{\bf r}) ,
\end{equation}
where the function $\Psi(t,{\bf r})$ is an arbitrary scalar function. The equation above can be deduced by differential vector analysis, as in p. 179-181 of \cite{Jackson}. The choice of this arbitrary function defines the gauge. The Coulomb's gauge is defined as $\Psi=0$.
Then, the vector potential in Coulomb's gauge, ${\bf A}_c$, is
\begin{equation}
\label{ACou}
{\bf A}_c(t,{\bf r})=\frac{\mu_0}{4\pi}\int_{V'} \frac{{\bf J}(t,{\bf r}')}{|{\bf r}-{\bf r}'|} \ \dvol ' .
\end{equation}
As a consequence, $\nabla\cdot {\bf A}_c=0$ because $\nabla\cdot {\bf J}=0$. In addition, ${\bf A}_c$ approaches 0 as $\bf r$ approaches infinity, except if there is current at infinity. Another property of $\bf A$ in the Coulomb's gauge is that it only depends on the current density.

The scalar potential is defined as follows. Faraday's law and ${\bf B}=\nabla\times{\bf A}$ imply that $\nabla\times ( {\bf E}+{\partial{\bf A}}/{\partial t})=0$. Then, ${\bf E}+{\partial{\bf A}}/{\partial t}$ can be written as the gradient of a scalar function, $\phi$, and hence ${\bf E}=-{\partial {\bf A}}/{\partial t}-\nabla \phi$. Inserting this equation into Gauss law, $\nabla\cdot{\bf E}=\rho/\epsilon_0$, and using equation (\ref{Agen}) results in 
$\nabla^2\left(\phi+{\partial \Psi}/{\partial t}\right)=-{\rho}/{\epsilon_0}$, where $\rho$ is the charge density and $\Psi$ is the function that defines the gauge of the vector potential. The unique solution of this Laplace equation, under the assumption that $\phi+\partial\Psi/\partial t$ vanishes at the infinity, is
\begin{equation}
\phi(t,{\bf r})=\frac{1}{4\pi\epsilon_0}\int_{V'} \frac{\rho({\bf r}')}{|{\bf r}-{\bf r}'|}\ \dvol ' - \frac{\partial\Psi}{\partial t}.
\end{equation}
The Coulomb's gauge for $\bf A$ is defined as $\Psi=0$. With this gauge, the scalar potential becomes the electrostatic potential
\begin{equation}
\phi_c(t,{\bf r})=\frac{1}{4\pi\epsilon_0}\int_{V'} \frac{\rho({\bf r}')}{|{\bf r}-{\bf r}'|}\ \dvol '.
\end{equation}

For infinitely long conducting wires, the vector potential in the Coulomb's gauge is the integral along the length of the wire of equation (\ref{ACou}), resulting in
\begin{equation}
{\bf A}_c(t,{\bf r})=\frac{\mu_0}{2\pi}\int_{V'} {\bf J}(t,{\bf r}') \ln{|{\bf r}-{\bf r}'|} \ \dif^2{\bf r} '.
\end{equation}
%

\section*{Acknowledgment}
Weijia Yuan would like to thank Prof. Archie Campbell, Dr. Sastry Pamidi and Dr. Jozef Kvitkovic for their helpful comments.

\ifCLASSOPTIONcaptionsoff
  \newpage
\fi





\begin{thebibliography}{100}
\providecommand{\url}[1]{#1}
\csname url@samestyle\endcsname
\providecommand{\newblock}{\relax}
\providecommand{\bibinfo}[2]{#2}
\providecommand{\BIBentrySTDinterwordspacing}{\spaceskip=0pt\relax}
\providecommand{\BIBentryALTinterwordstretchfactor}{4}
\providecommand{\BIBentryALTinterwordspacing}{\spaceskip=\fontdimen2\font plus
\BIBentryALTinterwordstretchfactor\fontdimen3\font minus
  \fontdimen4\font\relax}
\providecommand{\BIBforeignlanguage}[2]{{%
\expandafter\ifx\csname l@#1\endcsname\relax
\typeout{** WARNING: IEEEtran.bst: No hyphenation pattern has been}%
\typeout{** loaded for the language `#1'. Using the pattern for}%
\typeout{** the default language instead.}%
\else
\language=\csname l@#1\endcsname
\fi
#2}}
\providecommand{\BIBdecl}{\relax}
\BIBdecl

\bibitem{Bean:PRL62}
C.~P. Bean, ``{Magnetization of hard superconductors},'' \emph{Physical Review
  Letters}, vol.~8, no.~6, pp. 250--252, 1962.

\bibitem{London:PL63}
H.~London, ``{Alternating current losses in superconductors of the second
  kind},'' \emph{Physics Letters}, vol.~6, pp. 162--165, 1963.

\bibitem{Bean:RMP64}
C.~P. Bean, ``{Magnetization of High-Field Superconductors},'' \emph{Reviews of
  Modern Physics}, vol.~36, pp. 31--39, 1964.

\bibitem{Bossavit:TMAG94}
A.~Bossavit, ``{Numerical Modelling of Superconductors in Three Dimensions: a
  Model and a Finite Element Method},'' \emph{IEEE Transactions on Magnetics},
  vol.~30, no.~5, pp. 3363--3366, 1994.

\bibitem{Norris:JPDAP70}
W.~Norris, ``Calculation of hysteresis losses in hard superconductors carrying
  ac: isolated conductors and edges of thin sheets,'' \emph{Journal of Physics
  D: Applied Physics}, vol.~3, pp. 489--507, 1970.

\bibitem{Halse:JPDAP70}
M.~R. Halse, ``{AC face field losses in a type II superconductor},''
  \emph{Journal of Physics D: Applied Physics}, vol.~3, pp. 717--720, 1970.

\bibitem{Campbell:Cryo82}
A.~M. Campbell, ``{A general treatment of losses in multifilamentary
  superconductors},'' \emph{Cryogenics}, pp. 3--16, 1982.

\bibitem{Wilson83}
M.~N. Wilson, \emph{Superconducting Magnets}.\hskip 1em plus 0.5em minus
  0.4em\relax Clarendon Press Oxford, 1983.

\bibitem{Carr83}
W.~J. Carr, \emph{{AC Loss and Macroscopic Theory of Superconductors}}.\hskip
  1em plus 0.5em minus 0.4em\relax New York : Gordon \& Breach Sci. Publishers
  Inc., 1983.

\bibitem{Ashkin:JAP79}
M.~Ashkin, ``{Flux distribution and hysteresis loss in a round superconducting
  wire for the complete range of flux penetration},'' \emph{Journal of Applied
  Physics}, vol.~50, no.~11, pp. 7060--7066, 1979.

\bibitem{Yeshurun:PRL88}
Y.~Yeshurun and A.~P. Malozemoff, ``{Giant Flux Creep and Irreversibility in an
  Y-Ba-Cu-0 Crystal: An Alternative to the Superconducting-GlassModel},''
  \emph{Physical Review Letters}, vol.~60, no.~21, p. 2202, 1988.

\bibitem{Kim:PR65}
Y.~B. Kim, C.~F. Hempstead, and A.~R. Strnad, ``{Flux-flow resistance in
  type-II superconductors},'' \emph{Physical Review}, vol. 139, no.~4A, p.
  A1163, 1965.

\bibitem{Gomory:SST07}
F.~G{\"o}m{\"o}ry, J.~{\v S}ouc, M.~Vojen{\v c}iak, and B.~Klin{\v c}ok,
  ``{Phenomenological description of flux pinning in non-uniform
  high-temperature superconductors in magnetic fields }lower than the
  self-field,'' \emph{Superconductor Science and Technology}, vol.~20, no.~9,
  p. S271, 2007.

\bibitem{Majoros:SST01}
M.~Majoros, B.~A. Glowacki, and A.~M. Campbell, ``{Critical current anisotropy
  in $\rm Ag/(Pb,Bi)_2Sr_2Ca_2Cu_3O_{10+x}$ multifilamentary tapes: influence
  of self-magnetic field},'' \emph{Superconductor Science and Technology},
  vol.~14, no.~6, pp. 353--362, 2001.

\bibitem{Pardo:SST11}
E.~Pardo, M.~Vojen{\v c}iak, F.~G{\"o}m{\"o}ry, and J.~{\v S}ouc,
  ``{Low-magnetic-field dependence and anisotropy of the critical current
  density in coated conductors},'' \emph{Superconductor Science and
  Technology}, vol.~24, p. 065007, 2011.

\bibitem{Pashitski:APL95}
A.~E. Pashitski, A.~Polyanskii, A.~Gurevich, J.~Parrell, and D.~C.
  Larbalestier, ``{Suppression of magnetic granularity by transport current in
  {$\rm (Bi,Pb)_2Sr_2Ca_2Cu_3O_x$} tapes},'' \emph{Applied Physics Letters},
  vol.~67, no.~18, pp. 2720--2722, 1995.

\bibitem{Grasso:PhysC95}
G.~Grasso, B.~Hensel, A.~Jeremie, and R.~Fl{\"u}kiger, ``{Distribution of the
  transport critical current density in Ag sheathed $\rm
  (Bi,Pb)_2aSr_2Ca_2Cu_3O_x$ tapes produced by rolling},'' \emph{Physica C},
  vol. 241, no. 45-52, 1995.

\bibitem{Tsukamoto:SST05}
O.~Tsukamoto, ``{AC} losses in a type {II} superconductor strip with
  inhomogeneous critical current distribution,'' \emph{Superconductor Science
  and Technology}, vol.~18, pp. 596--605, 2005.

\bibitem{Nishioka:TAS05}
T.~Nishioka, N.~Amemiya, N.~Enomoto, Z.~Jiang, Y.Yamada, T.~Izumi, Y.~Shiohara,
  T.~Saitoh, Y.~Iijima, and K.~Kakimoto, ``{AC} loss of {YBCO} coated
  conductors fabricated by {IBAD/PLD} method,'' \emph{IEEE Transactions on
  Applied Superconductivity}, vol.~15, no.~2, pp. 2843--2846, 2005.

\bibitem{Grilli:TAS07a}
F.~Grilli, R.~Brambilla, and L.~Martini, ``{Modeling High-Temperature
  Superconducting Tapes by Means of Edge Finite Elements},'' \emph{IEEE
  Transactions on Applied Superconductivity}, vol.~17, no.~2, pp. 3155--3158,
  2007.

\bibitem{Carpenter:PIEE77}
C.~J. Carpenter, ``{Comparison of alternative formulations of 3-dimensional
  magnetic-field and eddy-current problems at power frequencies},''
  \emph{Proceedings of the Institute Electrical Engineers}, vol. 124, no.~11,
  pp. 1026--1034, 1977.

\bibitem{Preston:TMAG82}
T.~W. Preston and A.~B.~J. Reece, ``{Solution of 3-dimensional eddy current
  problems: The $\rm T-\Omega$ method},'' \emph{IEEE Transactions on
  Magnetics}, vol.~18, no.~2, pp. 486--491, 1982.

\bibitem{Brandt:PRB96}
E.~H. Brandt, ``{Superconductors of finite thickness in a perpendicular
  magnetic field: strips and slabs},'' \emph{Physical Review B}, vol.~54,
  no.~6, pp. 4246--4264, 1996.

\bibitem{Brandt:PRB98a}
------, ``{Superconductor disks and cylinders in an axial magnetic field. I.
  Flux penetration and magnetization curves},'' \emph{Physical Review B},
  vol.~58, no.~10, pp. 6506--6522, 1998.

\bibitem{Rhyner:PhysC98}
J.~Rhyner, ``Calculation of {AC} losses in {HTSC} wires with arbitrary current
  voltage characteristics,'' \emph{Physica C}, vol. 310, pp. 42--47, 1998.

\bibitem{Yazawa:PhysC98}
T.~Yazawa, J.~J. Rabbers, B.~ten Haken, H.~H.~J. ten Kate, and Y.~Yamada,
  ``Numerical calculation of current density distributions in high temperature
  superconducting tapes with finite thickness in self field and external
  field,'' \emph{Physica C}, vol. 310, no. 1-4, pp. 36--41, 1998.

\bibitem{Stavrev:TAS02}
S.~Stavrev, B.~Dutoit, and N.~Nibbio, ``Geometry considerations for use of
  {Bi-2223/Ag} tapes and wires with different models of {$\rm J_c(B)$},''
  \emph{IEEE Transactions on Applied Superconductivity}, vol.~3, pp.
  1857--1865, 12.

\bibitem{Kajikawa:TAS03}
K.~Kajikawa, T.~Hayashi, R.~Yoshida, M.~Iwakuma, and K.~Funaki, ``{Numerical
  Evaluation of {AC} Loss in High Temperature Superconducting Coilvaluation of
  AC losses in HTS wires with 2D FEM formulated by self magnetic field},''
  \emph{IEEE Transactions on Applied Superconductivity}, vol.~13, no.~2, pp.
  3630--3633, 2003.

\bibitem{Hong:SST06}
Z.~Hong, A.~M. Campbell, and T.~A. Coombs, ``{Numerical solution of critical
  state in superconductivity by finite element software},''
  \emph{Superconductor Science and Technology}, vol.~19, pp. 1246--1252, 2006.

\bibitem{Amemiya:PhysC98b}
N.~Amemiya, S.~Murasawa, N.~Banno, and K.~Miyamoto, ``{Numerical modelings of
  superconducting wires for AC loss calculations},'' \emph{Physica C}, vol.
  310, pp. 16--29, 1998.

\bibitem{Enomoto:PhysC04}
N.~Enomoto and N.~Amemiya, ``{Electromagnetic field analysis of rectangular
  high {$T_c$} superconductor with large aspect ratio},'' \emph{Physica C},
  vol. 412-414, pp. 1050--1055, 2004.

\bibitem{Grilli:TAS05a}
F.~Grilli, S.~Stavrev, Y.~{Le Floch}, M.~Costa-Bouzo, E.~Vinot, I.~Klutsch,
  G.~Meunier, P.~Tixador, and B.~Dutoit, ``{Finite-Element Method Modeling of
  Superconducors: from 2-D to 3-D},'' \emph{IEEE Transactions on Applied
  Superconductivity}, vol.~15, no.~1, pp. 17--25, 2005.

\bibitem{Stenvall:SST10a}
A.~Stenvall and T.~Tarhasaari, ``{Programming finite element method based
  hysteresis loss computation software using non-linear superconductor
  resistivity and {$T-\varphi$} formulation},'' \emph{Superconductor Science
  and Technology}, vol.~23, p. 075010, 2010.

\bibitem{Hauser:TMAG97}
A.~O. Hauser, ``{Calculation of superconducting magnetic bearings using a
  commercial FE-program (ANSYS)},'' \emph{IEEE Transactions on Magnetics},
  vol.~33, no.~2, pp. 1572--1575, 1997.

\bibitem{Tonsho:TAS03}
H.~Tonsho, S.~Fukui, T.~Sato, M.~Yamaguchi, S.~Torii, T.~Takao, and K.~Ueda,
  ``Theoretical and experimental study on {AC} loss in {HTS} tape in {AC}
  magnetic field carrying {AC} transport current,'' \emph{IEEE Transactions on
  Applied Superconductivity}, vol.~13, no.~2, pp. 2368--2371, 2003.

\bibitem{Stenvall:SST10b}
A.~Stenvall and T.~Tarhasaari, ``{An eddy current vector potential formulation
  for estimating hysteresis losses of superconductors with FEM},''
  \emph{Superconductor Science and Technology}, vol.~23, p. 125013, 2010.

\bibitem{Maslouh:TAS97}
M.~Maslouh, F.~Bouillault, A.~Bossavit, and J.~C. V{\'e}rit{\'e}, ``{From
  Bean's model to the HM characteristic of a superconductor: some numerical
  experiments},'' \emph{IEEE Transactions on Applied Superconductivity},
  vol.~7, no.~3, pp. 3797--3801, 1997.

\bibitem{Prigozhin:TAS97}
L.~Prigozhin, ``{Analysis of critical-state problems in type-II
  superconductivity},'' \emph{IEEE Transactions on Applied Superconductivity},
  vol.~7, no.~4, pp. 3866--3873, 1997.

\bibitem{Prigozhin:SST11}
L.~Prigozhin and V.~Sokolovsky, ``{Computing AC losses in stacks of
  high-temperature superconducting tapes},'' vol.~24, p. 075012, 2011.

\bibitem{Prigozhin:JCP96}
L.~Prigozhin, ``{The Bean Model in Superconductivity: Variational Formulation
  and Numerical Solution},'' \emph{Journal of Computational Physics}, vol. 129,
  no.~1, pp. 190--200, 1996.

\bibitem{Badia:PRB02}
A.~B. A and C.~L{\'o}pez, ``{Vector magnetic hysteresis of hard
  superconductors},'' \emph{Physical Review B}, vol.~65, p. 104514, 2002.

\bibitem{Pardo:SST07}
E.~Pardo, F.~G{\"o}m{\"o}ry, J.~{\v S}ouc, and J.~M. Ceballos, ``Current
  distribution and ac loss for a superconducting rectangular strip with
  in-phase alternating current and applied field,'' \emph{Superconductor
  Science and Technology}, vol.~20, no.~4, pp. 351--364, 2007.

\bibitem{Souc:SST09}
J.~{\v Souc}, E.~Pardo, M.~{Vojen\v ciak}, and F.~{G\"om\"ory}, ``{Theoretical
  and experimental study of AC loss in high temperature superconductor single
  pancake coils},'' \emph{Superconductor Science and Technology}, vol.~22, p.
  015006, 2009.

\bibitem{Sanchez:PRB01}
A.~Sanchez and C.~Navau, ``{Magnetic properties of finite superconducting
  cylinders. I. Uniform applied field},'' \emph{Physical Review B}, vol.~64, p.
  214506, 2001.

\bibitem{Gomory:SST09}
F.~G{\"o}m{\"o}ry, M.~Vojen\v{c}iak, E.~Pardo, and J.~{\v S}ouc, ``{Magnetic
  flux penetration and AC loss in a composite superconducting wire with
  ferromagnetic parts},'' \emph{Superconductor Science and Technology},
  vol.~22, no. 034017, 2009.

\bibitem{Gomory:SST10}
F.~G{\"o}m{\"o}ry, M.~Vojen{\v c}iak, E.~Pardo, M.~Solovyov, and J.~{\v S}ouc,
  ``{AC losses in coated conductors},'' \emph{Superconductor Science and
  Technology}, vol.~23, p. 034012, 2010.

\bibitem{Campbell:SST07}
A.~M. Campbell, ``{A new method of determining the critical state in
  superconductors},'' \emph{Superconductor Science and Technology}, vol.~20, p.
  292, 2007.

\bibitem{Campbell:SST09}
------, ``{A direct method for obtaining the critical state in two and three
  dimensions},'' \emph{Superconductor Science and Technology}, vol.~22, p.
  034005, 2009.

\bibitem{Barnes:SST99}
G.~Barnes, M.~McCulloch, and D.~Dew-Hughes, ``{Computer modelling of type II
  superconductors in applications},'' \emph{Superconductor Science and
  Technology}, vol.~12, p. 518, 1999.

\bibitem{Takeda:Cryo94}
N.~Takeda, M.~Uesaka, and K.~Miya, ``{Computation and experiments on the static
  and dynamic characteristics of high $T_c$ superconducting levitation},''
  \emph{Cryogenics}, vol.~34, no.~9, pp. 745--752, 1994.

\bibitem{IEEE-TAS_paper3}
F.~Sirois, ``{Numerical Methods for Modeling High-Temperature
  Superconductors},'' \emph{IEEE Transactions on Applied Superconductivity},
  2013.

\bibitem{Brandt:PRL95}
E.~H. Brandt, ``Square and rectangular thin superconductors in a transverse
  magnetic field,'' \emph{Physical Review Letters}, vol.~74, no.~15, pp.
  3025--3028, 1995.

\bibitem{Vestgarden:PRB08}
J.~I. Vestg{\aa}rden, D.~V. Shantsev, Y.~M. Galperin, and T.~H. Johansen,
  ``{Flux distribution in superconducting films with holes},'' \emph{Physical
  Review B}, vol.~77, no.~1, p. 014521, 2008.

\bibitem{Amemiya:JAP06}
N.~Amemiya, S.~Sato, and T.~Ito, ``{Magnetic flux penetration into twisted
  multifilamentary coated superconductors subjected to ac transverse magnetic
  fields},'' \emph{Journal of Applied Physics}, vol. 100, no.~12, pp.
  123\,907--123\,907, 2006.

\bibitem{Grilli:Cryo12}
F.~Grilli, R.~Brambilla, F.~Sirois, A.~Stenvall, and S.~Memiaghe,
  ``{Development of a three-dimensional finite-element model for
  high-temperature superconductors based on the $H$-formulation},''
  \emph{Cryogenics}, 2012.

\bibitem{Zhang:SST12}
M.~Zhang and T.~A. Coombs, ``{{3D} modeling of high{-$T_c$} superconductors by
  finite element software},'' \emph{Superconductor Science and Technology},
  vol.~25, p. 015009, 2012.

\bibitem{Prigozhin:JCP98}
L.~Prigozhin, ``{Solution of Thin Film Magnetization Problems in Type-II
  Superconductivity},'' \emph{Journal of Computational Physics}, vol. 144,
  no.~1, pp. 180--193, 1998.

\bibitem{Navau:JAP08}
C.~Navau, A.~Sanchez, N.~Del-Valle, and D.~X. Chen, ``{Alternating current
  susceptibility calculations for thin-film superconductors with regions of
  different critical-current densities},'' \emph{Journal of Applied Physics},
  vol. 103, p. 113907, 2008.

\bibitem{Takeuchi:SST11}
K.~Takeuchi, N.~Amemiya, T.~Nakamura, O.~Maruyama, and T.~Ohkuma, ``{Model for
  electromagnetic field analysis of superconducting power transmission cable
  comprising spiraled coated conductors},'' \emph{Superconductor Science and
  Technology}, vol.~24, p. 085014, 2011.

\bibitem{Nii:SST12}
M.~Nii, N.~Amemiya, and T.~Nakamura, ``{Three-dimensional model for numerical
  electromagnetic field analyses of coated superconductors and its application
  to Roebel cables},'' \emph{Superconductor Science and Technology}, vol.~25,
  no.~9, p. 095011, 2012.

\bibitem{Campbell:AP72}
A.~M. Campbell and J.~E. Evetts, ``{Flux vortices and transport currents in
  type II superconductors},'' \emph{Advances in Physics}, vol.~21, no.~90, pp.
  199--428, 1972.

\bibitem{RomeroSalazar:APL03}
C.~Romero-Salazar and F.~P{\'e}rez-Rodr{\'\i}guez, ``{Elliptic
  flux-line-cutting critical-state model},'' \emph{Applied Physics Letters},
  vol.~83, p. 5256, 2003.

\bibitem{BadiaMajos:PRB09}
A.~Bad{\'\i}a-Maj{\'o}s, C.~L. C, and H.~S. Ruiz, ``{General critical states in
  type-II superconductors},'' \emph{Physical Review B}, vol.~80, p. 144509,
  2009.

\bibitem{Clem:SST11}
J.~R. Clem, M.~Weigand, J.~H. Durrell, and A.~M. Campbell, ``{Theory and
  experiment testing flux-line cutting physics},'' \emph{Superconductor Science
  and Technology}, vol.~24, p. 062002, 2011.

\bibitem{Lousberg:SST09}
G.~P. Lousberg, M.~Ausloos, C.~Geuzaine, P.~Dular, P.~Vanderbemden, and
  B.~Vanderheyden, ``{Numerical simulation of the magnetization of
  high-temperature superconductors: a 3D finite element method using a single
  time-step iteration},'' \emph{Superconductor Science and Technology},
  vol.~22, p. 055005, 2009.

\bibitem{Zehetmayer:SST06}
M.~Zehetmayer, M.~Eisterer, and H.~W. Weber, ``{Simulation of the current
  dynamics in a superconductor induced by a small permanent magnet: application
  to the magnetoscan technique},'' \emph{Superconductor Science and
  Technology}, vol.~19, pp. S429--S437, 2006.

\bibitem{Blatter:RMP94}
G.~Blatter, M.~V. Feigelman, V.~B. Geshkenbein, A.~I. Larkin, and V.~M.
  Vinokur, ``{Vortices in high-temperature superconductors},'' \emph{Rev. Mod.
  Phys.}, vol.~66, p. 1125, 1994.

\bibitem{Brandt:RPP95}
E.~H. Brandt, ``The flux-line lattice in superconductors,'' \emph{Reports on
  Progress in Physics}, vol.~11, pp. 1465--1594, 1995.

\bibitem{Jackson}
J.~D. Jackson, \emph{Classical Electrodynamics}.\hskip 1em plus 0.5em minus
  0.4em\relax 3rd edition: John Wiley {\&} Sons {Inc.}, 1999.

\bibitem{Kotzler:PRB94}
J.~K{\"o}tzler, G.~Nakielski, M.~Baumann, R.~Behr, F.~Goerke, and E.~H. Brandt,
  ``{Universality of frequency and field scaling of the conductivity measured
  by ac susceptibility of a {YBa$_{2}$Cu$_{3}$O$_{7-\delta}$} film},''
  \emph{Physical Review B}, vol.~50, no.~5, p. 3384, 1994.

\bibitem{Rhyner:PhysC02}
J.~Rhyner, ``{Vector potential theory of AC losses in superconductors},''
  \emph{Physica C}, vol. 377, no. 1-2, pp. 56--66, 2002.

\bibitem{Brandt:EPL93}
E.~H. Brandt, M.~V. Indenbom, and A.~Forkl, ``{Type-II Superconducting Strip in
  Perpendicular Magnetic Field},'' \emph{Europhysics Letters}, vol.~22, pp.
  735--740, 1993.

\bibitem{landau}
L.~D. Landau, E.~M. Lifshitz, and L.~P. Pitaevskii, \emph{Electrodynamics of
  Continuous Media}.\hskip 1em plus 0.5em minus 0.4em\relax Amsterdam: Elsevier
  Butterworth Heinemann, 2008.

\bibitem{Ashworth:PhysC99a}
S.~P. Ashworth and M.~Suenaga, ``{Measurement of ac losses in superconductors
  due to ac transport currents in applied ac magnetic fields},'' \emph{Physica
  C}, vol. 313, pp. 175--187, 1999.

\bibitem{Vojenciak:SST06}
M.~Vojenciak, J.~{\v S}ouc, J.~Ceballos, F.~G{\"o}m{\"o}ry, B.~Klin{\v c}ok,
  E.~Pardo, and F.~Grilli, ``Study of ac loss in {Bi-2223/Ag} tape under the
  simultaneous action of ac transport current and ac magnetic field shifted in
  phase,'' \emph{Superconductor Science and Technology}, vol.~19, no.~4, pp.
  397--404, 2006.

\bibitem{Pardo:SST08}
E.~Pardo, ``Modeling of coated conductor pancake coils with a large number of
  turns,'' \emph{Superconductor Science and Technology}, vol.~21, p. 065014,
  2008.

\bibitem{Carr:SST06b}
W.~J. {Carr Jr.}, ``{Hysteresis loss in a coated conductor subject to a
  combined applied magnetic field and transport current},''
  \emph{Superconductor Science and Technology}, vol.~19, pp. 454--458, 2006.

\bibitem{levitation}
C.~Navau and A.~Sanchez, ``{Modeling magnetization and levitation force in high
  temperature superconductors},'' \emph{This issue}, 2012.

\bibitem{Goldfarb91}
R.~B. Goldfarb, M.~Lelental, and C.~Thompson{, edited by R. A. Hein, p. 49},
  \emph{Magnetic Susceptibility of Superconductors and Other Spin
  Systems}.\hskip 1em plus 0.5em minus 0.4em\relax New York: Plenum, 1991.

\bibitem{Clem91}
J.~R. Clem{, edited by R. A. Hein, p. 177}, \emph{{Magnetic Susceptibility of
  Superconductors and Other Spin Systems}}.\hskip 1em plus 0.5em minus
  0.4em\relax New York: Plenum, 1991.

\bibitem{Fietz:PR64}
W.~A. Fietz, M.~R. Beasley, J.~Silcox, and W.~W. Webb, ``{Magnetization of
  superconducting Nb-25\% Zr wire},'' \emph{Physical Review}, vol. 136, no.~2A,
  p. A335, 1964.

\bibitem{Watson:JAP68}
J.~H.~P. Watson, ``{Magnetization of Synthetic Filamentary Superconductors. B.
  The Dependence of the Critical Current Density on Temperature and Magnetic
  Field},'' \emph{Journal of Applied Physics}, vol.~39, no.~7, pp. 3406--3413,
  1968.

\bibitem{Chen:JAP91}
D.~X. Chen and A.~Sanchez, ``{Theoretical critical-state susceptibility spectra
  and their application to high-$T_c$ superconductors},'' \emph{Journal of
  Applied Physics}, vol.~70, no.~10, pp. 5463--5477, 1991.

\bibitem{Amemiya:TAS97}
N.~Amemiya, K.~Miyamoto, N.~Banno, and O.~Tsukamoto, ``{Numerical Analysis of
  AC Losses in High Tc Superconductors Based on $E-j$ Characteristics
  Represented with $n$-Value},'' \emph{IEEE Transactions on Applied
  Superconductivity}, vol.~7, no.~2, pp. 2110--2113, 1997.

\bibitem{Chen:APL06}
D.~X. Chen and E.~Pardo, ``{Power-law $E(J)$} characteristic converted from
  field-amplitude and frequency dependent ac susceptibility in
  superconductors,'' \emph{Applied Physics Letters}, vol.~88, p. 222505, 2006.

\bibitem{Brandt:PRB97}
E.~H. Brandt, ``{Susceptibility of superconductor disks and rings with and
  without flux creep},'' \emph{Physical Review B}, vol.~55, pp.
  14\,513--14\,526, 1997.

\bibitem{Pardo:SST04b}
E.~Pardo, D.-X. Chen, A.~Sanchez, and C.~Navau, ``{The transverse
  critical-state susceptibility of rectangular bars},'' \emph{Supercond. Sci.
  Technol.}, vol.~17, pp. 537--544, 2004.

\bibitem{Clem:PRB94}
J.~R. Clem and A.~Sanchez, ``{Hysteretic ac losses and susceptibility of thin
  superconducting disks},'' \emph{Physical Review B}, vol.~50, no.~13, pp.
  9355--9362, 1994.

\bibitem{Krasnov:PhysC91}
V.~M. Krasnov, V.~A. Larkin, and V.~V. Ryazanov, ``{The extended bean critical
  state model for superconducting 3-axes ellipsoid and its application for
  obtaining the bulk critical field $H_{c1}$ and the pinning current $J_c$ in
  {high-$T_c$} superconducting single crystals},'' \emph{Physica C}, vol. 174,
  no. 4-6, pp. 440--446, 1991.

\bibitem{Gomory:SST02a}
F.~{G\"om\"ory}, R.~Tebano, A.~Sanchez, E.~Pardo, C.~Navau, I.~Husek,
  F.~Strycek, and P.~Kovac, ``{Current profiles and ac losses of a
  superconducting strip with an elliptic cross-section in a perpendicular
  magnetic field},'' \emph{Superconductor Science and Technology}, vol.~15, p.
  1311, 2002.

\bibitem{Bhagwat:PhysC95}
K.~V. Bhagwat and P.~Chaddah, ``{Flux penetration in thin elliptic
  superconducting cylinders subjected to transverse magnetic fields},''
  \emph{Physica C}, vol. 254, no. 1-2, pp. 143--150, 1995.

\bibitem{Chen:SST05}
D.~X. Chen, E.~Pardo, and A.~Sanchez, ``{Transverse ac susceptibility of
  superconducting bars with elliptical cross-section and constant
  critical-current density},'' \emph{Superconductor Science and Technology},
  vol.~18, p. 997, 2005.

\bibitem{Amemiya:PhysC98a}
N.~Amemiya, K.~Miyamoto, S.~Murasawa, H.~Mukai, and K.~Ohmatsu, ``{Finite
  element analysis of AC loss in non-twisted Bi-2223 tape carrying AC transport
  current and/or exposed to DC or AC external magnetic field},'' \emph{Physica
  C}, vol. 310, pp. 30--35, 1998.

\bibitem{Nibbio:TAS01}
N.~Nibbio, S.~Stavrev, and B.~Dutoit, ``Finite element method simulation of
  {AC} loss in {HTS} tapes with {B}-dependent {E-J} power law,'' \emph{IEEE
  Transactions on Applied Superconductivity}, vol.~11, no.~1, pp. 2631--2634,
  2001.

\bibitem{Mawatari:PRB11}
Y.~Mawatari, ``{Superconducting tubular wires in transverse magnetic fields},''
  \emph{Physical Review B}, vol.~83, no.~13, p. 134512, 2011.

\bibitem{Fukumoto:APL95}
Y.~Fukumoto, H.~J. Wiesmann, M.~Garber, M.~Suenaga, and P.~Haldar,
  ``{Alternating-current losses in silver-sheathed
  (Bi,Pb)$_2$Sr$_2$Ca$_2$Cu$_3$O$_{10}$ tapes II: Role of interfilamentary
  coupling},'' \emph{Applied Physics Letters}, vol.~67, pp. 3180--3182, 1995.

\bibitem{Pardo:PRB03}
E.~Pardo, A.~Sanchez, and C.~Navau, ``Magnetic properties of arrays of
  superconducting strips in a perpendicular field,'' \emph{Physical Review B},
  vol.~67, p. 104517, 2003.

\bibitem{Mawatari:PRB96}
Y.~Mawatari, ``Critical state of periodically arranged superconducting-strip
  lines in perpendicular fields,'' \emph{Physical Review B}, vol.~54, no.~18,
  pp. 13\,215--13\,221, 1996.

\bibitem{Ainbinder:SST03}
R.~M. Ainbinder and G.~M. Maksimova, ``{Hysteretic characteristics of a double
  stripline in the critical state},'' \emph{Superconductor Science and
  Technology}, vol.~16, no.~8, pp. 871--878, 2003.

\bibitem{Tebano:PhysC02}
R.~Tebano, F.~G{\"o}m{\"o}ry, E.~Seiler, and F.~Strycek, ``{Numerical
  investigations of the mutual magnetic coupling in superconducting tapes in
  z-stack arrangement with external AC magnetic field},'' \emph{Physica C},
  vol. 372, pp. 998--1000, 2002.

\bibitem{Grilli:PhysC06}
F.~Grilli, S.~P. Ashworth, and S.~Stavrev, ``{Magnetization AC losses of stacks
  of YBCO coated conductors},'' \emph{Physica C}, vol. 434, pp. 185--190, 2006.

\bibitem{Stavrev:PhysC03}
S.~Stavrev, B.~Dutoit, and P.~Lombard, ``Numerical modelling and {AC} losses of
  multifilamentary {Bi-2223/Ag} conductors with various geometry and filament
  arrangement,'' \emph{Physica C}, vol. 384, pp. 19--31, 2003.

\bibitem{Grilli:SST10b}
F.~Grilli and E.~Pardo, ``{Simulation of ac loss in Roebel coated conductor
  cables},'' \emph{Superconductor Science and Technology}, vol.~23, p. 115018,
  2010.

\bibitem{Rostila:TAS11}
L.~Rostila, E.~Demenc{\v c}{\`\i}k, J.~{\v S}ouc, S.~Brisigotti, P.~Kov{\'a}{\v
  c}, M.~Polak, G.~Grasso, M.~Lyly, A.~Stenvall, A.~Tumino, and L.~Kopera,
  ``{Magnesium Diboride Wires With Nonmagnetic Matrices---AC Loss Measurements
  and Numerical Calculations},'' \emph{IEEE Transactions on Applied
  Superconductivity}, vol.~21, no.~3, pp. 3338--3341, 2011.

\bibitem{Fabbricatore:PRB00}
P.~Fabbricatore, S.~Farinon, S.~Innocenti, and F.~G{\"o}m{\"o}ry, ``{Magnetic
  flux shielding in superconducting strip arrays},'' \emph{Physical Review B},
  vol.~61, no.~9, p. 6413, 2000.

\bibitem{Mikitik:PRB04}
G.~P. Mikitik, E.~H. Brandt, and M.~Indenbom, ``{Superconducting strip in an
  oblique magnetic field},'' \emph{Physical Review B}, vol.~70, p. 014520,
  2004.

\bibitem{Brandt:PRB05}
E.~H. Brandt and G.~P. Mikitik, ``{Anisotropic superconducting strip in an
  oblique magnetic field},'' \emph{Physical Review B}, vol.~72, p. 024516,
  2005.

\bibitem{Ichiki:PhysC04}
Y.~Ichiki and H.~Ohsaki, ``Numerical analysis of {AC} losses in {YBCO} coated
  conductor in external magnetic field,'' \emph{Physica C}, vol. 412-414, pp.
  1015--1020, 2004.

\bibitem{Enomoto:TAS05}
N.~Enomoto, T.~Izumi, and N.~Amemiya, ``{Electromagnetic Field Analysis of
  Rectangular Superconductor With Large Aspect Ratio in Arbitrary Orientated
  Magnetic Fields},'' \emph{IEEE Transactions on Applied Superconductivity},
  vol.~15, no.~2, pp. 1574--1577, 2005.

\bibitem{Stavrev:SST05}
S.~Stavrev, F.~Grilli, B.~Dutoit, and S.~Ashworth, ``Comparison of the {AC}
  losses of {BSCCO} and {YBCO} conductors by means of numerical analysis,''
  \emph{Superconductor Science and Technology}, vol.~18, no.~10, pp.
  1300--1312, 2005.

\bibitem{Karmakar:PRB01}
D.~Karmakar and K.~V. Bhagwat, ``{Magnetization of hard superconductor samples
  subjected to oblique fields},'' \emph{Physical Review B}, vol.~65, no.~2, p.
  024518, 2001.

\bibitem{Pardo:SST12}
E.~Pardo and F.~Grilli, ``{Numerical simulations of the angular dependence of
  magnetization AC losses: coated conductors, Roebel cables and double pancake
  coils},'' \emph{Superconductor Science and Technology}, vol.~25, p. 014008,
  2012.

\bibitem{Zhu:PhysC93}
J.~Zhu, J.~Mester, J.~Locldaart, and J.~Turneaure, ``Critical states in {2D}
  disk-shaped {type-II} superconductors in periodic external magnetic field,''
  \emph{Physica C}, vol. 212, pp. 216--222, 1993.

\bibitem{Brandt:PRB98b}
E.~H. Brandt, ``{Superconductor disks and cylinders in an axial magnetic field:
  II. Nonlinear and linear ac susceptibilities},'' \emph{Physical Review B},
  vol.~58, no.~10, pp. 6523--6533, 1998.

\bibitem{Navau:PRB05}
C.~Navau, A.~Sanchez, E.~Pardo, D.~X. Chen, E.~Bartolom{\'e}, X.~Granados,
  T.~Puig, and X.~Obradors, ``{Critical state in finite {type-II}
  superconducting rings},'' \emph{Physical Review B}, vol.~71, no.~21, p.
  214507, 2005.

\bibitem{Navarro:PRB91}
R.~Navarro and L.~J. Campbell, ``{Magnetic-flux profiles of high-$T_{c}$
  superconducting granules: {Three-dimensional} critical-state-model
  approximation},'' \emph{Physical Review B}, vol.~44, no.~18, p. 10146, 1991.

\bibitem{Navarro:SST92}
------, ``{Three dimensional solution of critical state models: {AC} hysteresis
  losses},'' \emph{Superconductor Science and Technology}, vol.~5, pp.
  S200--S203, 1992.

\bibitem{Telschow:PRB94}
K.~L. Telschow and L.~S. Koo, ``{Integral-equation approach for the Bean
  critical-state model in demagnetizing and nonuniform-field geometries},''
  \emph{Physical Review B}, vol.~50, no.~10, 1994.

\bibitem{Clem:CJP96}
J.~R. Clem, M.~Benkraouda, and J.~McDonald, ``{Penetration of Magnetic Flux and
  Electrical Current Density into Superconducting Strips and Disks},''
  \emph{Chinese Journal of Physics}, vol.~34, no. 2-11, pp. 284--290, 1996.

\bibitem{Hancox:PIEE66}
R.~Hancox, ``{Calculation of ac losses in a type-II superconductor},''
  \emph{Proceedings of the Institute Electrical Engineers}, vol. 113, no.~7,
  pp. 1221--1228, 1966.

\bibitem{Chen:APL05}
D.~X. Chen and C.~Gu, ``{Alternating current loss in a cylinder with power-law
  current-voltage characteristic},'' \emph{Applied Physics Letters}, vol.~86,
  p. 252504, 2005.

\bibitem{Gomory:PhysC97}
F.~G{\"o}m{\"o}ry and L.~Gherardi, ``{Transport AC losses in round
  superconducting wire consisting of two concentric shells with different
  critical current density},'' \emph{Physica C}, vol. 280, no.~3, pp. 151--157,
  1997.

\bibitem{Kajikawa:SST04}
K.~Kajikawa, Y.~Mawatari, T.~Hayashi, and K.~Funaki, ``{AC loss evaluation of
  thin superconducting wires with critical current distribution along width},''
  \emph{Superconductor Science and Technology}, vol.~17, pp. 555--563, 2004.

\bibitem{Norris:JPDAP71}
W.~T. Norris, ``{Calculation of hysteresis losses in hard superconductors:
  polygonal-section conductors},'' \emph{Journal of Physics D: Applied
  Physics}, vol.~4, pp. 1358--1364, 1971.

\bibitem{Fukunaga:APL98}
T.~Fukunaga, R.~Inada, and A.~Oota, ``{Field-free core, current distribution,
  and alternating current losses in self fields for rectangular superconductor
  tapes},'' \emph{Applied Physics Letters}, vol.~72, p. 3362, 1998.

\bibitem{Daumling:SST98}
M.~D{\"a}umling, ``{Ac power loss for superconducting strips of arbitrary
  thickness in the critical state carrying a transport current},''
  \emph{Superconductor Science and Technology}, vol.~11, pp. 590--593, 1998.

\bibitem{Gu:TAS05}
C.~Gu and Z.~Han, ``{Calculation of AC losses in HTS tape with FEA program
  ANSYS},'' \emph{IEEE Transactions on Applied Superconductivity}, vol.~15,
  no.~2, pp. 2859--2862, 2005.

\bibitem{Fukunaga:TAS99}
T.~Fukunaga, R.~Inada, and A.~Oota, ``{Current Distributions and AC Losses in
  Self-Fields for Superconductor Tapes and Cables},'' \emph{IEEE Transactions
  on Applied Superconductivity}, vol.~9, no.~2, pp. 1057--1060, 1999.

\bibitem{Oota:PhysC03}
A.~Oota, R.~Inada, N.~Inagaki, P.~X. Zhang, and H.~Fujimoto, ``{A progress in
  reducing {AC} transport losses of {Ag}-sheathed {Bi2223} tapes by a
  rectangular deformation using two-axial rollers},'' \emph{Physica C}, vol.
  386, pp. 100--105, 2003.

\bibitem{Pardo:PRB05}
E.~Pardo, A.~Sanchez, D.-X. Chen, and C.~Navau, ``Theoretical analysis of the
  transport critical-state ac loss in arrays of superconducting rectangular
  strips,'' \emph{Physical Review B}, vol.~71, p. 134517, 2005.

\bibitem{Stavrev:TAS03b}
S.~Stavrev, B.~Dutoit, and P.~Lombard, ``{AC losses of multifilamentary
  {Bi-2223/Ag} conductors with different geometry and filament arrangement},''
  \emph{IEEE Transactions on Applied Superconductivity}, vol.~13, no.~2, pp.
  3561--3565, 2003.

\bibitem{Carr:TMAG79}
W.~J. {Carr Jr.}, ``{AC Loss from the combined action of transport current and
  applied field},'' \emph{IEEE Transactions on Magnetics}, vol.~15, no.~1, pp.
  240--243, 1979.

\bibitem{Brandt:PRB93}
E.~H. Brandt and M.~Indenbom, ``{Type-II-superconductor strip with current in a
  perpendicular magnetic field},'' \emph{Physical Review B}, vol.~48, no.~17,
  pp. 12\,893--12\,906, 1993.

\bibitem{Zeldov:PRB94}
E.~Zeldov, J.~Clem, M.~{McElfresh}, and M.~Darwin, ``Magnetization and
  transport currents in thin superconducting films,'' \emph{Physical Review B},
  vol.~49, no.~14, pp. 9802--9822, 1994.

\bibitem{Schonborg:JAP01}
N.~Schonborg, ``{Hysteresis losses in a thin high-temperature superconductor
  strip exposed to ac transport currents and magnetic fields},'' \emph{Journal
  of Applied Physics}, vol.~90, no.~6, pp. 2930--2933, 2001.

\bibitem{Zannella:TAS01}
S.~Zannella, L.~Montelatici, G.~Grenci, M.~Pojer, L.~Jansak, M.~Majoros,
  G.~Coletta, R.~Mele, R.~Tebano, and F.~Zanovello, ``{AC} losses in transport
  current regime in applied {AC} magnetic field: Experimental analysis and
  modeling,'' \emph{IEEE Transactions on Applied Superconductivity}, vol.~11,
  no.~1, pp. 2441--2444, 2001.

\bibitem{Takacs:SST07}
S.~Takacs, ``{Hysteresis losses in superconductors with an out-of-phase applied
  magnetic field and current: slab geometry},'' \emph{Superconductor Science
  and Technology}, vol.~20, pp. 1093--1096, 2007.

\bibitem{Mawatari:APL07}
Y.~Mawatari and K.~Kajikawa, ``{Hysteretic ac loss of superconducting strips
  simultaneously exposed to ac transport current and phase-different ac
  magnetic field},'' \emph{Applied Physics Letters}, vol.~90, p. 022506, 2007.

\bibitem{Kajikawa:TAS01}
K.~Kajikawa, A.~Takenaka, K.~Kawasaki, M.~Iwakuma, and K.~Funaki, ``{Numerical
  simulation for AC losses of HTS tapes in combined alternating transport
  current and external AC magnetic field with phase shift},'' vol.~11, no.~1,
  pp. 2240--2243, 2001.

\bibitem{Nguyen:JAP05}
D.~N. Nguyen, P.~V. Sastry, D.~C. Knoll, G.~Zhang, and J.~Schwartz,
  ``{Experimental and numerical studies of the effect of phase difference
  between transport current and perpendicular applied magnetic field on total
  ac loss in Ag-sheathed (Bi, Pb) SrCaCuO tape},'' \emph{Journal of Applied
  Physics}, vol.~98, p. 073902, 2005.

\bibitem{Vellego:SST95}
G.~Vellego and P.~Metra, ``An analysis of the transport losses measured on
  {HTSC} single-phase conductor prototypes,'' \emph{Superconductor Science and
  Technology}, vol.~8, pp. 476--483, 1995.

\bibitem{Grilli:TAS03}
F.~Grilli, S.~Stavrev, B.~Dutoit, and S.~Spreafico, ``{Numerical modeling of a
  HTS cable},'' \emph{IEEE Transactions on Applied Superconductivity}, vol.~13,
  no.~2, pp. 1886--1889, 2003.

\bibitem{Grilli:SST04}
F.~Grilli and M.~Sj{\"o}str{\"o}m, ``{Prediction of resistive and hysteretic
  losses in a multi-layer high-$\rm T_c$ superconducting cable},''
  \emph{Superconductor Science and Technology}, vol.~17, no.~3, pp. 409--416,
  2004.

\bibitem{Klincok:JOPCS06}
B.~Klin{\v c}ok and F.~G{\"o}m{\"o}ry, ``{Influence of gaps in monolayer
  superconducting cable on AC losses},'' \emph{Journal of Physics: Conference
  Series}, vol.~43, pp. 897--900, 2006.

\bibitem{Miyagi:PhysC07}
D.~Miyagi, M.~Umabuchi, N.~Takahashi, and O.~Tsukamoto, ``{Study of AC
  transport current loss of assembled HTS coated-conductors with ferromagnetic
  substrate using FEM},'' \emph{Physica C}, vol. 463, pp. 785--789, 2007.

\bibitem{Amemiya:PhysC08}
N.~Amemiya, Z.~Jiang, Z.~Li, M.~Nakahata, T.~Kato, M.~Ueyama, N.~Kashima,
  S.~Nagaya, and S.~Shiohara, ``{Transport losses in single and assembled
  coated conductors with textured-metal substrate with reduced magnetism},''
  \emph{Physica C}, vol. 468, no.~15, pp. 1718--1722, 2008.

\bibitem{Rostila:PhysC10}
L.~Rostila, S.~Suuriniemi, J.~Lehtonen, and G.~Grasso, ``{Numerical
  minimization of AC losses in coaxial coated conductor cables},''
  \emph{Physica C}, vol. 470, no.~3, pp. 212--217, 2010.

\bibitem{Jiang:SST08b}
Z.~Jiang, N.~Amemiya, and M.~Nakahata, ``Numerical calculation of {AC} losses
  in multi-layer superconducting cables composed of coated conductors,''
  \emph{Superconductor Science and Technology}, vol.~21, p. 025013, 2008.

\bibitem{Nakahata:SST08}
J.~Nakahata and N.~Amemiya, ``Electromagnetic field analyses of two-layer power
  transmission cables consisting of coated conductors with magnetic and
  non-magnetic substrates and ac losses in their superconductor layers,''
  \emph{Superconductor Science and Technology}, vol.~21, p. 015007, 2008.

\bibitem{Amemiya:SST10}
N.~Amemiya, M.~Nakahata, N.~Fujiwara, and Y.~Shiohara, ``{Ac losses in
  two-layer superconducting power transmission cables consisting of coated
  conductors with a magnetic substrate},'' \emph{Superconductor Science and
  Technology}, vol.~23, p. 014022, 2010.

\bibitem{Inada:JOPCS08}
R.~Inada, Y.~Nakamura, and A.~Oota, ``{Numerical analysis for AC losses in
  single-layer cables composed of rectangular superconducting strips with
  various lateral $J_c$ distributions},'' \emph{Journal of Physics: Conference
  Series}, vol.~97, p. 012324, 2008.

\bibitem{Mawatari:SST10}
Y.~Mawatari, A.~Malozemoff, T.~Izumi, K.~Tanabe, N.~Fujiwara, and Y.~Shiohara,
  ``Hysteretic ac losses in power transmission cables with superconducting
  tapes: effect of tape shape,'' \emph{Superconductor Science and Technology},
  vol.~23, p. 025031, 2010.

\bibitem{Ohmatsu:TAS04}
K.~Ohmatsu, K.~Muranaka, K.~Fujino, M.~Konishi, and K.~Yasuda, ``{Design study
  of model cable conductor by using HoBCO thin film tape},'' \emph{IEEE
  Transactions on Applied Superconductivity}, vol.~14, no.~2, pp. 620--625,
  2004.

\bibitem{Nakamura:TAS05}
T.~Nakamura, H.~Kanzaki, K.~Higashikawa, T.~Hoshino, and I.~Muta, ``{Analysis
  of shielding layers in {HTS} cable taking account of spiral structure},''
  \emph{IEEE Transactions on Applied Superconductivity}, vol.~15, no.~2, pp.
  1747--1750, 2005.

\bibitem{Siahrang:TAS10}
M.~Siahrang, F.~Sirois, D.~N. Nguyen, S.~Babic, and S.~P. Ashworth, ``{Fast
  Numerical Computation of Current Distribution and AC Losses in Helically
  Wound Thin Tape Conductors: Single-Layer Coaxial Arrangement},'' \emph{IEEE
  Transactions on Applied Superconductivity}, vol.~17, no.~6, pp. 2381--2389,
  2010.

\bibitem{Terzieva:SST10}
S.~Terzieva, M.~Vojenciak, E.~Pardo, F.~Grilli, A.~Drechsler, A.~Kling,
  A.~Kudymow, F.~G\"om\"ory, and W.~Goldacker, ``Transport and magnetization ac
  losses of {ROEBEL} assembled coated conductor cables: measurements and
  calculations,'' \emph{Superconductor Science and Technology}, vol.~23, p.
  014023, 2010.

\bibitem{Jiang:SST11}
Z.~Jiang, K.~P. Thakur, M.~Staines, R.~A. Badcock, N.~J. Long, R.~G. Buckley,
  A.~D. Caplin, and N.~Amemiya, ``{The dependence of AC loss characteristics on
  the spacing between strands in YBCO Roebel cables},'' \emph{Superconductor
  Science and Technology}, vol.~24, no.~6, p. 065005, 2011.

\bibitem{Zermeno:SST13}
V.~Zermeno, F.~Grilli, and F.~Sirois, ``{A full 3-D time-dependent
  electromagnetic model for Roebel cables},'' \emph{Superconductor Science and
  Technology}, p. In press, 2013.

\bibitem{Pitel:SST02}
J.~Pitel, A.~Korpela, J.~Lehtonen, and P.~Kovac, ``{Mathematical model of
  voltage--current characteristics of {Bi(2223)/Ag} magnets under an external
  magnetic field},'' \emph{Superconductor Science and Technology}, vol.~15, pp.
  1499--1506, 2002.

\bibitem{Oomen:SST03}
M.~P. Oomen, R.~Nanke, and M.~Leghissa, ``Modelling and measurement of ac loss
  in {BSCCO/Ag}-tape windings,'' \emph{Superconductor Science and Technology},
  vol.~16, pp. 339--354, 2003.

\bibitem{Kawagoe:TAS04}
A.~Kawagoe, F.~Sumiyoshi, T.~Mito, H.~Chikaraishi, T.~Baba, K.~Okumura,
  M.~Iwakuma, T.~Hemmi, K.~Hayashi, R.~Abe, T.~Ushiku, and K.~Miyoshi,
  ``{Winding techniques for conduction cooled LTS pulse coils for 100 kJ class
  UPS-SMES as a protection from momentary voltage drops},'' \emph{IEEE
  Transactions on Applied Superconductivity}, vol.~14, no.~2, pp. 727--730,
  2004.

\bibitem{Tonsho:TAS04}
H.~Tonsho, M.~Toyoda, S.~Fukui, M.~Yamaguchi, T.~Sato, M.~Furuse, H.~Tanaka,
  K.~Arai, , and M.~Umeda, ``Numerical evaluation of {AC} loss in high
  temperature superconducting coil,'' \emph{IEEE Transactions on Applied
  Superconductivity}, vol.~14, pp. 674--677, 2004.

\bibitem{Muller:PhysC97b}
K.-H. M{\"u}ller, ``{AC power losses in flexible thick-film superconducting
  tapes},'' \emph{Physica C}, vol. 281, pp. 1--10, 1997.

\bibitem{Brambilla:SST09}
R.~Brambilla, F.~Grilli, D.~N. Nguyen, L.~Martini, and F.~Sirois, ``{AC losses
  in thin superconductors: the integral equation method applied to stacks and
  windings},'' \emph{Superconductor Science and Technology}, vol.~22, no.~7, p.
  075018, 2009.

\bibitem{Grilli:SST07}
F.~Grilli and S.~P. Ashworth, ``{Measuring transport AC losses in YBCO-coated
  conductor coils},'' \emph{Superconductor Science and Technology}, vol.~20,
  no.~8, pp. 794--799, 2007.

\bibitem{Zhang:JAP12}
M.~Zhang, J.-H. Kim, S.~Pamidi, M.~Chudy, W.~Yuan, and T.~A. Coombs, ``{Study
  of second generation, high-temperature superconducting coils: Determination
  of critical current},'' \emph{Journal of Applied Physics}, vol. 111, no.~8,
  p. 083902, 2012.

\bibitem{Ainslie:SST11}
M.~D. Ainslie, V.~M. Rodriguez-Zermeno, Z.~Hong, W.~Yuan, T.~J. Flack, and
  T.~A. Coombs, ``{An improved FEM model for computing transport AC loss in
  coils made of RABiTS YBCO coated conductors for electric machines},''
  \emph{Superconductor Science and Technology}, vol.~24, p. 045005, 2011.

\bibitem{Yuan:SST09}
W.~Yuan, A.~M. Campbell, and T.~A. Coombs, ``A model for calculating the {AC}
  losses of second-generation high temperature superconductor pancake coils,''
  \emph{Superconductor Science and Technology}, vol.~22, p. 075028, 2009.

\bibitem{Claassen:APL06}
J.~H. Claassen, ``{An approximate method to estimate ac loss in tape-wound
  superconducting coils},'' \emph{Applied Physics Letters}, vol.~88, p. 122512,
  2006.

\bibitem{Clem:SST07}
J.~R. Clem, J.~H. Claassen, and Y.~Mawatari, ``{AC losses in a finite Z stack
  using an anisotropic homogeneous-medium approximation},''
  \emph{Superconductor Science and Technology}, vol.~20, pp. 1130--1139, 2007.

\bibitem{Shevchenko:PhysC98}
O.~A. Shevchenko, J.~J. Rabbers, A.~Godeke, B.~{ten Haken}, and H.~H.~J. {ten
  Kate}, ``{AC loss in a high-temperature superconducting coil},''
  \emph{Physica C}, vol. 310, no. 1-4, pp. 106--110, 1998.

\bibitem{Pardo:SST09}
E.~Pardo, J.~{\v Souc}, and M.~{Vojen\v ciak}, ``{AC loss measurement and
  simulation of a coated conductor pancake coil with ferromagnetic parts},''
  \emph{Superconductor Science and Technology}, vol.~22, p. 075007, 2009.

\bibitem{Ainslie:PhysC12}
M.~D. Ainslie, T.~J. Flack, and A.~M. Campbell, ``{Calculating transport AC
  losses in stacks of high temperature superconductor coated conductors with
  magnetic substrates using FEM},'' \emph{Physica C: Superconductivity}, vol.
  472, no.~1, pp. 50--56, 2012.

\bibitem{Clem:PRB08}
J.~R. Clem, ``{Field and current distributions and ac losses in a bifilar stack
  of superconducting strips},'' \emph{Physical Review B}, vol.~77, p. 134506,
  2008.

\bibitem{Nguyen:SST09}
D.~N. Nguyen, F.~Grilli, S.~P. Ashworth, and J.~O. Willis, ``Ac loss study of
  antiparallel connected {YBCO} coated conductors,'' \emph{Superconductor
  Science and Technology}, vol.~22, no.~5, p. 055014, 2009.

\bibitem{Souc:SST12}
J.~{\v S}ouc, F.~G{\"o}m{\"o}ry, and M.~Vojen{\v c}iak, ``{Coated conductor
  arrangement for reduced AC losses in a resistive-type superconducting fault
  current limiter},'' \emph{Superconductor Science and Technology}, vol.~25,
  no.~1, p. 014005, 2012.

\bibitem{Majoros:TAS07}
M.~Majoros, L.~Ye, A.~M. Campbell, T.~A. Coombs, M.~D. Sumption, and E.~W.
  Collings, ``{Modeling of Transport AC Losses in Superconducting Arrays
  Carrying Anti-Parallel Currents},'' \emph{IEEE Transactions on Applied
  Superconductivity}, vol.~17, no.~2, pp. 1803--1806, 2007.

\bibitem{Mawatari:APL99}
Y.~Mawatari and H.~Yamasaki, ``{Alternating current loss in coplanar arrays of
  superconducting strips with bidirectional currents},'' \emph{Applied physics
  letters}, vol.~75, p. 406, 1999.

\bibitem{BadiaMajos:PRB07}
A.~Bad{\'\i}a-Maj{\'o}s and C.~L{\'o}pez, ``{Critical-state analysis of
  orthogonal flux interactions in pinned superconductors},'' \emph{Physical
  Review B}, vol.~76, no.~5, p. 054504, 2007.

\bibitem{Vanderbemden:SST07}
P.~Vanderbemden, Z.~Hong, T.~A. Coombs, M.~Ausloos, N.~H. Babu, D.~A. Cardwell,
  and A.~M. Campbell, ``{Remagnetization of bulk high-temperature
  superconductors subjected to crossed and rotating magnetic fields},''
  \emph{Superconductor Science and Technology}, vol.~20, pp. S174--S183, 2007.

\bibitem{Clem:PRB82}
J.~R. Clem, ``{Flux-line-cutting losses in type-II superconductors},''
  \emph{Physical Review B}, vol.~26, no.~5, p. 2463, 1982.

\bibitem{Clem:PRB84}
J.~R. Clem and A.~Perez-Gonzalez, ``{Flux-line-cutting and flux-pinning losses
  in type-II superconductors in rotating magnetic fields},'' \emph{Physical
  Review B}, vol.~30, pp. 5041--5047, 1984.

\bibitem{Ruiz:SST10}
H.~Ruiz and A.~Bad{\'\i}a-Maj{\'o}s, ``{Smooth double critical state theory for
  type-II superconductors},'' \emph{Superconductor Science and Technology},
  vol.~23, p. 105007, 2010.

\bibitem{Karmakar:PhysC04}
D.~Karmakar and K.~V. Bhagwat, ``{Critical state behaviour under rotating
  magnetic field: minimum flux-change technique},'' \emph{Physica C}, vol. 406,
  no.~3, pp. 210--222, 2004.

\bibitem{Meerovich:SST96}
V.~Meerovich, M.~Sinder, V.~Sokolovsky, S.~Goren, G.~Jung, G.~E. Shter, and
  G.~S. Grader, ``{Penetration dynamics of a magnetic field pulse into
  high-$T_c$ superconductors},'' \emph{Superconductor Science and Technology},
  vol.~9, pp. 1042--1047, 1996.

\bibitem{Takacs:SST05}
S.~Tak{\'a}cs and F.~G{\"o}m{\"o}ry, ``{Hysteresis and coupling losses of
  superconducting cables at additional change of the applied magnetic field},''
  \emph{Superconductor Science and Technology}, vol.~18, pp. 340--345, 2005.

\bibitem{Kajikawa:TAS08}
K.~Kajikawa, R.~Yokoo, K.~Tomachi, K.~Enpuku, K.~Funaki, H.~Hayashi, and
  H.~Fujishiro, ``{Numerical evaluation of pulsed field magnetization in a bulk
  superconductor using energy minimization technique},'' \emph{IEEE
  Transactions on Applied Superconductivity}, vol.~18, no.~2, pp. 1557--1560,
  2008.

\bibitem{Fujishiro:SST10}
H.~Fujishiro and T.~Naito, ``{Simulation of temperature and magnetic field
  distribution in superconducting bulk during pulsed field magnetization},''
  \emph{Superconductor Science and Technology}, vol.~23, p. 105021, 2010.

\bibitem{Kawagoe:TAS03}
A.~Kawagoe, F.~Sumiyoshi, M.~Nakanishi, T.~Mito, and T.~Kawashima, ``{A new
  winding method to reduce AC losses in stable LTS pulse coils},'' vol.~13,
  no.~2, pp. 2404--2407, 2003.

\bibitem{Lee:TAS06}
S.~Lee, Y.~Chu, W.~H. Chung, S.~J. Lee, S.~M. Choi, S.~H. Park, H.~Yonekawa,
  S.~H. Baek, J.~S. Kim, K.~W. Cho, K.~R. Park, B.~S. Lim, Y.~K. Oh, K.~Kim,
  J.~S. Bak, and G.~S. Lee, ``{AC loss characteristics of the KSTAR CSMC
  estimated by pulse test},'' \emph{IEEE Transactions on Applied
  Superconductivity}, vol.~16, no.~2, pp. 771--774, 2006.

\bibitem{Hong:TAS11b}
Z.~Hong, W.~Yuan, M.~Ainslie, Y.~Yan, R.~Pei, and T.~Coombs, ``{AC Losses of
  Superconducting Racetrack Coil in Various Magnetic Conditions},'' \emph{IEEE
  Transactions on Applied Superconductivity}, vol.~21, no.~3, pp. 2466--2469,
  2011.

\bibitem{Bottura:TAS07}
L.~Bottura, P.~Bruzzone, J.~B. Lister, C.~Marinucci, and A.~Portone,
  ``{Computations of AC Loss in the ITER Magnets During Fast Field
  Transients},'' \emph{IEEE Transactions on Applied Superconductivity},
  vol.~17, no.~2, pp. 2438--2441, 2007.

\bibitem{Giordano:SST06}
J.~L. Giordano, J.~Luzuriaga, A.~Bad{\'\i}a-Maj{\'o}s, G.~Nieva, and
  I.~Ruiz-Tagle, ``{Magnetization collapse in polycrystalline YBCO under
  transport current cycles},'' \emph{Superconductor Science and Technology},
  vol.~19, pp. 385--391, 2006.

\bibitem{Ogasawara:Cryo76}
T.~Ogasawara, K.~Yasuk{\"o}chi, S.~Nose, and H.~Sekizawa, ``Effective
  resistance of current-carrying superconducting wire in oscillating magnetic
  fields 1: Single core composite conductor,'' \emph{Cryogenics}, vol.~16,
  no.~1, pp. 33--38, 1976.

\bibitem{Oomen:SST99}
M.~P. Oomen, J.~Rieger, M.~Leghissa, B.~{ten Haken}, and H.~H.~J. {ten Kate},
  ``{Dynamic resistance in a slab-like superconductor with $Jc(B)$
  dependence},'' \emph{Superconductor Science and Technology}, vol.~12, pp.
  382--387, 1999.

\bibitem{Schonborg:TAS01}
N.~Sch{\"o}nborg and S.~P. H{\"o}rnfeldt, ``Losses in a high-temperature
  superconductor exposed to {AC} and {DC} transport currents and magnetic
  fields,'' \emph{IEEE Transactions on Applied Superconductivity}, vol.~11,
  no.~3, pp. 4086--4089, 2001.

\bibitem{Wang:TAS11}
Y.~Wang, Y.~Zheng, H.~Liuliu, S.~Dai, H.~Zhang, X.~Guan, Y.~Teng, L.~Zhao,
  J.~Xue, and L.~Lin, ``{A Novel Approach for Design of DC HTS Cable},''
  \emph{IEEE Transactions on Applied Superconductivity}, no.~99, pp. 1--4,
  2011.

\bibitem{Fisher:PhysC97}
L.~M. Fisher, A.~V. Kalinov, S.~E. Savel'ev, I.~F. Voloshin, V.~A. Yampol'skii,
  M.~A.~R. LeBlanc, and S.~Hirscher, ``{Collapse of the magnetic moment in a
  hard superconductor under the action of a transverse ac magnetic field},''
  \emph{Physica C}, vol. 278, no. 3-4, pp. 169--179, 1997.

\bibitem{Badia:PRL01}
A.~Bad{\'\i}a and C.~L{\'o}pez, ``{Critical state theory for nonparallel flux
  line lattices in type-II superconductors},'' \emph{Physical Review Letters},
  vol.~87, no.~12, p. 127004, 2001.

\bibitem{Brandt:PRL02}
E.~H. Brandt and G.~P. Mikitik, ``{Why an ac magnetic field shifts the
  irreversibility line in type-II superconductors},'' \emph{Physical Review
  Letters}, vol.~89, no.~2, p. 27002, 2002.

\bibitem{Vanderbemden:PRB07}
P.~Vanderbemden, Z.~Hong, T.~A. Coombs, S.~Denis, M.~Ausloos, J.~Schwartz,
  I.~B. Rutel, N.~H. Babu, D.~A. Cardwell, and A.~M. Campbell, ``{Behavior of
  bulk high-temperature superconductors of finite thickness subjected to
  crossed magnetic fields: Experiment and model},'' \emph{Physical Review B},
  vol.~75, no.~17, p. 174515, 2007.

\bibitem{Mikitik:PRB03}
G.~P. Mikitik and E.~H. Brandt, ``{Theory of the longitudinal vortex-shaking
  effect in superconducting strips},'' \emph{Physical Review B}, vol.~67,
  no.~10, p. 104511, 2003.

\bibitem{Kriezis:PIEEE92}
E.~E. Kriezis, T.~D. Tsiboukis, S.~M. Panas, and J.~A. Tegopoulos, ``{Eddy
  Currents: Theory and Applications},'' \emph{Proceedings of the IEEE},
  vol.~80, no.~10, pp. 1559--1589, 1992.

\bibitem{Nedelec:SIAMJNA78}
J.~C. Nedelec, ``{Computation of Eddy Currents on a Surface in $R^3$ by Finite
  Element Methods},'' \emph{SIAM Journal on Numerical Analysis}, vol.~15,
  no.~3, pp. 580--594, 1978.

\bibitem{Verite:IJEPES79}
J.~C. V{\'e}rit{\'e}, ``{Computation of eddy currents on the alternator output
  conductors by a finite element method},'' \emph{International Journal of
  Electrical Power and Energy Systems}, vol.~1, no.~3, pp. 193--198, 1979.

\bibitem{Chari:TPAS74}
M.~V.~K. Chari, ``{Finite-Element Solution of the Eddy-Current Problem in
  Magnetic Structures},'' \emph{IEEE Transactions on Power Apparatuses and
  Systems}, vol.~1, pp. 62--72, 1974.

\bibitem{McWhirter:TMAG79}
J.~H. McWhirter, R.~J. Duffin, P.~J. Brehm, and J.~J. Oravec, ``{Computational
  Methods for Solving Static Field and Eddy Current Problems via Fredholm
  Integral Equations},'' \emph{IEEE Transactions on Magnetics}, vol.~15, no.~3,
  pp. 1075--1084, 1979.

\bibitem{Bossavit:TMAG82}
A.~Bossavit and J.-C. {V\'erit\'e}, ``{A mixed FEM-BIEM method to solve 3-D
  eddy-current problems},'' \emph{IEEE Transactions on Magnetics}, vol.~18,
  no.~2, pp. 431--435, 1982.

\bibitem{Coulomb:TMAG81}
J.-L. Coulomb, ``{Finite elements three dimensional magnetic field
  computation},'' \emph{IEEE Transactions on Magnetics}, vol.~17, no.~6, pp.
  3241--3246, 1981.

\bibitem{Biro:TMAG90}
O.~{B\'ir\'o} and K.~Preis, ``{Finite element analysis of 3-D eddy currents},''
  \emph{IEEE Transactions on Magnetics}, vol.~26, no.~2, pp. 418--423, 1990.

\bibitem{Bossavit:CMAME81}
A.~Bossavit, ``{On the numerical analysis of eddy-current problems},''
  \emph{Computer Methods in Applied Mechanics and Engineering}, vol.~27, no.
  303-318, 1981.

\bibitem{Bossavit:IEOT82}
------, ``{Parallel eddy-currents: relevance of the boundary-operator
  approach},'' \emph{Integral Equations and Operator Theory}, vol.~5, pp.
  447--457, 1982.

\bibitem{Trowbridge:TMAG82}
C.~W. Trowbridge, ``{Three-Dimensional Field Computation},'' \emph{IEEE
  Transactions on Magnetics}, vol.~18, no.~1, pp. 293--297, 1982.

\bibitem{Carpenter:TMAG75}
C.~J. Carpenter, ``A network approach to the numerical solution of eddy-current
  problems,'' \emph{IEEE Transactions on Magnetics}, vol.~11, no.~5, pp.
  1517--1522, 1975.

\bibitem{Yamazaki_PIEMD99}
K.~Yamazaki, ``{Modeling and Analysis of Canned Motors for Hermetic Compressors
  Using Combination of 2D and 3D Finite Element Method},'' \emph{Proceedings of
  the International Conference IEMD 1999}, pp. 377--399, 1999.

\bibitem{Biblecombe:TMAG82}
C.~S. Biddlecombe, E.~A. Heighway, J.~Simkin, and C.~W. Trowbridge, ``{Methods
  for eddy current computation in three dimensions},'' \emph{IEEE Transactions
  on Magnetics}, vol.~18, no.~2, pp. 492--497, 1982.

\bibitem{Kettunen:TMAG92}
L.~Kettunen and L.~R. Turner, ``{A Volume Integral Formulation for Nonlinear
  Magnetostatics and Eddy Currents Using Edge Elements},'' \emph{IEEE
  Transactions on Magnetics}, vol.~28, no.~2, pp. 1639--1642, 1992.

\bibitem{Lari:TMAG83}
R.~J. Lari and L.~R. Turner, ``{Survey of eddy current programs},'' \emph{IEEE
  Transactions on Magnetics}, vol.~19, no.~6, pp. 2474--2477, 1983.

\bibitem{Salon:TMAG82}
S.~J. Salon and J.~M. Schneider, ``{A hybrid finite element-boundary integral
  formulation of the eddy current problem},'' \emph{IEEE Transactions on
  Magnetics}, vol.~18, no.~2, pp. 461--466, 1982.

\bibitem{Tarhasaari:TMAG98}
T.~Tarhasaari, A.~Koski, K.~Forsman, and L.~Kettunen, ``{Hybrid Formulations
  for Eddy Current Problem with Moving Objects},'' \emph{IEEE Transactions on
  Magnetics}, vol.~34, no.~5, pp. 2660--2663, 1998.

\bibitem{Elliott:IMAJNA06}
C.~M. Elliott and Y.~Kashima, ``{A finite-element analysis of critical-state
  models for type-II superconductivity in 3D},'' \emph{IMA Journal of Numerical
  Analysis}, vol.~27, pp. 293--331, 2007.

\bibitem{Biro:CMAME99}
O.~B{\'\i}r{\'o}, ``{Edge element formulations of eddy current problems},''
  \emph{Computer Methods in Applied Mechanics and Engineering}, vol. 169, pp.
  391--405, 1999.

\bibitem{Bossavit:IEEEPA88}
A.~Bossavit, ``{Whitney forms: a class of finite elements for three-dimensional
  computations in electromagnetism},'' \emph{IEE Proceedings A - Physical
  Science, Measurement and Instrumentation, Management and Education -
  Reviews}, vol. 135, no.~8, pp. 493--500, 1988.

\bibitem{Bossavit:TMAG88}
------, ``{A rationale for ``edge-elements'' in 3-D fields computations},''
  \emph{IEEE Transactions on Magnetics}, vol.~24, no.~1, pp. 74--79, 1988.

\bibitem{Webb:TMAG90}
J.~P. Webb and B.~Forghani, ``{A scalar-vector method for 3D eddy current
  problems using edge elements},'' \emph{IEEE Transactions on Magnetics},
  vol.~26, no.~5, pp. 2367--2369, 1990.

\bibitem{Kameari:TMAG90}
A.~Kameari, ``{Calculation of transient 3D eddy current using edge-elements},''
  \emph{IEEE Transactions on Magnetics}, vol.~26, no.~2, pp. 466--469, 1990.

\bibitem{Preis:TMAG92}
K.~Preis, I.~Bardi, O.~Biro, C.~Magele, G.~Vrisk, and K.~R. Richter,
  ``{Different Finite Element Formulations of 3D Magnetostatic Fields},''
  \emph{IEEE Transactions on Magnetics}, vol.~28, no.~2, pp. 1056--1059, 1992.

\bibitem{Brambilla:SST07}
R.~Brambilla, F.~Grilli, and L.~Martini, ``{Development of an edge-element
  model for AC loss computation of high-temperature superconductors},''
  \emph{Superconductor Science and Technology}, vol.~20, no.~1, pp. 16--24,
  2007.

\bibitem{Flux}
Electromagnetic and thermal finite element analysis software {Flux}.
  http://www.cedrat.com.

\bibitem{Opera}
Electromagnetic design software Opera 2D and Opera 3D. http://www.cobham.com.

\bibitem{Comsol}
Finite-element software package {Comsol Multiphysics}. http://www.comsol.com.

\bibitem{Geuzaine:IJNME09}
C.~Geuzaine and J.-F. Remacle, ``{Gmsh: A 3-D finite element mesh generator
  with built-in pre- and post-processing facilities},'' \emph{International
  Journal for Numerical Methods in Engineering}, vol.~79, pp. 1309--1331, 2009.

\bibitem{Launay:JPCS59}
J.~de~Launay, R.~L. Dolecek, and R.~T. Webber, ``{Magnetoresistance of
  copper},'' \emph{Journal of Physics and Chemistry of Solids}, vol.~11, pp.
  37--42, 1959.

\bibitem{Lengeler_JLTP70}
B.~Lengeler, W.~Schilling, and H.~Wenzl, ``{Deviations from Matthiessen's Rule
  and Longitudinal Magnetoresistance in Copper},'' \emph{Journal of Low
  Temperature Physics}, vol.~2, no.~1, pp. 59--86, 1970.

\bibitem{IEEE-TAS_paper1}
P.~Masson, ``{Modeling High-Temperature Superconductors: Needs, Issues and
  Challenges},'' \emph{IEEE Transactions on Applied Superconductivity}, 2012.

\bibitem{Takagi:TMAG90}
T.~Takagi, M.~Hashimoto, S.~Arita, S.~Norimatsu, T.~Sugiura, and K.~Miya,
  ``{Experimental verification of 3D eddy current analysis code using
  T-method},'' \emph{IEEE Transactions on Magnetics}, vol.~26, no.~2, pp.
  474--477, 1990.

\bibitem{Wang:TMAG99}
Z.~Wang, W.~Huang, W.~Jia, Q.~Zhao, Y.~Wang, and W.~Yan, ``{3D Multifields FEM
  Computation of Transverse Flux Induction Heating for Moving-Strips},''
  \emph{IEEE Transactions on Magnetics}, vol.~35, no.~3, pp. 1642--1645, 1979.

\bibitem{Ferreira:TEDU88}
J.~A. Ferreina, ``{Application of the Poynting Vector for Power Conditioning
  and Conversion},'' \emph{IEEE Transactions on Education}, vol.~31, no.~4, pp.
  257--264, 1988.

\bibitem{Oomen:Thesis00}
M.~Oomen, ``{AC loss in superconducting tapes and cables},'' Ph.D.
  dissertation, University of Twente, available online:
  http://purl.utwente.nl/publications/23468, 2000.

\bibitem{Kalimov:TMAG97}
A.~Kalimov, S.~Vaznov, and T.~Voronina, ``{Eddy Current Calculation Using
  Finite Element Method with Boundary Conditions of Integral Type},''
  \emph{IEEE Transactions on Magnetics}, vol.~33, no.~2, pp. 1326--1329, 1997.

\bibitem{Salon:TMAG85}
S.~J. Salon, ``{The hybrid finite element-boundary element method in
  electromagnetics},'' \emph{IEEE Transactions on Magnetics}, vol.~21, no.~5,
  pp. 1829--1834, 1985.

\bibitem{Darve:JCP00}
E.~Darve, ``{The Fast Multipole Method: Numerical Implementation},''
  \emph{Journal of Computational Physics}, vol. 160, pp. 195--240, 2000.

\bibitem{Albanese:TM88}
R.~Albanese and G.~Rubinacci, ``{Solution of three dimensional eddy current
  problems by integral and differential methods},'' \emph{IEEE Transactions on
  Magnetics}, vol.~24, no.~1, pp. 98--101, 1988.

\bibitem{Barnes:Cryo05}
P.~N. Barnes, M.~D. Sumption, and G.~L. Rhoads, ``{Review of high power density
  superconducting generators: Present state and prospects for incorporating
  YBCO windings},'' \emph{Cryogenics}, vol.~45, pp. 670--686, 2005.

\bibitem{Vase:SST00}
P.~Vase, R.~Fl{\"u}kiger, M.~Leghissa, and B.~Glowacki, ``{Current status of
  high-$T_c$ wire},'' \emph{Superconductor Science and Technology}, vol.~13,
  pp. R71--R84, 2000.

\bibitem{Stavrev:PhysC98b}
S.~Stavrev and B.~Dutoit, ``Frequency dependence of {AC} loss in
  {Bi(2223)/Ag-sheathed} tapes,'' \emph{Physica C}, vol. 310, pp. 86--89, 1998.

\bibitem{Stavrev:PhysC98a}
S.~Stavrev, B.~Dutoit, N.~Nibbio, and L.~{Le Lay}, ``Eddy current self-field
  loss in {Bi-2223} tapes with {a.c.} transport current,'' \emph{Physica C},
  vol. 307, pp. 105--116, 1998.

\bibitem{Paasi:TMAG96}
J.~Paasi, M.~Laforest, D.~Aized, S.~Fleshler, G.~Snitchler, and A.~P.
  Malozemoff, ``{AC Losses in Multifilamentary Bi-2223/Ag Superconducting
  Tapes},'' \emph{IEEE Transactions on Magnetics}, vol.~32, no.~4, pp.
  2792--2795, 1996.

\bibitem{Magnusson:PhysC01}
N.~Magnusson, ``{Semi-empirical model of the losses in HTS tapes carrying AC
  currents in AC magnetic fields applied parallel to the tape face},''
  \emph{Physica C}, vol. 349, pp. 225--234, 2001.

\bibitem{Majoros:SST07}
M.~Majoros, L.~Ye, A.~V. Velichko, T.~A. Coombs, M.~D. Sumption, and E.~W.
  Colling, ``{Transport AC losses in YBCO coated conductors},''
  \emph{Superconductor Science and Technology}, vol.~20, pp. S299--S304, 2007.

\bibitem{Duckworth:TAS05b}
R.~C. Duckworth, M.~J. Gouge, J.~W. Lue, C.~L.~H. Thieme, and D.~T. Verebelyi,
  ``{Substrate and Stabilization Effects on the Transport AC Losses in YBCO
  Coated Conductors},'' \emph{IEEE Transactions on Applied Superconductivity},
  vol.~15, no.~2, pp. 1583--1586, 2005.

\bibitem{Stavrev:Thesis02}
S.~Stavrev, ``{Modelling of high temperature superconductors for ac power
  applications},'' Ph.D. dissertation, \'Ecole Polytechnique F\'ed\'erale de
  Lausanne, available online: http://dx.doi.org/10.5075/epfl-thesis-2579, 2002.

\bibitem{Sumption:SST05}
M.~Sumption, E.~Collings, and P.~Barnes, ``{AC} loss in striped (filamentary)
  {YBCO} coated conductors leading to designs for high frequencies and
  field-sweep amplitudes,'' \emph{Superconductor Science and Technology},
  vol.~18, pp. 122--134, 2005.

\bibitem{Friend:TAS99}
C.~M. Friend, C.~Beduz, B.~Dutoit, R.~Navarro, E.~Cereda, and
  J.~Alonso-Llorente, ``{A European Project on the AC Losses of Bi-2223 Tapes
  for Power Applications},'' \emph{IEEE Transactions on Applied
  Superconductivity}, vol.~9, no.~2, pp. 1165--1168, 1999.

\bibitem{Takacs:SST97}
S.~Tak{\'a}cs, ``{AC losses in superconducting cables and their expected values
  in magnetic systems},'' \emph{Superconductor Science and Technology},
  vol.~10, pp. 733--748, 1997.

\bibitem{Hlasnik:TMAG81}
I.~Hl{\'a}snik, ``{Review on AC losses in superconductors},'' \emph{IEEE
  Transactions on Magnetics}, vol.~17, no.~5, pp. 2261--2269, 1981.

\bibitem{Gomory:SST06a}
F.~G{\"o}m{\"o}ry, J.~{\v S}ouc, M.~Vojen{\v c}iak, E.~Seiler, B.~Klin{\v c}ok,
  J.~M. Ceballos, E.~Pardo, A.~Sanchez, C.~Navau, S.~Farinon, and
  P.~Fabbricatore, ``{Predicting AC loss in practical superconductors},''
  \emph{Superconductor Science and Technology}, vol.~19, pp. S60--S66, 2006.

\bibitem{Amemiya:TAS07b}
N.~Amemiya, F.~Kimura, and T.~Ito, ``{Total AC Loss in Twisted Multifilamentary
  Coated Conductors Carrying AC Transport Current in AC Transverse Magnetic
  Field},'' \emph{IEEE Transactions on Applied Superconductivity}, vol.~17,
  no.~2, pp. 3183--3186, 2007.

\bibitem{Zola:SST04}
D.~Zola, F.~G{\"o}m{\"o}ry, M.~Polichetti, F.~Str{\'y}{\v c}ek, E.~Seiler,
  I.~Hu{\v s}ek, P.~Kov{\'a}{\v c}, and S.~Pace, ``{A study of coupling loss on
  bi-columnar BSCCO/Ag tapes through ac susceptibility measurements},''
  \emph{Superconductor Science and Technology}, vol.~17, pp. 501--511, 2004.

\bibitem{Campbell:JSNM11}
A.~M. Campbell, ``{An Introduction to Numerical Methods in Superconductors},''
  \emph{Journal of Superconductivity and Novel Magnetism}, vol.~24, pp. 27--33,
  2011.

\bibitem{Costa:TAS03}
M.~Costa, E.~Martinez, C.~Beduz, Y.~Yang, F.~Grilli, B.~Dutoit, E.~Vinot, and
  P.~Tixador, ``{3D Modeling of Coupling Between Superconducting Filaments via
  Resistive Matrix in AC Magnetic Field},'' \emph{IEEE Transactions on Applied
  Superconductivity}, vol.~13, no.~2, pp. 3634--3637, 2003.

\bibitem{CostaBouzo:SST04}
M.~{Costa Bouzo}, F.~Grilli, and Y.~Yang, ``{Modelling of coupling between
  superconductors of finite length using an integral formulation},''
  \emph{Superconductor Science and Technology}, vol.~17, no.~10, pp.
  1103--1112, 2004.

\bibitem{Grilli:SST03}
F.~Grilli, M.~C. Bouzo, Y.~Yang, C.~Beduz, and B.~Dutoit, ``{Finite element
  method analysis of the coupling effect between superconducting filaments of
  different aspect ratio},'' \emph{Superconductor Science and Technology},
  vol.~16, no.~10, pp. 1228--1234, 2003.

\bibitem{Lousberg:PICACOMEN08}
G.~P. Lousberg, M.~Ausloos, C.~Geuzaine, P.~Dular, P.~Vanderbemden, and
  B.~Vanderheyden, ``{Simulation of the highly non linear properties of bulk
  superconductors: finite element approach with a backward Euler method and a
  single time step},'' \emph{Proceedings of the Fourth International
  International Conference on Advanced Computational Methods in Engineering},
  2008.

\bibitem{Mikitik:LTP10}
G.~P. Mikitik, ``{Critical states in thin planar type-II superconductors in a
  perpendicular or inclined magnetic field (Review)},'' \emph{Low Temperature
  Physics}, vol.~36, no.~1, pp. 17--49, 2010.

\bibitem{Wilson:Cryo08}
M.~N. Wilson, ``{NbTi superconductors with low ac loss: A review},''
  \emph{Cryogenics}, vol.~48, pp. 381--395, 2008.

\bibitem{Abraimov:SST08}
D.~Abraimov, A.~Gurevich, A.~Polyanskii, X.~Y. Cai, A.~Xu, S.~Pamidi,
  D.~Larbalestier, and C.~L.~H. Thieme, ``{Significant reduction of AC losses
  in YBCO patterned coated conductors with transposed filaments},''
  \emph{Superconductor Science and Technology}, vol.~21, p. 082004, 2008.

\bibitem{Ashworth:SST06}
S.~P. Ashworth and F.~Grilli, ``{A strategy for the reduction of ac losses in
  YBCO coated conductors},'' \emph{Superconductor Science and Technology},
  vol.~19, no.~2, pp. 227--232, 2006.

\bibitem{Carr:TAS99}
W.~J. {Carr Jr.} and C.~E. Oberly, ``{Filamentary YBCO Conductors For AC
  applications},'' \emph{IEEE Transactions on Applied Superconductivity},
  vol.~9, no.~2, pp. 1475--1478, 1999.

\bibitem{Tsukamoto:TAS05b}
O.~Tsukamoto, N.~Sekine, M.~Ciszek, and J.~Ogawa, ``{A Method to Reduce
  Magnetization Losses in Assembled Conductors Made of YBCO Coated
  Conductors},'' \emph{IEEE Transactions on Applied Superconductivity},
  vol.~15, no.~2, pp. 2823--2826, 2005.

\bibitem{Polak:SST07}
M.~Polak, J.~Kvitkovic, P.~Mozola, E.~Usak, P.~N. Barnes, and G.~A. Levin,
  ``{Frequency dependence of hysteresis loss in YBCO tapes},''
  \emph{Superconductor Science and Technology}, vol.~20, pp. S293--S298, 2007.

\bibitem{Seo:Cryo01}
K.~Seo, K.~Fukuhara, and M.~Hasegawa, ``{Analyses for inter-strand coupling
  loss in multi-strand superconducting cable with distributed contact
  resistance between strands},'' \emph{Cryogenics}, vol.~41, pp. 131--137,
  2001.

\bibitem{Goldacker_SST09}
W.~Goldacker, A.~Frank, A.~Kudymow, R.~Heller, A.~Kling, S.~Terzieva, and
  C.~Schmidt, ``{Status of high transport current ROEBEL assembled coated
  conductor cables},'' \emph{Superconductor Science and Technology}, vol.~22,
  p. 034003, 2009.

\bibitem{Pasztor:TAS04}
G.~Pasztor, P.~Bruzzone, A.~Anghel, and B.~Stepanov, ``{An Alternative CICC
  Design Aimed at Understanding Critical Performance Issues in $\rm Nb_3Sn$
  Conductors for ITER},'' \emph{IEEE Transactions on Applied
  Superconductivity}, vol.~14, no.~2, pp. 1527--1530, 2004.

\bibitem{Kasai:TMAG05}
S.~Kasai and N.~Amemiya, ``{Numerical Analysis of Magnetization Loss in
  Finite-Length Multifilamentary YBCO Coated Conductors},'' \emph{IEEE
  Transactions on Applied Superconductivity}, vol.~15, no.~2, pp. 2855--2858,
  2005.

\bibitem{Paasi:TAS97}
J.~Paasi and M.~Lahtinen, ``{Computational Comparison of AC Losses in Different
  Kinds of HTS Composite Conductors},'' \emph{IEEE Transactions on Applied
  Superconductivity}, vol.~7, no.~2, pp. 322--325, 1997.

\bibitem{Genenko:PRB00}
Y.~A. Genenko, A.~Snezhko, and H.~C. Freyhardt, ``{Overcritical states of a
  superconductor strip in a magnetic environment},'' \emph{Physical Review B},
  vol.~62, no.~5, pp. 3453--3472, 2000.

\bibitem{Genenko:JAP02}
Y.~A. Genenko and A.~Snezhko, ``{Superconductor strip near a magnetic wall of
  finite thicknes},'' \emph{Journal of Applied Physics}, vol.~92, no.~1, pp.
  357--360, 2002.

\bibitem{Genenko:PhysC04}
Y.~A. Genenko, ``{Strong reduction of ac losses in a superconductor strip
  located between superconducting ground plates},'' \emph{Physica C}, vol. 401,
  pp. 210--213, 2004.

\bibitem{Mawatari:PRB08}
Y.~Mawatari, ``Magnetic field distributions around superconducting strips on
  ferromagnetic substrates,'' \emph{Physical Review B}, vol.~77, p. 104505,
  2008.

\bibitem{Suenaga:PhysC08}
M.~Suenaga, M.~Iwakuma, T.~Sueyoshi, T.~Izumi, M.~Mimura, Y.~Takahashi, and
  Y.~Aoki, ``{Effects of a ferromagnetic substrate on hysteresis losses of a
  $\rm YBa_2Cu_3O_7$ coated conductor in perpendicular ac applied magnetic
  fields},'' \emph{Physica C}, vol. 468, pp. 1714--1717, 2008.

\bibitem{Nguyen:JAP09b}
D.~N. Nguyen, S.~P. Ashworth, and J.~O. Willis, ``{Experimental and
  finite-element method studies of the effects of ferromagnetic substrate on
  the total ac loss in a rolling-assisted biaxially textured substrate $\rm
  YBa_2Cu_3O_7$ tape exposed to a parallel ac magnetic field},'' \emph{Journal
  of Applied Physics}, vol. 106, p. 093913, 2009.

\bibitem{Nguyen:SST10}
D.~N. Nguyen, S.~P. Ashworth, J.~O. Willis, F.~Sirois, and F.~Grilli, ``{A new
  finite-element method simulation model for computing AC loss in roll assisted
  biaxially textured substrate YBCO tapes},'' \emph{Superconductor Science and
  Technology}, vol.~23, p. 025001, 2010.

\bibitem{Miyagi:PhysC08}
D.~Miyagi, Y.~Yunoki, M.~Umabuchi, N.~Takahashi, and O.~Tsukamoto,
  ``Measurement of magnetic properties of {Ni}-alloy substrate of {HTS} coated
  conductor in {{\rm $LN_2$}},'' \emph{Physica C}, vol. 468, pp. 1743--1746,
  2008.

\bibitem{Nguyen:SST11}
D.~N. Nguyen, J.~Y. Coulter, J.~O. Willis, S.~P. Ashworth, H.~P. Kraemer,
  W.~Schmidt, B.~Carter, and A.~Otto, ``{AC loss and critical current
  characterization of a noninductive coil of two-in-hand RABiTS YBCO tape for
  fault current limiter applications},'' \emph{Superconductor Science and
  Technology}, vol.~24, p. 035017, 2011.

\bibitem{Zhang:SST12b}
M.~Zhang, J.~Kvitkovic, S.~V. Pamidi, and T.~A. Coombs, ``{Experimental and
  numerical study of a YBCO pancake coil with a magnetic substrate},''
  \emph{Superconductor Science and Technology}, vol.~25, p. 125020, 2012.

\bibitem{Zhang:APL12}
M.~Zhang, J.~Kvitkovic, J.-H. Kim, C.-H. Kim, S.~V. Pamidi, and T.~A. Coombs,
  ``{Alternating current loss of second-generation high-temperature
  superconducting coils with magnetic and non-magnetic substrate},''
  \emph{Applied Physics Letters}, vol. 101, no.~10, p. 102602, 2012.

\bibitem{Nguyen:unpublished}
D.~Nguyen, S.~Ashworth, and S.~Fleshler, ``{Numerical calculation for AC losses
  in a three phase tri-axial cable of RABiTS YBCO tapes},'' \emph{Unpublished}.

\bibitem{Miyagi:TAS07}
D.~Miyagi, Y.~Amadutsumi, N.~Takahashi, and O.~Tsukamoto, ``{FEM} analysis of
  effect of magnetism of substrate on {AC} transport current loss of {HTS}
  conductor with ferromagnetic substrate,'' \emph{IEEE Transactions on Applied
  Superconductivity}, vol.~17, no.~2, pp. 3167--3170, 2007.

\bibitem{Miyagi:TAS08}
D.~Miyagi, M.~Umabuchi, N.~Takahashi, and O.~Tsukamoto, ``Analysis of effect of
  nonlinear magnetic property of magnetic substrate on {AC} transport current
  loss of {HTS} coated conductor using {FEM},'' \emph{IEEE Transactions on
  Applied Superconductivity}, vol.~18, no.~2, pp. 1297--1300, 2008.

\bibitem{Gomory:TAS12}
F.~{G\"om\"ory} and F.~Inanir, ``{AC Losses in Coil Wound From Round Wire
  Coated by a Superconducting Layer},'' \emph{IEEE Transactions on Applied
  Superconductivity}, vol.~22, no.~3, p. 4704704, 2012.

\bibitem{Genenko:SST09}
Y.~A. Genenko, H.~Rauh, P.~Kr{\"u}ger, and N.~Narayanan, ``{Finite-element
  simulations of overcritical states of a magnetically shielded superconductor
  strip},'' \emph{Superconductor Science and Technology}, vol.~22, no.~5, p.
  055001, 2009.

\bibitem{Genenko:APL11}
Y.~A. Genenko, H.~Rauh, and P.~Kr{\"u}ger, ``{Finite-element simulations of
  hysteretic ac losses in a bilayer superconductor/ferromagnet heterostructure
  subject to an oscillating transverse magnetic field},'' \emph{Applied Physics
  Letters}, vol.~98, no.~15, p. 152508, 2011.

\bibitem{Farinon:SST10}
S.~Farinon, P.~Fabbricatore, and F.~G{\"o}m{\"o}ry, ``{Critical state and
  magnetization loss in multifilamentary superconducting wire solved through
  the commercial finite element code ANSYS},'' \emph{Superconductor Science and
  Technology}, vol.~23, p. 115004, 2010.

\bibitem{Farinon:JSNM12}
S.~Farinon, P.~Fabbricatore, F.~Grilli, and P.~A.~C. Kr{\"u}ger,
  ``{Applicability of the adaptive resistivity method to describe the critical
  state of complex superconducting systems},'' \emph{Journal of
  Superconductivity and Novel Magnetism}, 2012.

\bibitem{PriceLN2}
\BIBentryALTinterwordspacing
 [Online]. Available: \url{http://hypertextbook.com/facts/2007/KarenFan.shtml}
\BIBentrySTDinterwordspacing

\bibitem{Woodcraft:07}
\BIBentryALTinterwordspacing
A.~L. Woodcraft. (2007) {An introduction to cryogenics}. [Online]. Available:
  \url{http://uk.lowtemp.org/1-Woodcraft.pdf}
\BIBentrySTDinterwordspacing

\bibitem{Gouge:CryoAssessRpt02}
M.~J. Gouge, J.~A. Demko, B.~W. McConnell, and J.~M. Pfotenhauer, ``{Cryogenics
  Assessment Report},'' Oak Ridge National Laboratory,
  http://www.ornl.gov/sci/htsc/documents/pdf/CryoAssessRpt.pdf, Tech. Rep.,
  2002.

\bibitem{Hassenzahl:PIEEE04}
W.~V. Hassenzahl, D.~W. Hazelton, B.~K. Johnson, P.~Komarek, M.~Noe, and C.~T.
  Reis, ``{Electric power applications of superconductivity},''
  \emph{Proceedings of the IEEE}, vol.~92, no.~10, pp. 1655--1674, 2004.

\bibitem{Zushi:Cryo05}
Y.~Zushi, I.~Asaba, J.~Ogawa, K.~Yamagishi, and O.~Tsukamoto, ``{AC losses in
  HTS bulk and their influence on trapped magnetic field},'' \emph{Cryogenics},
  vol.~45, no.~1, pp. 17--22, 2005.

\bibitem{Amemiya:SST04b}
N.~Amemiya, S.~Kasai, K.~Yoda, Z.~Jiang, G.~A. Levin, P.~N. Barnes, and C.~E.
  Oberly, ``{AC loss reduction of YBCO coated conductors by multifilamentary
  structure},'' \emph{Superconductor Science and Technology}, vol.~17, pp.
  1464--1471, 2004.

\bibitem{Sumption:TAS05}
M.~D. Sumption, P.~N. Barnes, and E.~W. Collings, ``{AC losses of coated
  conductors in perpendicular fields and concepts for twisting},'' \emph{IEEE
  Transactions on Applied Superconductivity}, vol.~15, no.~2, pp. 2815--2818,
  2005.

\bibitem{Marchevsky:TAS09}
M.~Marchevsky, E.~Zhang, Y.~Xie, V.~Selvamanickam, and P.~G. Ganesan, ``{AC
  Losses and Magnetic Coupling in Multifilamentary 2G HTS Conductors and Tape
  Arrays},'' \emph{IEEE Transactions on Applied Superconductivity}, vol.~19,
  no.~3, pp. 3094--3097, 2009.

\bibitem{Ainslie:COMP11}
M.~D. Ainslie, T.~J. Flack, Z.~Hong, and T.~A. Coombs, ``{Comparison of first-
  and second-order 2D finite element models for calculating AC loss in high
  temperature superconductor coated conductors},'' \emph{COMPEL: The
  International Journal for Computation and Mathematics in Electrical and
  Electronic Engineering}, vol.~30, no.~2, pp. 762--774, 2011.

\bibitem{Ainslie:PhysC10}
M.~D. Ainslie, Y.~Jiang, W.~Xian, Z.~Hong, W.~Yuan, R.~Pei, T.~J. Flack, and
  T.~A. Coombs, ``{Numerical analysis and finite element modelling of an HTS
  synchronous motor},'' \emph{Physica C}, vol. 470, no.~20, pp. 1752--1755,
  2010.

\bibitem{Sugimoto:TAS07}
H.~Sugimoto, T.~Tsuda, T.~Morishita, Y.~Hondou, T.~Takeda, H.~Togawa, T.~Oota,
  K.~Ohmatsu, and S.~Yoshida, ``{Development of an Axial Flux Type PM
  Synchronous Motor With the Liquid Nitrogen Cooled HTS Armature Windings},''
  \emph{IEEE Transactions on Applied Superconductivity}, vol.~17, no.~2, pp.
  1637 --1640, 2007.

\bibitem{Chen:TAS12}
Y.~Chen, W.~Yuan, M.~Zhang, and T.~A. Coombs, ``{The Experiment to Evaluate the
  AC Loss of 2G HTS Windings in the Application of Rotating Electric
  Machines},'' \emph{IEEE Transactions on Applied Superconductivity}, To be
  published 2012.

\bibitem{Nick:PhysC12}
W.~Nick, J.~Grundmann, and J.~Frauenhofer, ``{Test Results from Siemens
  Low-Speed, High-Torque HTS Machine and Description of further Steps towards
  Commercialization of HTS Machines},'' \emph{Physica C}, 2012.

\bibitem{Gouge:TAS05}
M.~Gouge, D.~Lindsay, J.~Demko, R.~Duckworth, A.~Ellis, P.~Fisher, D.~James,
  J.~Lue, M.~Roden, I.~Sauers, J.~Tolbert, C.~Traeholt, and D.~Willen, ``{Tests
  of tri-axial HTS cables},'' \emph{IEEE Transactions on Applied
  Superconductivity}, vol.~15, no.~2, pp. 1827--1830, 2005.

\bibitem{Noji:SST97}
H.~Noji, ``{AC} loss of a high-{$\rm T_c$} superconducting power-cable
  conductor,'' \emph{Superconductor Science and Technology}, vol.~10, pp.
  552--556, 1997.

\bibitem{Noji:PhysC03}
H.~Noji, K.~Haji, and T.~Hamada, ``{AC} loss analysis of 114 {MVA} high-{$\rm
  T_c$} superconducting model cable,'' \emph{Physica C}, vol. 392-396, pp.
  1134--1139, 2003.

\bibitem{Clem:SST10}
J.~R. Clem and A.~P. Malozemoff, ``{Theory of ac loss in power transmission
  cables with second generation high temperature superconductor wires},''
  \emph{Superconductor Science and Technology}, vol.~23, p. 034014, 2010.

\bibitem{Malozemoff:TAS09}
A.~Malozemoff, G.~Snitchler, and Y.~Mawatari, ``Tape-width dependence of {AC}
  losses in {HTS} cables,'' \emph{IEEE Transactions on Applied
  Superconductivity}, vol.~19, no.~3, pp. 3115--3118, 2009.

\bibitem{Fukui:TAS06b}
S.~Fukui, R.~Kojima, J.~Ogawa, M.~Yamaguchi, T.~Sato, and O.~Tsukamoto,
  ``{Numerical Analysis of AC Loss Characteristics of Cable Conductor Assembled
  by HTS Tapes in Polygonal Arrangement},'' \emph{IEEE Transactions on Antennas
  and Propagation}, vol.~16, no.~2, pp. 143--146, 2006.

\bibitem{Li:SST10}
Q.~Li, N.~Amemiya, K.~Takeuchi, T.~Nakamura1, and N.~Fujiwara, ``{AC loss
  characteristics of superconducting power transmission cables: gap effect and
  {$\rm J_c$} distribution effect},'' \emph{Superconductor Science and
  Technology}, vol.~23, p. 115003, 2010.

\bibitem{Takayasu:SST12}
{M. Takayasu and L. Chiesa and L. Bromberg and J. V. Minervini}, ``{HTS twisted
  stacked-tape cable conductor},'' \emph{Superconductor Science and
  Technology}, vol.~25, no.~1, p. 014011, 2012.

\bibitem{Staines:SST12}
M.~Staines, N.~Glasson, M.~Pannu, K.~P. Thakur, R.~Badcock, N.~Allpress,
  P.~D'Souza, and E.~Talantsev, ``{The development of a Roebel cable based 1
  MVA HTS transformer},'' \emph{Superconductor Science and Technology},
  vol.~25, no.~1, p. 014002, 2012.

\bibitem{Schlachter:TAS11}
S.~I. Schlachter, W.~Goldacker, F.~Grilli, R.~Heller, and A.~Kudymow, ``{Coated
  Conductor Rutherford Cables (CCRC) for High-Current Applications: Concept and
  Properties},'' \emph{IEEE Transactions on Applied Superconductivity},
  vol.~21, no.~3, pp. 3021--3024, 2011.

\bibitem{Noe:TAS01}
M.~Noe, K.-P. Juengst, F.~Werfel, L.~Cowey, A.~Wolf, and S.~Elschner,
  ``{Investigation of high-Tc bulk material for its use in resistive
  superconducting fault current limiters},'' \emph{IEEE Transactions on Applied
  Superconductivity}, vol.~11, no.~1, pp. 1960--1963, 2001.

\bibitem{Kudymow:TAS09}
A.~Kudymow, C.~Schacherer, M.~Noe, and W.~Goldacker, ``{Experimental
  Investigation of Parallel Connected YBCO Coated Conductors for Resistive
  Fault Current Limiters},'' \emph{IEEE Transactions on Applied
  Superconductivity}, vol.~19, no.~3, pp. 1806--1809, 2009.

\bibitem{Hong:TAS11a}
Z.~Hong, Z.~Jin, M.~Ainslie, J.~Sheng, W.~Yuan, and T.~A. Coombs, ``{Numerical
  Analysis of the Current and Voltage Sharing Issues for Resistive Fault
  Current Limiter Using YBCO Coated Conductors},'' \emph{IEEE Transactions on
  Applied Superconductivity}, vol.~21, no.~3, pp. 1198--1201, 2011.

\bibitem{Noe:SST07}
M.~Noe and M.~Steurer, ``High-temperature superconductor fault current
  limiters: concepts, applications, and development status,''
  \emph{Superconductor Science and Technology}, vol.~20, pp. R15--R29, 2007.

\bibitem{Dommerque:SST10}
R.~Dommerque, S.~Kr{\"a}mer, A.~Hobl, R.~B{\"o}hm, M.~Bludau, J.~Bock,
  D.~Klaus, H.~Piereder, A.~Wilson, T.~Kr{\"u}ger, G.~Pfeiffer, K.~Pfeiffer,
  and S.~Elschner, ``{First commercial medium voltage superconducting
  fault-current limiters: production, test and installation},''
  \emph{Superconductor Science and Technology}, vol.~23, no.~3, p. 034020,
  2010.

\bibitem{Heydari:SST08}
H.~Heydari, F.~Faghihi, and R.~B. Aligholizadeh, ``{A new approach for AC loss
  reduction in HTS transformer using auxiliary windings, case study: 25 kA HTS
  current injection transformer},'' \emph{Superconductor Science and
  Technology}, vol.~21, no.~1, p. 015009, 2008.

\bibitem{AlMosawi:TAS01}
M.~K. Al-Mosawi, C.~Beduz, Y.~Yang, M.~Webb, and A.~Power, ``{The effect of
  flux diverters on AC losses of a 10 kVA high temperature superconducting
  demonstrator transformer},'' \emph{IEEE Transactions on Applied
  Superconductivity}, vol.~11, no.~1, pp. 2800--2803, 2001.

\bibitem{Funaki:Cryo98}
K.~Funaki, M.~Iwakuma, K.~Kajikawa, M.~Takeo, J.~Suehiro, M.~Hara, K.~Yamafuji,
  M.~Konno, Y.~Kasagawa, K.~Okubo, Y.~Yasukawa, S.~Nose, M.~Ueyama, K.~Hayashi,
  and K.~Sato, ``{Development of a 500 kVA-class oxide-superconducting power
  transformer operated at liquid-nitrogen temperature},'' \emph{Cryogenics},
  vol.~38, no.~2, pp. 211-- 220, 1998.

\bibitem{Iwakuma:TAS01}
M.~Iwakuma, K.~Funaki, K.~Kajikawa, H.~Tanaka, T.~Bohno, A.~Tomioka, H.~Yamada,
  S.~Nose, M.~Konno, Y.~Yagi, H.~Maruyama, T.~Ogata, S.~Yoshida, K.~Ohashi,
  K.~Tsutsumi, and K.~Honda, ``{AC loss properties of a 1 MVA single-phase HTS
  power transformer},'' \emph{IEEE Transactions on Applied Superconductivity},
  vol.~11, no.~1, pp. 1482--1485, 2001.

\bibitem{Tixador:TAS05}
P.~Tixador, B.~Bellin, M.~Deleglise, J.~Vallier, C.~Bruzek, S.~Pavard, and
  J.~Saugrain, ``{Design of a 800 kJ HTS SMES},'' \emph{IEEE Transactions on
  Applied Superconductivity}, vol.~15, no.~2, pp. 1907--1910, 2005.

\bibitem{Kim:TAS11a}
K.~Kim, A.-R. Kim, J.-G. Kim, M.~Park, I.-K. Yu, M.-H. Sohn, B.-Y. Eom, K.~Sim,
  S.~Kim, H.-J. Kim, J.-H. Bae, and K.-C. Seong, ``{Analysis of Operational
  Loss Characteristics of 10 kJ Class Toroid-Type SMES},'' \emph{IEEE
  Transactions on Applied Superconductivity}, vol.~21, no.~3, pp. 1340--1343,
  2011.

\bibitem{Zhu:TAS11}
J.~Zhu, Q.~Cheng, B.~Yang, W.~Yuan, T.~Coombs, and M.~Qiu, ``{Experimental
  Research on Dynamic Voltage Sag Compensation Using 2G HTS SMES},'' \emph{IEEE
  Transactions on Applied Superconductivity}, vol.~21, no.~3, pp. 2126--2130,
  2011.

\bibitem{Zhu:PhysC11}
J.~Zhu, W.~Yuan, T.~Coombs, and Q.~Ming, ``{Simulation and experiment of a YBCO
  SMES prototype in voltage sag compensation},'' \emph{Physica C}, vol. 471,
  no. 5-6, pp. 199--204, 2011.

\bibitem{Yuan:TAS10}
W.~Yuan, W.~Xian, M.~Ainslie, Z.~Hong, Y.~Yan, R.~Pei, Y.~Jiang, and T.~A.
  Coombs, ``{Design and Test of a Superconducting Magnetic Energy Storage
  (SMES) Coil},'' \emph{IEEE Transactions on Applied Superconductivity},
  vol.~20, no.~3, pp. 1379--1382, 2010.

\bibitem{Park:TAS07}
M.-J. Park, S.-Y. Kwak, W.-S. Kim, S.-W. Lee, J.-K. Lee, J.-H. Han, K.-D. Choi,
  H.-K. Jung, K.-C. Seong, and S.-Y. Hahn, ``{AC Loss and Thermal Stability of
  HTS Model Coils for a 600 kJ SMES},'' \emph{IEEE Transactions on Applied
  Superconductivity}, vol.~17, no.~2, pp. 2418--2421, 2007.

\bibitem{Park:Cryo07}
M.-J. Park, S.-Y. Kwak, W.-S. Kim, S.-W. Lee, J.-K. Lee, K.-D. Choi, H.-K.
  Jung, K.-C. Seong, and S.-Y. Hahn, ``{Analysis of magnetic field distribution
  and AC losses of a 600 kJ SMES},'' \emph{Cryogenics}, vol.~47, no. 7-8, pp.
  391--396, 2007.

\bibitem{Kolacek:PRL01}
J.~Kol{\'a}{\v{c}}ek, P.~Lipavsk{\`y}, and E.~H. Brandt, ``{Charge profile in
  vortices},'' \emph{Physical Review Letters}, vol.~86, no.~2, pp. 312--315,
  2001.

\bibitem{Pearl:APL64}
J.~Pearl, ``{Current distribution in superconducting films carrying quantized
  fluxoids},'' \emph{Applied Physics Letters}, vol.~5, pp. 65--66, 1964.

\end{thebibliography}
\end{document}